\pdfoutput=1
\documentclass[10pt]{amsart}
\usepackage[utf8]{inputenc}

\usepackage{amsmath, amssymb, amsthm, nicefrac, mathtools, url, tcolorbox}

\usepackage{comment}
\usepackage{verbatim}
\usepackage{wasysym}
\usepackage{graphicx}
\usepackage{caption}
\usepackage{import}
\usepackage{latexsym}
\usepackage{cite}
\usepackage{enumerate}
\usepackage{float}
\usepackage{amsthm}
\usepackage{upgreek}
\usepackage{nicefrac}
\usepackage{multicol}
\usepackage{rotating}
\usepackage{array, multirow}
\usepackage{tabularx}
\usepackage{lipsum}
\usepackage{microtype}
\usepackage{vwcol} 

\usepackage{etoolbox}
\makeatletter
\pretocmd{\chapter}{\addtocontents{toc}{\protect\addvspace{15\p@}}}{}{}
\pretocmd{\section}{\addtocontents{toc}{\protect\addvspace{5\p@}}}{}{}
\makeatother
\let\oldtocsection=\tocsection
\let\oldtocsubsection=\tocsubsection
\let\oldtocsubsubsection=\tocsubsubsection
\renewcommand{\tocsection}[2]{\hspace{0em}\oldtocsection{#1}{#2}}
\renewcommand{\tocsubsection}[2]{\hspace{1.8em}\oldtocsubsection{#1}{#2}}
\renewcommand{\tocsubsubsection}[2]{\hspace{4.4em}\oldtocsubsubsection{#1}{#2}}

\usepackage[svgnames]{xcolor}
\definecolor{linkcolor}{HTML}{e88d67} 
\definecolor{citecolor}{HTML}{e88d67} 
\definecolor{urlcolor}{HTML}{e88d67} 
\definecolor{myNewColorA}{HTML}{fec3a6}
\definecolor{myNewColorB}{HTML}{ffaf80}
\definecolor{myNewColorC}{HTML}{fb8f67}
\definecolor{seagreen}{HTML}{337180}
\definecolor{mseagreen}{HTML}{369673}
\definecolor{darksalmon}{HTML}{e88d67}
\definecolor{silver}{HTML}{bbbbbb}
\definecolor{flowerblue}{HTML}{4e77fc}
\definecolor{tomato}{HTML}{ff6347}
\definecolor{orange}{HTML}{f2b13d}
\definecolor{darkgray}{HTML}{939393}

\usepackage[
  bookmarks = true,
  bookmarksopen = true,
  colorlinks = , 
  urlcolor = urlcolor, 
  linkcolor = linkcolor, 
  citecolor = citecolor, 
  linktoc = all,
  pagebackref = true
]{hyperref} 
\mathtoolsset{showonlyrefs,showmanualtags}

\DeclareMathOperator{\Mat}{Mat}

\DeclareMathOperator{\Pic}{Pic}
\DeclareMathOperator{\nc}{nc}

\DeclareMathOperator{\spn}{Span}
\DeclareMathOperator{\GL}{GL}
\DeclareMathOperator{\PGL}{PGL}

\newcommand{\br}[1]{\left( #1 \right)}
\newcommand{\pbr}[1]{\left\{ #1 \right\}}
\newcommand{\lbr}[1]{\left[ #1 \right]}
\newcommand{\abr}[1]{\langle #1 \rangle}

\newcommand{\Painleve}{Painlev{\'e} }

\newcommand{\tomato}[1]{{\color{tomato} #1}}

\theoremstyle{plain}
\newtheorem{thm}{Theorem}[]

\newtheorem{lem}{Lemma}[]

\newtheorem{prop}{Proposition}[]

\newtheorem{corl}{Corollary}[]

\theoremstyle{definition}
\newtheorem{defn}{Definition}[]

\newtheorem{exmp}{Example}[]

\theoremstyle{remark}
\newtheorem{rem}{Remark}

\usepackage{chngcntr}
\counterwithin*{thm}{section} 
\counterwithin*{lem}{section} 
\counterwithin*{claim}{section} 
\counterwithin*{prop}{section} 
\counterwithin*{corl}{section} 
\counterwithin*{defn}{section} 
\counterwithin*{exmp}{section} 
\counterwithin*{ex}{section} 
\counterwithin*{rem}{section} 

\usepackage{apptools}
\AtAppendix{\counterwithin{thm}{section}}
\AtAppendix{\counterwithin{lem}{section}}
\AtAppendix{\counterwithin{claim}{section}}
\AtAppendix{\counterwithin{prop}{section}}
\AtAppendix{\counterwithin{corl}{section}}
\AtAppendix{\counterwithin{defn}{section}}
\AtAppendix{\counterwithin{exmp}{section}}
\AtAppendix{\counterwithin{ex}{section}}
\AtAppendix{\counterwithin{rem}{section}}

\allowdisplaybreaks

\usepackage{tikz}

\begin{document}

    \title[On a non-commutative sixth $q$-Painlevé system: from discrete system to surface theory]{On a non-commutative sixth $q$-Painlevé system: 
    \\
    from discrete system to surface theory}

    \author{Irina Bobrova}
    \noindent\address{\noindent 
    Max-Planck-Institute for Mathematics in the Sciences, 04103, Leipzig, Germany
    }
    \email{irrrina.bobrova@gmail.com}

    
    \maketitle
    
    \begin{flushright}
    \textit{
    ``Good theory starts with good examples''
    \\
    {\rm(V.V. Sokolov)}
    }
    \end{flushright}
    
    \begin{abstract}
    In this paper, we describe the non-commutative formal geometry underlying a certain class of discrete integrable systems. Our main example is a non-commutative analog, labeled \ref{eq:qPA3}, of the sixth $q$-Painlevé equation. The system \ref{eq:qPA3} is constructed by postulating an extended birational representation of the extended affine Weyl group $\widetilde{W}$ of type $D_5^{(1)}$ and by selecting the same translation element in $\widetilde{W}$ as in the commutative case. Starting from this non-commutative discrete system, we develop a non-commutative version of Sakai’s surface theory, which allows us to derive the same birational representation that we initially postulated. Moreover, we recover the well-known cascade of multiplicative discrete Painlevé equations rooted in \ref{eq:qPA3} and establish a connection between \ref{eq:qPA3} and the non-commutative $d$-Painlevé systems introduced in \cite{bobrova2024affine}.
    \end{abstract}
    
    \tableofcontents
    
    \section{Introduction}

This paper presents a first attempt at constructing an approach for obtaining birational representations of non-commutative versions of the discrete Painlevé equations. Since, in the commutative setting, such representations can be derived using Sakai’s surface theory, we propose a naïve generalization of this theory in the non-commutative framework and apply it to a non-commutative analog of the well-known sixth $q$-Painlevé equation. This analog is labeled here as \ref{eq:qPA3} and is written in the form
\begin{gather}
    \tag*{\ref{eq:qPA3}}
    \begin{gathered}
        q
        = (b_1 \, b_2 \, b_7 \, b_8)^{\frac14} \, (b_3 \, b_4 \, b_5 \, b_6)^{- \frac14}
        ,
        \\[1mm]
    \begin{aligned}
        \begin{aligned}
        \underline{f} \, f
        = b_7 b_8 \, (g + b_6) \, &(g + b_8)^{-1} 
        \\[1mm]
        &(g + b_5) \, (g + b_7)^{-1}
        ,
        \end{aligned}&
        &&&&&
        &
        \begin{aligned}
        \bar g \, g
        = b_3 b_4 \, (f + b_2) \, &(f + b_4)^{-1}
        \\[1mm]
        &(f + b_1) \, (f + b_3)^{-1}
        ,
        \end{aligned}
    \end{aligned}
    \end{gathered}
\end{gather}
where the parameters $b_j$ evolve according to the following rules:
\begin{align}
    &&&&
        \bar b_i 
        &= q^2 \, b_i,
        &&&&
        i 
        = 1, 2, 5, 6,&
        &&&&&
        &
        \bar b_j 
        &= q^{-2} \, b_j,
        &&&&&
        j 
        &= 3, 4, 7, 8.
    &&&&
\end{align}
Here, elements $f$ and $g$ belong to a division ring $\mathcal{R}$ equipped with a shift operator $T\br{f, g} = \br{\bar f, \bar g}$, while all the elements $b_k$, $k = 1, \dots, 8$ lie in the center $\mathcal{Z}(\mathcal{R})$ of $\mathcal{R}$
(a detailed description is given in Subsection \ref{sec:ncdsys}). 
Note that when $f \, g = g \, f$, this system becomes the sixth $q$-Painlevé equation \cite{jimbo1996q}. 
This non-commutative analog was first obtained using the affine Weyl group approach (see Theorem \ref{thm:WD5_nc} in Subsection~\ref{sec:qPA3_affW}), originally developed in the commutative case in \cite{noumi1998affine} and extended to the non-commutative setting in \cite{bobrova2024affine}. Since constructing a discrete dynamical system requires an extended birational representation of the corresponding affine Weyl group, this representation is typically postulated. However, Sakai’s surface theory offers a systematic method for constructing it. Thus, we have aimed to generalize this theory in order to derive the representation that we initially postulated.
\medskip

The celebrated Sakai surface theory \cite{sakai2001rational} was inspired by a series of papers by K. Okamoto, in which the \textit{space of initial conditions} for the differential Painlevé equations was studied \cite{okam1}, \cite{okam2}, \cite{okam3}, \cite{okam4}. Considering the space $\mathbb{P}^1 \times \mathbb{P}^1$ with coordinates $(f,g)$, K. Okamoto noticed that the dynamics of the differential Painlevé equations is not defined on the entire space due to the emergence of inaccessible points. Hence, a blow-up procedure was required. By performing a sequence of blow-ups on $\mathbb{P}^1 \times \mathbb{P}^1$, he obtained a rational surface $\mathcal{X}$, which contains inaccessible points on a divisor $\mathcal{D}$. The irreducible components $\mathcal{D}_i$ of $\mathcal{D}$ are sometimes called \textit{vertical leaves}. The space of initial conditions $\mathcal{X}\setminus\mathcal{D}$ is obtained from $\mathcal{X}$ by removing the inaccessible divisor $\mathcal{D}$. Thus, the equation becomes regular and globally defined on the surface $\mathcal{X}\setminus\mathcal{D}$.
Following this idea, H. Sakai observed that the space of initial conditions of a Painlevé system corresponds to a rational surface obtained by blowing up eight points on $\mathbb{P}^1 \times \mathbb{P}^1$ (or nine points on $\mathbb{P}^2$). With such a surface, one can associate a Picard lattice $\Pic(\mathcal{X})$ equipped with the intersection form. Taking the $(-2)$-curves $\mathcal{D}_i$, one can construct a generalized Cartan matrix of type $R$, and, therefore, the rational surface $\mathcal{X}$ can be encoded by a Dynkin diagram $\Gamma(R)$. The type $R$ determines the \textit{surface type} of the Painlevé equation. Considering the orthogonal complement $Q(R^\bot)$ of the root lattice $Q(R)$ in $\Pic(\mathcal{X})$, one obtains the \textit{symmetry type} $R^\bot$ of the equation. The extended affine Weyl group $\widetilde W(R^\bot)$ acts by reflections on the Picard lattice as Cremona isometries. This action can, in fact, be lifted to the variables $f$ and $g$, which define the dynamics This lift yields an extended birational representation of the extended affine Weyl group $\widetilde W(R^\bot)$, which is the main object of interest in our study. We provide a brief overview of this theory in Section \ref{sec:sakai}, while detailed expositions can be found in \cite{sakai2001rational}, \cite{kajiwara2017geometric}.
\medskip

The well-known differential Painlevé equations first appeared in the early 20th century during the classification of second-order ordinary differential equations whose solutions exhibit properties generalizing those of elliptic functions. P. Painlevé introduced the so-called Painlevé property, which requires that the locations of any essential singularities \cite{conte2008painleve} do not depend on the initial conditions.  
\begin{table}[h!]
\centering
    \begin{tabular}{c||cc}
        & critical
        & non-critical
        \\
        \hline
        \hline
        movable
        & \tomato{$\quad (z - z_0)^{\nicefrac12 \dfrac{}{}}$}
        & {$e^{\frac{1}{z - z_0}}$}
        \\
        fixed 
        & $\ln{(z - 1)}$
        & $\dfrac{1}{(z - 2)^{3}}$
    \end{tabular}
\end{table}
This property ensures that solutions behave in a controlled and predictable way, making them suitable for defining new transcendental functions. P. Painlevé classified 50 families of such equations \cite{painleve1900memoire}, \cite{painleve1902equations}, but only six of them define new special functions \cite{gambier1910equations}.

By the end of the 20th century, these equations had become one of the central objects of study in mathematics, largely due to their ubiquitous appearance in mathematical and theoretical physics (for details, see, e.g., \cite{fokas2006painleve}). In particular, discrete analogs of the Painlevé equations were discovered by applying a singularity confinement test proposed in \cite{grammaticos1991integrable}, which plays a similar, but not the same, role to the Painlevé property in identifying well-behaved equations. As these discrete versions had been appearing chaotically, a systematic framework for their derivation became necessary. Sakai’s theory not only establishes a deep connection between algebraic geometry and dynamical systems, but also provides a unified approach for classifying discrete equations of Painlevé type. Namely, he classified sublattices $Q\br{R}$ and $Q\br{R^\bot}$ in the $E_8^{(1)}$-root lattice, which is the orthogonal complement of the anti-canonical divisor in the Picard group $\Pic\br{\mathcal{X}}$ with respect to the intersection form. As a result, he obtained 22 classes of discrete Painlevé systems of either elliptic ($ell$-), multiplicative ($q$-), or additive ($d$-) types. 
Since an elliptic curve passes through eight points in general position on $\mathbb{P}^1 \times \mathbb{P}^1$, the master equation in his classification is a discrete elliptic Painlevé equation, which can be reduced to lower systems via a coalescence procedure.
The differential Painlevé equations arise as continuous limits of the 
$d$-Painlevé systems, and thus the results of K. Okamoto are recovered (see Table~\ref{tab:ssdifP}).
\begin{table}[H]
    \centering
    \begin{tabular}{c||cccccccc}
      & PVI & PV & PVI & PIII & PII & PI
      \\
      \hline \hline
      surface type $R$
      &  $D_4^{(1)}$
      &  $D_5^{(1)}$
      &  $E_6^{(1)}$
      &  $D_6^{(1)}$
      &  $E_7^{(1)}$
      &  $E_8^{(1)}$
      \\[2mm]
      symmetry type $R^{\bot}$
      &  $D_4^{(1)}$
      &  $A_3^{(1)}$
      &  $A_2^{(1)}$
      &  $\br{2 A_1}^{(1)}$
      &  $A_1^{(1)}$
      &  $-$
    \end{tabular}
    \caption{The surface/symmetry types of the differential Painlevé equations}
    \label{tab:ssdifP}
\end{table}

Matrix or, more general, non-commutative integrable systems have emerged as natural generalizations of the classical ones and appear in various applications in mathematical and quantum physics (see,~e.g.,~\cite{faddeev1994quantum}, \cite{landi2002introduction}, \cite{castellani2000non}, \cite{asakawa2000noncommutative}, \cite{szabo2003quantum}). Today, they are among the central objects of study in these fields. The Painlevé equations are one of the remarkable examples of such phenomena. They include quantum versions \cite{nagoya2004quantum}, matrix~differential equations \cite{Kawakami_2015},  non-commutative ``differential'' equations---defined in the setting of a unital associative algebra $\mathcal{A}$ equipped with a derivation---\cite{bobrova2023classification}, as well as matrix discrete equations \cite{cassatella2014singularity}. Several of these non-commutative Painlevé equations are related to integrable non-commutative analogs of partial differential equations \cite{OS_1998a}, \cite{Adler_Sokolov_2020_1} and partial discrete equations \cite{adler2020}, \cite{bobrova2023non}, the matrix singularity confinement property \cite{cassatella2014singularity}, the matrix Riemann-Hilbert problem \cite{cafasso2014non}, matrix orthogonal polynomials \cite{cafasso2018toda}, multipartite Calogero-type systems \cite{bertola2018noncommutative}. 

\medskip
In particular, a matrix version of the sixth $q$-Painlevé equation was derived by H. Kawakami \cite{kawakami2020q} using matrix discrete $q$-Schlesinger systems. His version of the sixth $q$-Painlevé equation can be written as
\begin{align}
    \label{eq:qPA3_mat}
    \begin{aligned}
    \bar F \, K \, F
    &= \theta_1 \theta_2 \br{\kappa_1 a_1 a_2}^{-1} \,
    \br{\bar G - a_1 a_2 \theta_1^{-1} t \, \mathbb{I}} \,
    \br{\bar G - a_1 a_2 \theta_2^{-1} t \, \mathbb{I}} \,
    \br{\bar G - (q \kappa_1)^{-1} \, \mathbb{I}}^{-1} \,
    \br{\bar G - \rho \, \mathbb{I}}^{-1}
    ,
    \\
    \bar G \, K \, G
    &= \br{q \kappa_1}^{-1} \,
    \br{F - a_1 t \, \mathbb{I}} \, 
    \br{F - a_2 t \, \mathbb{I}} \, 
    \br{F - a_3 \, \mathbb{I}}^{-1} \, 
    \br{F - a_4 \, \mathbb{I}}^{-1}
    ,
    \end{aligned}
\end{align}
where $F = F(t)$, $G = G (t)$, $\bar F = F(q t)$, $\bar G = G(q t)$, the matrix $K$ is the subject of the following relation
\begin{align}
    &&
    \label{eq:qPA3_matK}
    F^{-1} \, G \, F \, G^{-1}
    &= \rho \, K
    ,
    &
    \rho
    &:= a_1 a_2 a_3 a_4 \kappa_1 \br{\theta_1 \theta_2}^{-1}
    ,
    &&
\end{align}
and $a_1$, $a_2$, $a_3$, $\theta_{1}$, $\theta_2$, $\kappa_1$, $\kappa_2$ are scalar parameters associated with the corresponding isomonodromic problem (see Theorem 5.2 in \cite{kawakami2020q}). Its quantum version, i.e. when $F \, G = \lambda \, G \, F$ with some central element $\lambda$, is presented in \cite{hasegawa2011quantizing} (see Theorem 5 therein).

\begin{rem}
Another non-commutative version of the $q$-Painlevé VI equation was obtained in \cite{doliwa2013non} by using a similarity reduction of the non-isospectral non-autonomous lattice
non-commutative mKdV equation:
\begin{align}
    &&
    \bar F
    &= \br{G + t \, c_1^{-1}} \, 
    \br{G + c_2^{-1}}^{-1} \, F^{-1} \,
    \br{G + t \, c_1} \, 
    \br{G + c_2}^{-1}
    ,
    &
    &\bar t
    = q^4 \, t,
    &&
    \\
    &&
    \bar G
    &= \br{\bar F + t \, c_3^{-1}} \, 
    \br{\bar F + c_4^{-1}}^{-1} \, 
    G^{-1} \,
    \br{\bar F + t \, c_3} \, 
    \br{\bar F + c_4}^{-1}
    ,
    &
    &c_1, c_2, c_3, c_4, q, t 
    \in \mathbb{C}
    .
    &&
\end{align}
\end{rem}
\medskip

Our system \ref{eq:qPA3} can be viewed as a generalization of the Kawakami system \eqref{eq:qPA3_mat}, owing to the properties of non-commutative rational functions (see Lemma \ref{thm:ncpoly} and Corollary \ref{thm:ncrat}) and the existence of the element $I (f, g) = f^{-1} \, g \, f \, g^{-1}$ preserved under the double iteration (see Proposition \ref{thm:firstint_qcase} and Remark \ref{rem:firstint_qcase}), as well as the quantum version presented in \cite{hasegawa2011quantizing}. We also note that a matrix analog of the fifth $q$-Painlevé equation has been derived in \cite{kawakami2023q}, and its non-commutative generalization is obtained here via a coalescence procedure (see \ref{eq:qPA4} and Subsection~\ref{sec:coal}). This procedure enables us to derive non-commutative versions of the lower $q$-Painlevé equations (systems \ref{eq:qPA4} -- \ref{eq:qPA7'}) and to connect the \ref{eq:qPA3} system with the non-commutative $d$-Painlevé systems constructed in \cite{bobrova2024affine}, thanks to the limit \ref{eq:qPA3} $\to$ \ref{eq:dPD4}. 

\medskip
While the present paper focuses on developing the surface theory in order to describe a formal geometry beyond a non-commutative version of the sixth $q$-Painlevé equation, a broader goal is to understand a wide range of non-commutative analogs of the differential Painlevé equations classified in the paper \cite{bobrova2023classification}.
The~author of the current paper expects that a non-commutative generalization of the Sakai surface theory could play a key role in addressing this problem. 
As a first step, we attempt to adapt Sakai's theory to the non-commutative setting. To do so, we begin with the \ref{eq:qPA3} system, obtained via the affine Weyl group approach (see Subsection \ref{sec:qPA3_affW} and Theorem \ref{thm:WD5_nc} therein), and investigate its surface type (see Subsection \ref{sec:qPA3surf_nc}) by following the same steps as in the commutative case, with appropriate modifications to the non-commutative context. This led us to the construction of the non-commutative theory presented in Subsection \ref{sec:ncgeom}. 

Another related approach is presented in \cite{rains2025noncommutative}, which introduces a notion of non-commutative birational geometry for discrete equations within more abstract framework. It proposes a classification scheme based on isomonodromic Lax pairs. However, that work does not provide any examples of such equations in explicit coordinates. A key difference between Rain's theory and the present work lies in the starting point and methodology: E. Rains adopts a global, categorical approach based on non-commutative surfaces and their auto-equivalences, whereas our approach is more explicit, rooted in the concrete structure of a specific non-commutative Painlevé equation and its surface-type classification via blow-ups and Picard lattices, closely mirroring Sakai’s original paper. 

At this stage, it remains unclear how our theory could be applied to produce a non-commutative version of the master discrete Painlevé equation, which is of elliptic type. In contrast, by using Sklyanin-type algebras, the authors of \cite{okounkov2015noncommutative} construct a non-commutative analog of the elliptic discrete Painlevé equation---though, again, without any explicit coordinate realizations. This issue could be solved once a non-commutative version of the elliptic function is presented. 
Note that a generalizations of the elliptic Painlevé equation to the~$\mathbb{P}^3$ case was obtained by T. Takenawa in \cite{takenawa2004discrete}, while four-dimensional analogues of certain other discrete Painlevé systems also were constructed in \cite{carstea2019space}, \cite{stokes2025geometry}.

\subsubsection*{Structure of the paper}
Although the starting point of the present paper is based on Subsection \ref{sec:qPA3_affW}, we chose to describe first the non-commutative geometry and then its application to the \ref{eq:qPA3} system. Thus, the paper is organized as follows. In Section \ref{sec:sakai}, we briefly recall all the key definitions and constructions of the Sakai surface theory, which serve as the foundation for our non-commutative generalization considered in Section~\ref{sec:ncsurfandsys}. The latter is divided into two parts. One of them, Subsection \ref{sec:ncgeom} presents a non-commutative version of the Sakai surface theory, including definitions of the non-commutative versions of the projective lines, Möbius transformation, pencil of biquadratic curves and its base points, a blow-up procedure and the associated Picard lattice. We also describe how to derive discrete dynamical systems from the surface theory. The second part, Subsection \ref{sec:ncdsys}, focuses on one-dimensional discrete dynamical systems. There, we provide the necessary definitions (see Subsection \ref{sec:ncodde}), and in Subsection \ref{sec:afftosys}, we describe a method for deriving such systems from extended birational representations of affine Weyl groups. Additionally, in Subsection \ref{sec:firstint}, we investigate first integrals of certain discrete systems, and, in particular, first integrals of the non-commutative $d$-Painlevé systems obtained in \cite{bobrova2024affine} (see Appendix A therein) are derived (see Propositions \ref{thm:firstint_dcase} and~\ref{thm:firstint_qdP}). 
Section~\ref{sec:qPA3} presents our main example, the \ref{eq:qPA3} system. Thus, in Subsection \ref{sec:qPA3_affW}, we construct this system, by postulating an extended birational representation of the extended affine Weyl group of type $D_5^{(1)}$, following the method initially developed in \cite{noumi1998affine} and extended to the non-commutative case in \cite{bobrova2024affine}. Subsection~\ref{sec:qPA3surf_nc} then explores the associated surface theory, which is used in Subsection~\ref{sec:qPA3surftoW_nc} to reconstruct the same extended birational representation from the point of view of formal birational geometry described in Subsection~\ref{sec:ncgeom}. 
Thanks to the fact that the root variables are central elements, a coalescence cascade starting at the \ref{eq:qPA3} can be interpreted via a coalescence of the point configurations (see Figure \ref{pic:qPA3deg}). 
This leads to non-commutative analogs of the lower $q$-Painlevé equations. Moreover, by taking a suitable limit from \ref{eq:qPA3} to \ref{eq:dPD4}, we connect our \ref{eq:qPA3} system with the non-commutative $d$-Painlevé equations previously derived in \cite{bobrova2024affine}. 
We conclude this paper with a set of open questions (see Section \ref{sec:openq}) and two appendices listing the non-commutative analogs of the $q$- and $d$-Painlevé equations (Appendices~\ref{app:qP}~and~\ref{app:dP}, respectively). The list of $d$-Painlevé equations is simplified, by using the first integrals discussed in Subsection~\ref{sec:firstint}. 

\subsubsection*{Notation}
We fix the following notation throughout the paper.
\begin{itemize}
\item ${\rm f} = \pbr{f_1, f_2, \dots, f_k}$ is a finite set of non-commutative elements, called \textit{functions}.
\vspace{1mm}


\item $\mathcal{A} = \mathbb{C} \abr{{\rm f}} = \mathbb{C} \abr{f_1, f_2, \dots, f_k}$ is the free unital associative algebra over $\mathbb{C}$ generated by the set ${\rm f}$. Elements of $\mathcal{A}$ are \textit{non-commutative polynomials}.
\vspace{1mm}


\item $\mathcal{R}$ is a division ring over $\mathbb{C}$ generated by the set ${\rm f}$. By definition, every nonzero element of $\mathcal{R}$ is invertible. Elements of $\mathcal{R}$ are called \textit{non-commutative rational functions}.
\vspace{1mm}

\item $\mathcal{Z}(\mathcal{A})$
and $\mathcal{Z}(\mathcal{R})$ denote the centers of $\mathcal{A}$
and $\mathcal{R}$, respectively.
\vspace{1mm}

\item ${\rm b} = \pbr{b_1, b_2, \dots, b_l}$ is a finite set of central elements, called \textit{parameters}.

\item All greek letters denote the central elements of $\mathcal{R}$.
\vspace{1mm}

\item For functions given on a lattice, we set $T(f_n) = f_{n + 1} =: \bar f$ and $T^{-1}(f_n) = f_{n - 1} =: \underline f$.
\end{itemize}

\subsubsection*{Acknowledgments}
The author is deeply grateful to Anton Dzhamay and Ivan Sechin for fruitful discussions. The author is also thankful to the CEP program for financial support of author's visit to BIMSA in July~2025. 
The author is also grateful to the referees for their comments, which have significantly improved the paper.

\section{Sakai's surface theory: a brief review}
\label{sec:sakai}

The Sakai surface theory \cite{sakai2001rational} provides a uniform geometric framework for
the discrete Painlevé equations. {The starting point is 
a compact rational surface $\mathcal{X}$ that admits a unique anti-canonical divisor $-\mathcal{K}_{\mathcal{X}}$.} 
The rational surface $\mathcal{X}$ can be obtained by blowing up eight points on~$\mathbb{P}^1 \times \mathbb{P}^1$ (or nine points on~$\mathbb{P}^2$), which naturally give rise to the Picard lattice $\Pic\br{\mathcal{X}}$. 
Within $\Pic\br{\mathcal{X}}$, the orthogonal complement of $-\mathcal{K}_\mathcal{X}$ becomes a root lattice of type $E_8^{(1)}$. 
This orthogonal complement has two sublattices of types $R$ and $R^{\bot}$. The root lattice $Q(R)$ is related to the classes of irreducible components of $-\mathcal{K}_\mathcal{X}$. 

The extended affine Weyl group $\widetilde W\br{R^{\bot}}$ plays a key role since it acts on the basis of $\Pic\br{\mathcal{X}}$ as \emph{Cremona isometries}, i.e. lattice
automorphisms that preserve the intersection form, the anti‐canonical
class, and the effectiveness of each effective divisor of $\Pic(\mathcal{X})$. This action can be lifted to a birational action of the families of surfaces $\mathcal{X}$. The translation elements in $\widetilde W(R^{\bot})$ give rise to discrete Painlevé equations.

\medskip
Below we recall the basic notions from the
commutative case, which we will extend to a non-commutative setting in Section~\ref{sec:ncgeom}.

\medskip
\paragraph{\textbf{Setup}}
{Let $X_0 := \mathbb{P}^{1} \times \mathbb{P}^1$} carry affine coordinates $(f, g)$. {Let $P = P(f,g)$ be a biquadratic homogeneous polynomial in coordinates $(f,g)$.} 
Recall that a polynomial is \textit{homogeneous of degree $d$} if every monomial term has the same total degree (the sum of exponents of the variables). This leads to the \textit{scaling property}: 
\begin{align}
    \label{eq:scpr}
    \exists \,\, d \in \mathbb{Z}, \, \lambda \neq 0
    \,\, | \,\,
    P(\lambda \, f, \lambda \, g) = \lambda^d \, P(f,g)
    .
\end{align}

\begin{defn}
The set $\mathcal{C}$ of points satisfying $P(f,g) = 0$ is called a \textit{curve of degree $2$}, or \textit{$2$-curve}:
\begin{align}
    \mathcal{C}
    &= \pbr{
    \br{f,g} \in \mathbb{P}^1 \times \mathbb{P}^1 \,\, \big| \,\,
    P(f,g) = 0
    }
    .
\end{align}    
\end{defn}

Suppose that $P$ factors into $k$ irreducible homogeneous polynomials $P_j = P_j (f, g)$, $j = 1, \dots, k$, i.e. $P = P_1^{m_1} \cdot P_2^{m_2} \cdot \dots \cdot P_k^{m_k}$. Then we write
\begin{align}
    \label{eq:irrcurves}
    \mathcal{C}
    &= m_1 \mathcal{C}_1 + m_2 \mathcal{C}_2 + 
    \dots + m_k \mathcal{C}_k,
\end{align}
where each $\mathcal{C}_j$ is the curve defined by $P_j(f, g) = 0$. 
\begin{defn}
\label{def:irrcomp}
The curves $\mathcal{C}_j$ in \eqref{eq:irrcurves} are the \textit{irreducible components} of the curve $\mathcal{C}$. If $P = P(f,g)$ is itself irreducible, then $\mathcal{C}$ is called an \textit{irreducible curve}. 
\end{defn}

Consider two homogeneous polynomials $P_1(f,g)$, $P_2 (f,g) \in \mathbb{C}[f,g]$ of degree 2, with no common~factors. 
\begin{defn}
\label{def:pencil}
The set $\mathcal{P}$ is called a \textit{pencil of biquadratic curves}, or \textit{$(2,2)$-curves}:
\begin{align}
    \mathcal{P}
    &= \pbr{
    \lambda \, P_1(f,g) + \mu \, P_2(f,g) = 0 \,\, \big| \,\,
    \lbr{\lambda : \mu} \in \mathbb{P}^1
    }
    .
\end{align}
\end{defn}
\begin{rem}
Note that, by rescaling, one may set $\lambda = 1$ without loss of generality.
\end{rem}

\hspace{-5mm}
\begin{minipage}[l]{0.58\linewidth}
\hspace{4mm}
\begin{defn}
Consider the set $\mathcal{B}$ consisting of the intersection of the zero loci of $P_1$ and $P_2$ of a given pencil $\mathcal{P}$:
\begin{align}
    \mathcal{B}
    &= \pbr{
    \br{f,g} \in \mathbb{P}^1 \times \mathbb{P}^1 \,\, \big| \,\,
    P_1(f,g) = 0 \,\, \wedge \,\, P_2 (f,g) = 0
    }
    .
\end{align}
The elements of $\mathcal{B}$ are called the \textit{base points} of the pencil $\mathcal{P}$.
\end{defn}

To determine the base points, it is convenient to work in the four standard affine charts of $\mathbb{P}^{1}~\times~\mathbb{P}^1$, represented schematically~as in~Figure~\ref{pic:affcharts}.
\begin{align}
    \\[-2mm]
    &&
    \pbr{\br{f, g}},
    &&&
    \pbr{\br{f^{-1}, g}},
    &&
    \\[4mm]
    &&
    \pbr{\br{f, g^{-1}}},
    &&&
    \pbr{\br{f^{-1}, g^{-1}}}
    &&
\end{align}
\end{minipage}
\hspace{-2mm}
\begin{minipage}[c]{0.47\linewidth}
\begin{figure}[H]
    \centering
    \scalebox{0.85}{\tikzset{every picture/.style={line width=0.75pt}} 

\begin{tikzpicture}[x=0.75pt,y=0.75pt,yscale=-1,xscale=1]

\draw    (90.8,20.92) -- (90.8,210.7) ;
\draw    (220.38,21.42) -- (220.38,211.2) ;
\draw    (250.33,50.5) -- (60.55,50.5) ;
\draw    (250.33,180.75) -- (60.55,180.75) ;
\draw [color={rgb, 255:red, 255; green, 99; blue, 71 }  ,draw opacity=1 ][line width=1.5]    (90.8,50.41) -- (90.8,81.11) ;
\draw [shift={(90.8,85.11)}, rotate = 270] [fill={rgb, 255:red, 255; green, 99; blue, 71 }  ,fill opacity=1 ][line width=0.08]  [draw opacity=0] (11.07,-5.32) -- (0,0) -- (11.07,5.32) -- (7.35,0) -- cycle    ;
\draw [color={rgb, 255:red, 255; green, 99; blue, 71 }  ,draw opacity=1 ][line width=1.5]    (220.4,50.12) -- (220.4,80.82) ;
\draw [shift={(220.4,84.82)}, rotate = 270] [fill={rgb, 255:red, 255; green, 99; blue, 71 }  ,fill opacity=1 ][line width=0.08]  [draw opacity=0] (11.07,-5.32) -- (0,0) -- (11.07,5.32) -- (7.35,0) -- cycle    ;
\draw [color={rgb, 255:red, 255; green, 99; blue, 71 }  ,draw opacity=1 ][line width=1.5]    (90.8,180.73) -- (90.8,149.54) ;
\draw [shift={(90.8,145.54)}, rotate = 90] [fill={rgb, 255:red, 255; green, 99; blue, 71 }  ,fill opacity=1 ][line width=0.08]  [draw opacity=0] (11.07,-5.32) -- (0,0) -- (11.07,5.32) -- (7.35,0) -- cycle    ;
\draw [color={rgb, 255:red, 255; green, 99; blue, 71 }  ,draw opacity=1 ][line width=1.5]    (220.4,181.23) -- (220.4,150.04) ;
\draw [shift={(220.4,146.04)}, rotate = 90] [fill={rgb, 255:red, 255; green, 99; blue, 71 }  ,fill opacity=1 ][line width=0.08]  [draw opacity=0] (11.07,-5.32) -- (0,0) -- (11.07,5.32) -- (7.35,0) -- cycle    ;
\draw [color={rgb, 255:red, 255; green, 99; blue, 71 }  ,draw opacity=1 ][line width=1.5]    (90.01,180.8) -- (120.71,180.8) ;
\draw [shift={(124.71,180.8)}, rotate = 180] [fill={rgb, 255:red, 255; green, 99; blue, 71 }  ,fill opacity=1 ][line width=0.08]  [draw opacity=0] (11.07,-5.32) -- (0,0) -- (11.07,5.32) -- (7.35,0) -- cycle    ;
\draw [color={rgb, 255:red, 255; green, 99; blue, 71 }  ,draw opacity=1 ][line width=1.5]    (90.01,50.5) -- (120.71,50.5) ;
\draw [shift={(124.71,50.5)}, rotate = 180] [fill={rgb, 255:red, 255; green, 99; blue, 71 }  ,fill opacity=1 ][line width=0.08]  [draw opacity=0] (11.07,-5.32) -- (0,0) -- (11.07,5.32) -- (7.35,0) -- cycle    ;
\draw [color={rgb, 255:red, 255; green, 99; blue, 71 }  ,draw opacity=1 ][line width=1.5]    (220.29,180.8) -- (189.93,180.8) ;
\draw [shift={(185.93,180.8)}, rotate = 360] [fill={rgb, 255:red, 255; green, 99; blue, 71 }  ,fill opacity=1 ][line width=0.08]  [draw opacity=0] (11.07,-5.32) -- (0,0) -- (11.07,5.32) -- (7.35,0) -- cycle    ;
\draw [color={rgb, 255:red, 255; green, 99; blue, 71 }  ,draw opacity=1 ][line width=1.5]    (221.29,50.5) -- (190.93,50.5) ;
\draw [shift={(186.93,50.5)}, rotate = 360] [fill={rgb, 255:red, 255; green, 99; blue, 71 }  ,fill opacity=1 ][line width=0.08]  [draw opacity=0] (11.07,-5.32) -- (0,0) -- (11.07,5.32) -- (7.35,0) -- cycle    ;

\draw (45.53,154.52) node [anchor=north west][inner sep=0.75pt]  [font=\Large]  {$( f,g)$};
\draw (28.53,58.77) node [anchor=north west][inner sep=0.75pt]  [font=\Large]  {$\left( f ,g^{-1}\right)$};
\draw (229.29,154.02) node [anchor=north west][inner sep=0.75pt]  [font=\Large]  {$\left( f^{-1} ,g\right)$};
\draw (228.54,59.27) node [anchor=north west][inner sep=0.75pt]  [font=\Large]  {$\left( f^{-1} ,g^{-1}\right)$};

\end{tikzpicture}}
    \caption{Affine charts in $\mathbb{P}^1 \times \mathbb{P}^1$}
    \label{pic:affcharts}
\end{figure}
\end{minipage}
\\[2mm]

\medskip
\paragraph{\textbf{Blow-ups and divisors}}
Through a base point, several curves pass. In order to separate them, one applies the blow-up procedure. Below we define it algebraically as a change of coordinates. 

{\begin{defn}
\label{def:blups}
Let $p = \br{f_0, g_0} \in \mathcal{B}$ be given in an affine chart $U$. The \textit{blow up} of $X_0$ at $p$ replaces the chart $U$ by two charts with coordinated $(f_1, g_1)$ and $(f_2, g_2)$ defined by
\begin{align}
    \left\{
    \begin{array}{lcl}
         f_1
         &=& \br{f - f_0} \, \br{g - g_0}^{-1}
         ,
         \\[2mm]
         g_1 
         &=& g - g_0, 
    \end{array}
    \right.
    &&&
    \left\{
    \begin{array}{lcl}
         f_2
         &=& f - f_0
         ,
         \\[2mm]
         g_2 
         &=& \br{f - f_0}^{-1} \, \br{g - g_0}
         .
    \end{array}
    \right.
\end{align}
Under this transformation, the base point $p$ is replaced by the \textit{exceptional set} $E$ defined by
\begin{align}
    {E}
    &= \pbr{g_1 = 0} \cup \pbr{f_2 = 0}
    .
\end{align}
\end{defn}}

\begin{defn}
\label{def:div}
A \textit{divisor} ${D}$ is a formal $\mathbb{Z}$-linear combination of irreducible curves $\mathcal{C}_i${\rm:}
\begin{align}
    {D}
    &= m_1 \, \mathcal{C}_1 
    + m_2 \, \mathcal{C}_2 
    + \dots + m_k \, \mathcal{C}_k
    .
\end{align}
It is called \textit{effective} if all $m_i \geq 0$.
\end{defn}

\begin{defn}
\label{def:divclass}
Two divisors $D_1$ and $D_2$ are \textit{linearly equivalent}, $D_1 \sim D_2$, if their defining curves belong to the same linear space of bihomogeneous polynomials. The equivalence class is called the \textit{divisor class} $\mathcal{D}$.
\end{defn}

\begin{exmp}
All curves $\mathbb{P}^1\times \pbr{pt}$ form the  divisor class $\mathcal{H}_1$, while vertical curves $\pbr{pt} \times \mathbb{P}^1$ form class~$\mathcal{H}_2$. 
\end{exmp}

\medskip
\paragraph{\textbf{Picard groups and Cremona isometries}}
Consider the surface $\mathcal{X}$ obtained after blowing up $n$ points on $\mathbb{P}^1 \times \mathbb{P}^1$. It contains the set $\mathcal{D}$ of lines appeared during the blow-up process. 

\medskip
Consider a free $\mathbb{Z}$-module of rank $n + 2$ generated by $\mathcal{H}_1$, $\mathcal{H}_2$, and $\mathcal{E}_i$, $i = 1, \dots, n$,
\begin{align}
    \Lambda
    &= \mathbb{Z} \mathcal{H}_1 \oplus 
    \mathbb{Z} \mathcal{H}_2 \oplus 
    \mathbb{Z} \mathcal{E}_1 \oplus \dots \oplus 
    \mathbb{Z} \mathcal{E}_n
    ,
\end{align}
equipped with the symmetric bilinear form $\br{- \, \cdot \, -}: \, \Lambda \times \Lambda \to \mathbb{Z}$ defined by
\begin{align}
    \label{eq:intform}
    &&&&
    \br{\mathcal{H}_i \cdot \mathcal{H}_j}
    &= 1 - \delta_{ij},
    &
    \br{\mathcal{H}_i \cdot \mathcal{E}_j}
    &= 0,
    &
    \br{\mathcal{E}_i \cdot \mathcal{E}_j}
    &= - \delta_{ij}
    ,
    &&&&
\end{align}
where $\delta_{ij}$ denotes the standard Kronecker delta.

\begin{defn}
\label{def:picgr}
Let $\mathcal{X}$ be a surface obtained from $\mathbb{P}^1 \times \mathbb{P}^1$ by $n$ blow-ups and let ${H}_1$ and ${H}_2$ stand for the divisors corresponding to the lines $f = 0$ and $g = 0$, respectively. Then $\Lambda$ can be identified with the \textit{Picard lattice $\Pic\br{\mathcal{X}}$} associated with $\mathcal{X}$, and the symmetric bilinear form \eqref{eq:intform} induces a symmetric bilinear form on $\Pic\br{\mathcal{X}} \times \Pic \br{\mathcal{X}}$, called the \textit{intersection form}.
\end{defn}

Let $V = \Lambda \otimes \mathbb{Q}$ be a vector space and $\abr{-,-} : V \times V \to \mathbb{Q}$ be an inner product such that $\abr{\lambda, \mu} = - \br{\lambda, \mu}$ for any $\lambda$, $\mu \in V$. Consider the simple roots
\begin{align}
    \delta_0
    &= \mathcal{E}_1 - \mathcal{E}_2
    ,
    &
    \delta_1
    &= \mathcal{H}_1 - \mathcal{H}_2
    ,
    &
    \delta_2 
    &= \mathcal{H}_2 - \mathcal{E}_{1} - \mathcal{E}_{2}
    ,
    &
    \delta_i
    &= \mathcal{E}_{i - 1} - \mathcal{E}_i
    ,
    &
    i = 3, \dots, n
    .
\end{align}
The matrix $C = \br{\abr{\delta_i, \delta_j}}_{0 \leq i,j \leq n}$ is the generalized Cartan matrix of type $E_n^{(1)}$. The null root reads as
\begin{align}
    \label{eq:antidiv}
    -\mathcal{K}_{\mathcal{X}} 
    &= 2 \mathcal{H}_1 + 2 \mathcal{H}_2 - \mathcal{E}_1 - \dots - \mathcal{E}_n
    .
\end{align}
Thus, we have obtained the root lattice $Q(E_n^{(1)}) \subset \Lambda \subset V$. Note that for $n = 8$, $- \mathcal{K}_{\mathcal{X}}$ can be decomposed as
\begin{align}
    \label{eq:ddec}
    - \mathcal{K}_{\mathcal{X}}
    &= 3 \delta_0 
    + 2 \delta_1 
    + 4 \delta_2
    + 6 \delta_3 
    + 5 \delta_4 
    + 4 \delta_5 
    + 3 \delta_6 
    + 2 \delta_7 
    + \delta_8 
    .
\end{align}

\begin{defn}
The null root $-\mathcal{K}_{\mathcal{X}} $ is called the \textit{anti-canonical divisor class}.
\end{defn}

\begin{defn}
A divisor class $\mathcal{D}$ is of \textit{canonical type} if, for every irreducible component $\mathcal{D}_i$ of $\mathcal{D}$, we have 
\begin{align}
    \br{- \mathcal{K}_{\mathcal{X}} \cdot \mathcal{D}_i}
    = 0
    .
\end{align}
\end{defn}

\begin{defn}
The surface $\mathcal{X}$ is called a \textit{generalized Halphen surface of index zero} if it has a unique anti-canonical divisor of canonical type.
\end{defn}

\begin{rem}
The bilinear form on the Picard lattice $\Lambda$ can be interpreted as the intersection form of divisors. The anti-canonical divisor $-\mathcal{K}_{\mathcal{X}}$ given in \eqref{eq:antidiv} corresponds to the class of curves on $\mathbb{P}^1\times\mathbb{P}^1$ of bidegree $(2,2)$ passing through the blow-up points $p_i$ with with multiplicity $1$.
\end{rem}

\begin{defn}
An automorphism of $\Pic\br{\mathcal{X}}$ that preserves the intersection form on $\Pic\br{\mathcal{X}}$, the anti-canonical class $-\mathcal{K}_{\mathcal{X}}$, and the semi-group of effective divisors of $\Pic\br{\mathcal{X}}$ is called a \textit{Cremona isometry}.
\end{defn}

\medskip
\paragraph{\textbf{From surfaces to dynamics}}

Given an explicit set of base points and a resulting surface $\mathcal{X}$ with a unique anti-canonical divisor class $- \mathcal{K}_{\mathcal{X}}$ of canonical type, H. Sakai showed how discrete Painlevé equations can be constructed from the Cremona isometries. 
More precisely, one considers a rational surface $\mathcal{X}$ obtained by blowing up eight points on $\mathbb{P}^1 \times \mathbb{P}^1$ (or nine points on $\mathbb{P}^2$). According to the discussion above, this leads to the Picard lattice $\Pic\br{\mathcal{X}}$, which contains a lattice of type  $E_8^{(1)}$. The latter has two important sublattices $Q\br{R} = \spn_{\mathbb{Z}}\pbr{\delta_i}$ and $Q\br{R^\bot} = \spn_{\mathbb{Z}}\pbr{\alpha_j}$, encoding the \textit{surface} and \textit{symmetry types} of $\mathcal{X}$.
Associated to $R$ and $R^{\bot}$ types affine Weyl groups acts by \textit{reflections} on $\Pic(\mathcal{X})$. The reflection $s$ in the root $\alpha_j$ is given~by 
\begin{align}
    \label{eq:refl}
    s_{\alpha_j} \br{\mathcal{L}}
    &= \mathcal{L} + \br{\alpha_j \, \cdot \, \mathcal{L}} \, \alpha_j
    ,
\end{align}
where $\br{- \, \cdot \, -}$ stands for the intersection form. The group
$\widetilde W \br{R^\bot}$ acts on $\Pic\br{\mathcal{X}}$ as Cremona isometries. They can be lifted to the coordinates~$(f,g)$. A translation element $T$ in $\widetilde W\br{R^{\bot}}$ gives discrete dynamics
\begin{align}
    T \br{f, g; \alpha_j}
    &= \br{\bar f, \, \bar g; \, \bar \alpha_j}
    .
\end{align}

\begin{defn}
A \textit{discrete Painlevé equation} is a discrete dynamical system on the family of surfaces $\mathcal{X}$ induced
by a translation in the affine symmetry sublattice $Q(R^\bot)$ of the corresponding surface.
\end{defn}

Note that the Painlevé equations are non-autonomous systems. In the discrete case, we have three types of iterations: elliptic, multiplicative, and additive (see Figure \ref{pic:iter}). The non-autonomous behavior is closely linked to the variation of the base points under the $\widetilde W (R^{\bot})$ action.
\begin{figure}[H]
    \centering
    \begin{minipage}{0.3\textwidth}
    \centering
    \scalebox{.9}{\tikzset{every picture/.style={line width=0.75pt}} 

\begin{tikzpicture}[x=0.75pt,y=0.75pt,yscale=-1,xscale=1]

\draw  [draw opacity=0] (211.98,143.91) .. controls (207.2,147.35) and (201.79,149.29) .. (196.06,149.29) .. controls (176.75,149.29) and (161.1,127.29) .. (161.1,100.16) .. controls (161.1,73.03) and (176.75,51.04) .. (196.06,51.04) .. controls (201.79,51.04) and (207.2,52.98) .. (211.98,56.42) -- (196.06,100.16) -- cycle ; \draw   (211.98,143.91) .. controls (207.2,147.35) and (201.79,149.29) .. (196.06,149.29) .. controls (176.75,149.29) and (161.1,127.29) .. (161.1,100.16) .. controls (161.1,73.03) and (176.75,51.04) .. (196.06,51.04) .. controls (201.79,51.04) and (207.2,52.98) .. (211.98,56.42) ;  
\draw    (211.98,143.91) .. controls (251.98,113.91) and (258.76,156.79) .. (260.5,169.94) ;
\draw    (211.98,56.42) .. controls (251.16,85.55) and (258.01,49.84) .. (260.06,30.82) ;
\draw [color={rgb, 255:red, 155; green, 155; blue, 155 }  ,draw opacity=1 ] [dash pattern={on 0.84pt off 2.51pt}]  (156.22,69.2) -- (274.79,39.2) ;
\draw [color={rgb, 255:red, 155; green, 155; blue, 155 }  ,draw opacity=1 ] [dash pattern={on 0.84pt off 2.51pt}]  (209.76,39.01) -- (209.86,162.14) ;
\draw  [fill={rgb, 255:red, 0; green, 0; blue, 0 }  ,fill opacity=1 ] (206.9,144.58) .. controls (206.9,142.78) and (208.36,141.33) .. (210.15,141.33) .. controls (211.94,141.33) and (213.4,142.78) .. (213.4,144.58) .. controls (213.4,146.37) and (211.94,147.82) .. (210.15,147.82) .. controls (208.36,147.82) and (206.9,146.37) .. (206.9,144.58) -- cycle ;
\draw  [fill={rgb, 255:red, 0; green, 0; blue, 0 }  ,fill opacity=1 ] (254.33,43.17) .. controls (254.33,41.37) and (255.79,39.92) .. (257.58,39.92) .. controls (259.37,39.92) and (260.82,41.37) .. (260.82,43.17) .. controls (260.82,44.96) and (259.37,46.41) .. (257.58,46.41) .. controls (255.79,46.41) and (254.33,44.96) .. (254.33,43.17) -- cycle ;
\draw  [fill={rgb, 255:red, 0; green, 0; blue, 0 }  ,fill opacity=1 ] (167.76,65.29) .. controls (167.76,63.49) and (169.21,62.04) .. (171,62.04) .. controls (172.8,62.04) and (174.25,63.49) .. (174.25,65.29) .. controls (174.25,67.08) and (172.8,68.53) .. (171,68.53) .. controls (169.21,68.53) and (167.76,67.08) .. (167.76,65.29) -- cycle ;

\draw (219.56,149.44) node [anchor=north west][inner sep=0.75pt]  [font=\Large]  {$p_{3}$};
\draw (264.99,49.02) node [anchor=north west][inner sep=0.75pt]  [font=\Large]  {$p_{2}$};
\draw (150,45.64) node [anchor=north west][inner sep=0.75pt]  [font=\Large]  {$p_{1}$};

\end{tikzpicture}}
    \end{minipage}
    \begin{minipage}{0.3\textwidth}
    \centering
    \scalebox{.9}{\tikzset{every picture/.style={line width=0.75pt}} 

\begin{tikzpicture}[x=0.75pt,y=0.75pt,yscale=-1,xscale=1]

\draw    (227.44,46.08) .. controls (269.41,24.3) and (329.47,58.09) .. (330.51,96.45) ;
\draw    (227.31,46.29) .. controls (180.07,73.24) and (205.66,132.57) .. (257.05,123.79) ;
\draw    (253.65,79.71) .. controls (287.7,89.17) and (284.87,118.32) .. (257.05,123.79) ;

\draw  [fill={rgb, 255:red, 0; green, 0; blue, 0 }  ,fill opacity=1 ] (287.12,48.13) .. controls (287.12,46.34) and (288.57,44.89) .. (290.36,44.89) .. controls (292.16,44.89) and (293.61,46.34) .. (293.61,48.13) .. controls (293.61,49.93) and (292.16,51.38) .. (290.36,51.38) .. controls (288.57,51.38) and (287.12,49.93) .. (287.12,48.13) -- cycle ;
\draw  [fill={rgb, 255:red, 0; green, 0; blue, 0 }  ,fill opacity=1 ] (216.22,51.27) .. controls (216.22,49.48) and (217.67,48.03) .. (219.47,48.03) .. controls (221.26,48.03) and (222.71,49.48) .. (222.71,51.27) .. controls (222.71,53.07) and (221.26,54.52) .. (219.47,54.52) .. controls (217.67,54.52) and (216.22,53.07) .. (216.22,51.27) -- cycle ;
\draw  [fill={rgb, 255:red, 0; green, 0; blue, 0 }  ,fill opacity=1 ] (217.76,118.2) .. controls (217.76,116.41) and (219.21,114.96) .. (221,114.96) .. controls (222.8,114.96) and (224.25,116.41) .. (224.25,118.2) .. controls (224.25,120) and (222.8,121.45) .. (221,121.45) .. controls (219.21,121.45) and (217.76,120) .. (217.76,118.2) -- cycle ;

\draw (290.78,25.99) node [anchor=north west][inner sep=0.75pt]  [font=\Large]  {$q \, z$};
\draw (203.46,35.63) node [anchor=north west][inner sep=0.75pt]  [font=\Large]  {$z$};
\draw (182,118.56) node [anchor=north west][inner sep=0.75pt]  [font=\Large]  {$q^{-1} z$};

\end{tikzpicture}}
    \end{minipage}
    \begin{minipage}{0.3\textwidth}
    \centering
    \scalebox{.9}{\tikzset{every picture/.style={line width=0.75pt}} 

\begin{tikzpicture}[x=0.75pt,y=0.75pt,yscale=-1,xscale=1]

\draw    (19.35,134.48) -- (121.48,23.93) ;
\draw  [fill={rgb, 255:red, 0; green, 0; blue, 0 }  ,fill opacity=1 ] (67.76,79.59) .. controls (67.76,77.79) and (69.21,76.34) .. (71.01,76.34) .. controls (72.8,76.34) and (74.25,77.79) .. (74.25,79.59) .. controls (74.25,81.38) and (72.8,82.83) .. (71.01,82.83) .. controls (69.21,82.83) and (67.76,81.38) .. (67.76,79.59) -- cycle ;
\draw  [fill={rgb, 255:red, 0; green, 0; blue, 0 }  ,fill opacity=1 ] (100.61,43.39) .. controls (100.61,41.59) and (102.06,40.14) .. (103.86,40.14) .. controls (105.65,40.14) and (107.1,41.59) .. (107.1,43.39) .. controls (107.1,45.18) and (105.65,46.63) .. (103.86,46.63) .. controls (102.06,46.63) and (100.61,45.18) .. (100.61,43.39) -- cycle ;
\draw  [fill={rgb, 255:red, 0; green, 0; blue, 0 }  ,fill opacity=1 ] (35.11,114.54) .. controls (35.11,112.74) and (36.56,111.29) .. (38.36,111.29) .. controls (40.15,111.29) and (41.6,112.74) .. (41.6,114.54) .. controls (41.6,116.33) and (40.15,117.78) .. (38.36,117.78) .. controls (36.56,117.78) and (35.11,116.33) .. (35.11,114.54) -- cycle ;

\draw (78.42,83.45) node [anchor=north west][inner sep=0.75pt]  [font=\Large]  {$z$};
\draw (111.27,47.25) node [anchor=north west][inner sep=0.75pt]  [font=\Large]  {$z+n$};
\draw (45.77,118.4) node [anchor=north west][inner sep=0.75pt]  [font=\Large]  {$z-n$};

\end{tikzpicture}}
    \end{minipage}
    \caption{Three types of iterations: elliptic, multiplicative, and additive}
    \label{pic:iter}
\end{figure}

Moreover, the Möbius group $\PGL_2(\mathbb{C})$ plays a key role in the theory by acting on the rational variables through fractional linear transformations. This action corresponds to changes of coordinates in the projective plane and enables normalization procedures that simplify the representation of base point configurations and the resulting dynamical systems.

\medskip
In the non-commutative setting, we will follow a parallel path. Since all the definitions and contractions here were given in a purely algebraic way, they will be transformed to the non-commutative case with minor changes. In Subsection \ref{sec:qPA3surf_nc}, we apply this non-commutative theory to the non-commutative analogue of the sixth $q$-Painlevé equation, obtained by postulating an extended birational representation of the extended affine Weyl group of type $D_5^{(1)}$ (see Theorem \ref{thm:WD5_nc}). 

\section{Non-commutative surface theory and discrete systems}
\label{sec:ncsurfandsys}

In this section, we lay out the algebraic groundwork necessary for studying non-commutative analogues of discrete Painlevé equations. Subsection~\ref{sec:ncgeom} introduces a formal version of Sakai’s surface theory over a division ring $\mathcal{R}$, while Subsection~\ref{sec:ncdsys} presents a general framework for non-commutative discrete systems, categorized into autonomous, additive, and multiplicative types. We explore the existence of first integrals and illustrate how affine Weyl group actions can generate discrete dynamics through extended birational representations. While such representations alone offer limited geometric insight, they become meaningful when combined with the surface-theoretic tools developed in Subsection~\ref{sec:ncgeom}. Their interplay is exemplified in the non-commutative analogue \ref{eq:qPA3} of the sixth $q$-Painlevé equation, serving as the central case study for this formalism.

\subsection{Surface theory: a naïve description}
\label{sec:ncgeom}

This subsection outlines a non-commutative adaptation of Sakai surface theory, aiming to describe the formal geometry underlying discrete Painlevé dynamics. At this stage, we do not attempt to address the classification of discrete Painlevé equations in the non-commutative setting; rather, we aim to develop tools that can be used to investigate extended birational representations of affine Weyl groups that generate non-commutative Painlevé-type discrete dynamics.
\medskip

We begin by defining the non-commutative projective line $\mathbb{P}_{\nc}^1$, Möbius transformations, and their action on non-commutative coordinates. We then introduce formal biquadratic curves in $\mathbb{P}_{\nc}^1 \times \mathbb{P}_{\nc}^1$, define base points, and describe blow-ups viewed as a collection of affine charts. Considering non-commutative analogs of divisors as formal 1-curves, we define non-commutative Picard groups, equipped with an intersection form and an anti-canonical class. Cremona isometries are then recovered in this formal setting.

However, this non-commutative framework has notable limitations: it lacks an underlying topological or analytic structure, the notion of singularities is not well developed, and the treatment of elliptic Painlevé equations remains unclear. In addition, the formalism is algebraic and symbolic in nature, with no direct geometric interpretation beyond. Nevertheless, it provides a setting to study some examples (see Subsection \ref{sec:qPA3surf_nc}) such as the non-commutative sixth $q$-Painlevé equation obtained in Subsection \ref{sec:qPA3_affW} via the method suggested in \cite{bobrova2024affine}.
\medskip

Let us note that the approach in \cite{okounkov2015noncommutative} differs from ours: it treats a more abstract elliptic setting but does not provide any explicit dynamics. Meanwhile, \cite{rains2025noncommutative} adopts a module-theoretic formalism, focusing on Lax pairs rather than the discrete equations themselves. In contrast, our approach emphasizes explicit algebraic realizations in concrete coordinates for discrete systems.

\subsubsection{Projective lines and Möbius transformations}
We follow the exposition given in \cite{bongaarts1999mobius}.

Let $\mathcal{R}$ be a division ring and $\mathcal{R}^{\times} := \mathcal{R} \setminus \pbr{0}$. Consider the right action of $\mathcal{R}^\times$ on the space $\mathcal{R}\times\mathcal{R}$:
\begin{align}
    \begin{aligned}
    (x, y)
    &\mapsto (x \, \lambda, y \, \lambda),
    &
    \lambda
    &\in \mathcal{R}^{\times}
    .
    \end{aligned}
\end{align}
We define an equivalence relation on $\mathcal{R} \times \mathcal{R} \setminus \{(0,0)\}$ by
\begin{align}
    \begin{aligned}
    (x_1, y_1)
    &\sim (x_2, y_2)
    &\Leftrightarrow&
    &
    \exists \, \lambda 
    &\in \mathcal{R}^\times
    &
    \text{such that}&
    &
    (x_1, y_1)
    &= (x_2 \, \lambda, y_2 \, \lambda)
    .
    \end{aligned}
\end{align}

\begin{defn}
The \textit{projective line} $\mathbb{P}^1_{\nc}$ over $\mathcal{R}$ is the quotient space ${\br{\mathcal{R}\times\mathcal{R} \setminus \{(0,0)\}}}/{\sim}$.
\end{defn}

\begin{defn}
The \textit{finite part} of $\mathbb{P}^1_{\nc}$, denoted ${\mathbb{P}_{\nc, f}^{1}}$, consists of all equivalence classes $\lbr{\br{x, y}}$ with $y \in \mathcal{R}^{\times}$.
\end{defn}

\begin{rem}
If $\mathcal{R} = \mathbb{C}$, then $\mathbb{P}^1_{\nc}$ is obtained from $\mathbb{P}^1_{\nc, f}$ by adding a single point at infinity.
\end{rem}

Since $(x, y) \sim (x \, y^{-1}, 1)$ in $\mathbb{P}^1_{\nc, f}$ whenever $y$ is non-zero, there is a bijection between $\mathbb{P}^1_{\nc, f}$ and $\mathcal{R}$ given~by
\begin{align}
    (x, y) 
    &\mapsto x \, y^{-1} =: z
    .
\end{align}
The latter identifies $\mathbb{P}^1_{\nc, f}$ with $\mathcal{R}$. 

\medskip
Similarly, let us introduce the algebra $\Mat_2 (\mathcal{R})$ of $2 \times 2$ matrices over $\mathcal{R}$ and its subgroup $\GL_2 (\mathcal{R})$ of invertible elements in $\Mat_2 (\mathcal{R})$. Note that both of them have a natural left action on the space $\mathcal{R} \times \mathcal{R}$ thanks to the matrix multiplication: 
\begin{align}
    \begin{pmatrix}
        a & b \\[0.9mm]
        c & d
    \end{pmatrix}
    \, 
    \begin{pmatrix}
        z_1 \\[0.9mm] z_2
    \end{pmatrix}
    = 
    \begin{pmatrix}
        a \, z_1 + b \, z_2 \\[0.9mm]
        c \, z_1 + d \, z_2
    \end{pmatrix}
    .
\end{align}
Note also that this left action commutes with the right $\mathcal{R}^\times$ diagonal action, since
\begin{align}
    \begin{pmatrix}
        (a \, z_1) \, \lambda + (b \, z_2) \, \lambda \\[0.9mm]
        (c \, z_1) \, \lambda + (d \, z_2) \, \lambda
    \end{pmatrix}
    = 
    \begin{pmatrix}
        a \, (z_1 \, \lambda) + b \, (z_2 \, \lambda) \\[0.9mm]
        c \, (z_1 \, \lambda) + d \, (z_2 \, \lambda)
    \end{pmatrix}
    .
\end{align}
Thus, the action of $\GL_2 (\mathcal{R})$ descends to the projective line $\mathbb{P}^1_{\nc}$. 

Recall that $\mathcal{Z}\br{\mathcal{R}}$ denote the center of $\mathcal{R}$. Define a normal subgroup $N \subset \GL_2\br{\mathcal{R}}$ of the group $\GL_2 (\mathcal{R})$~by
\begin{align}
    N
    &:= \pbr{
    \begin{pmatrix}
        a & 0 \\[0.9mm]
        0 & a
    \end{pmatrix}
    \,\,
    \Big|
    \,\,\,
    a 
    \in \mathcal{Z} (\mathcal{R}) \cap \mathcal{R}^\times
    }.
\end{align} 

\begin{defn}
The \textit{projective linear group} $\PGL_2(\mathcal{R})$ over $\mathcal{R}$ is the following quotient
\begin{align}
    \PGL_2(\mathcal{R})
    &:= {\GL_2(\mathcal{R})}/{N}
    .
\end{align}
\end{defn}
Note that the projective group $\PGL_2(\mathcal{R})$ acts \textit{effectively} on the projective line $\mathbb{P}^1_{\nc}$. 

\medskip
Now, we can introduce the Möbius transformation. Let $G \in \PGL_2 (\mathcal{R})$. Then on $\mathbb{P}^1_{\nc, f}$, it acts via
\begin{align}
    \begin{pmatrix}
        a & b \\[0.9mm]
        c & d
    \end{pmatrix}
    \, 
    \begin{pmatrix}
        z \\[0.9mm] 1
    \end{pmatrix}
    = 
    \begin{pmatrix}
        a \, z + b \\[0.9mm]
        c \, z + d
    \end{pmatrix}
    .
\end{align}
The latter is an element of $\mathbb{P}^1_{\nc, f}$ iff
\begin{align}
    \begin{pmatrix}
        a \, z + b \\[0.9mm]
        c \, z + d
    \end{pmatrix}
    \sim
    \begin{pmatrix}
        (a \, z + b) \, (c \, z + d)^{-1} \\[0.9mm]
        1
    \end{pmatrix}
    .
\end{align}

\begin{defn}
\label{def:mobius}
A \textit{Möbius transformation} over $\mathcal{R}$ is a map of the form
\begin{align}
    z 
    &\mapsto (a \, z + b) \, (c \, z + d)^{-1}
    ,    
\end{align}
where $\small\begin{pmatrix}a & b \\ c & d \end{pmatrix}\in \GL_2(\mathcal{R})$.
\end{defn}

As in the commutative case, it forms a group under the composition.
\begin{prop}
Let $G_1$, $G_2 \in \PGL_2 (\mathcal{R})$. Then $G_1 G_2 = G_{12} \in \PGL_2 (\mathcal{R})$.
\end{prop}
\begin{proof}
Let
\begin{align}
&&
    G_1
    &= 
    \begin{pmatrix}
        a_1 & b_1 \\[0.9mm]
        c_1 & d_1
    \end{pmatrix}
    ,
    &
    G_2
    &= 
    \begin{pmatrix}
        a_2 & b_2 \\[0.9mm]
        c_2 & d_2
    \end{pmatrix}
    .
    &&
\end{align}
Then
\begin{align}
    G_1 \, G_2 (z)
    &= 
    \br{
    a_1 (a_2 \, z + b_2) + b_1
    } \, \br{
    c_1 (c_2 \, z + d_2) + d_1
    }^{-1}
    \\
    &= \br{
    (a_1 a_2 + b_1 c_2) \, z + (a_1 b_2 + b_1 d_2)
    } \, \br{
    (c_1 a_2 + d_1 c_2) \, z + (c_1 b_2 + d_1 d_2)
    }^{-1}
    .
\end{align}
This coincides with the Möbius transformation defined by the matrix product $G_{12} = G_1 G_2 \in \PGL_2\br{\mathcal{R}}$:
\begin{align}
    G_1 G_2
    \equiv
    \begin{pmatrix}
        a_1 & b_1 \\[0.9mm]
        c_1 & d_1
    \end{pmatrix}
    \, 
    \begin{pmatrix}
        a_2 & b_2 \\[0.9mm]
        c_2 & d_2
    \end{pmatrix}
    = 
    \begin{pmatrix}
        a_1 a_2 + b_1 c_2 & 
        a_1 b_2 + b_1 d_2 \\[0.9mm]
        c_1 a_2 + d_1 c_2 &
        c_1 b_2 + d_1 d_2
    \end{pmatrix}
    \equiv G_{12}
    .
\end{align}
\end{proof}

\subsubsection{Non-commutative polynomials and formal curves}
Since in the Sakai surface theory, we are working with the $(2,2)$-curves, we are going to define them in the non-commutative setting. Let us consider the space $\mathbb{P}_{\nc}^1 \times \mathbb{P}_{\nc}^1$ carrying the non-commutative affine coordinates $\br{f, g}$. It has four affine charts: $\pbr{\br{f, g}}$, $\pbr{\br{f^{-1}, g}}$, $\pbr{\br{f, g^{-1}}}$, and $\pbr{\br{f^{-1}, g^{-1}}}$, which can be represented schematically as in Figure \ref{pic:affcharts}.
\medskip

Consider a unital free associative algebra $\mathcal{A} = \mathbb{C}\langle f,g\rangle$ over $\mathbb{C}$. Its elements $P = P(f,g) \in \mathcal{A}$ are called \textit{non-commutative polynomials in the variables} $(f,g)$. The degree of a non-commutative polynomial is defined as the maximum of the total degrees of its monomial terms. We say that $P(f,g)$ is \textit{homogeneous} if all its monomial terms have the same total degree. A scaling property similar to \eqref{eq:scpr}, with a central element $\lambda$, also holds in the non-commutative case.
\begin{exmp}
$P (f,g) = f^2 g + 2 f g f + g f^2$ is homogeneous of degree $3$.
\end{exmp}

\begin{defn}
A polynomial $P = P(f, g)$ is called \textit{irreducible} if the condition
\begin{align}
    P(f, g)
    &= P_1(f, g) \cdot P_2(f, g)
\end{align}
implies that either $P_1=P_1(f, g)$ or $P_2 = P_2(f, g)$ is a unit.
\end{defn}
\begin{exmp}
$P(f, g) = f g - \lambda \, g f$ is an irreducible non-commutative polynomial, which is reducible in commutative case since $P(f, g) = (1 - \lambda) f g$. 
\end{exmp}

\medskip
Let $P = P(f,g)$ be a non-commutative biquadratic polynomial in $(f,g)$. For our further purposes, it is convenient to define the polynomial $P (f, g)$ by using the matrix $M~=~\br{m_{ij}}_{1 \leq i,j \leq 3} \in \GL_3\br{\mathcal{Z}(\mathcal{R})}$ and two monomial vectors $\mathbf{f} := \begin{pmatrix} f^2 & f & 1 \end{pmatrix}^T$ and $\mathbf{g} := \begin{pmatrix} g^2 & g & 1 \end{pmatrix}^T$ as {follows}:
\begin{gather}
    P(f,g)
    := 
    \begin{pmatrix} f^2 & f & 1 \end{pmatrix} \, \begin{pmatrix} 
    m_{00} & m_{01} & m_{02} \\ 
    m_{10} & m_{11} & m_{12} \\
    m_{20} & m_{21} & m_{22}
    \end{pmatrix}
    \, 
    \begin{pmatrix} g^2 \\ g \\ 1 \end{pmatrix}
    .
\end{gather}
Explicitly, this expression reads as
\begin{gather}
    \label{eq:curve}
    P(f,g)
    = 
    m_{00} \, f^2 \, g^2
    + m_{01} \, f^2 \, g
    + m_{02} \, f^2
    + m_{10} \, f \, g^2
    + m_{11} \, f \, g
    + m_{12} \, f
    + m_{20}  \, g^2
    + m_{21} \, g
    + m_{22}
    .
\end{gather}

Now we can formulate a non-commutative version of a biquadratic curve. 
\begin{defn}
\label{def:nccurve}
A \textit{biquadratic {\rm(}formal{\rm)} curve} $\mathcal{C}$ is given by the set $\mathcal{C} = \pbr{(f, g) \in \mathbb{P}_{\nc}^1 \times \mathbb{P}_{\nc}^1 \,\, \big| \,\, P(f, g) = 0}$.
\end{defn}

To examine the curve in different coordinate charts, one rewrites the equation accordingly. For instance, in the chart $\br{F, g} := \br{f^{-1}, g}$, the curve becomes
\begin{gather}
    m_{00} \, g^2
    + m_{01} \, g
    + m_{02}
    + m_{10} \, F \, g^2
    + m_{11} \, F \, g
    + m_{12} \, F
    + m_{20} \, F^2 \, g^2
    + m_{21} \, F^2 \, g
    + m_{22} \, F^2
    = 0
    .
\end{gather}
\begin{rem}
Definition \ref{def:nccurve} can be easily adapted to the curves of other bidegrees $(k, l)$, with $k, l \in \mathbb{Z}_{\geq 0}$ by considering the matrix $M \in \Mat_{k \times l} \br{\mathcal{Z}(\mathcal{R}})$ and the monomial vectors $\mathbf{f}$ and $\mathbf{g}$ of sizes $k$ and $l$ respectively.
\end{rem}

Suppose that a polynomial $P = P(f, g)$ can be factorized into irreducible homogeneous polynomials $P_1 = P_1(f, g)$, $P_2 = P_2(f, g)$, $\dots$, $P_k = P_k(f, g)$ of the multiplicities $m_1$, $m_2$, $\dots$, $m_k$ respectively. Note that $P = P(f, g)$ might have the form
\begin{align}
    P
    &= P_1 \cdot P_2^{m_2} \cdot P_1^{(m_1 - 1)} \cdot P_3^{m_3} \cdot \dots \cdot P_k^{m_k}
    .
\end{align}
Then, we say that the formal curve decomposes into \textit{irreducible components} $\mathcal{C}_1$, $\mathcal{C}_2$, $\dots$, $\mathcal{C}_k$, and write
\begin{align}
    \mathcal{C}
    &= m_1 \mathcal{C}_1 + m_2 \mathcal{C}_2 + \dots 
    + m_k \mathcal{C}_k.
\end{align}

\medskip

Let us define the base points of a pencil of formal curves similarly to the commutative case. Let~$P_1 = P_1 (f,g)$, $P_2 = P_2(f,g) \in \mathbb{C} \abr{f,g}$ be two non-commutative homogeneous polynomials without common factors. 
\begin{exmp}
$P_1(f,g) = 1 + f \, g$ and $P_2 (f,g) = 1 + g \, f$ have no common factors.
\end{exmp}
\begin{defn}
\label{def:ncpencil}
The set $\mathcal{P} = \pbr{\lambda \, P_1(f,g) + \mu \, P_2(f,g) = 0 \,\, \big| \,\, \lbr{\lambda : \mu} \in \mathbb{P}^1}$ is called a \textit{non-commutative pencil of formal biquadratic curves}, or \textit{$(2,2)$-curves}.
\end{defn}

\begin{defn}
\label{def:ncbasept}
Let $\mathcal{P}$ be a given non-commutative pencil of formal biquadratic curves. Elements of the set
\begin{align}
    \mathcal{B}
    &= \pbr{
    \br{f,g} \in \mathbb{P}^1_{\nc} \times \mathbb{P}^1_{\nc} \,\, \big| \,\,
    P_1(f,g) = 0 \,\, \wedge \,\, P_2 (f,g) = 0
    }
\end{align}
are called \textit{base points} of the non-commutative pencil $\mathcal{P}$.
\end{defn}

This definition is analogous to the commutative geometric case: base points are those that lie on all members of a linear system, i.e. they are common solutions to a family of curves defined by varying coefficients. 

\begin{rem}
The notion of singular points of a non-commutative curve can be defined analogously to the commutative case using non-commutative partial derivatives, such as those introduced by M. Kontsevich in \cite{kontsevich1993formal}. However, we focus only on base points and omit a formal definition of singularities. 
\end{rem}

\subsubsection{Blow-ups and divisors}
In Definition~\ref{def:blups}, we defined the blow-up at a point using affine charts; this construction naturally extends to the non-commutative setting.

\medskip
Let $X_0 = \mathbb{P}_{\nc}^1 \times \mathbb{P}_{\nc}^1$ carry the coordinates $(f,g)$. Consider transformations of this space which are (invertible) algebraic changes of these non-commutative variables.

\begin{defn}
\label{def:ncblup}
Let $p = \br{f_0, g_0} \in \mathcal{B}$ be a point in $X_0$ given in one of the affine charts $U$. A \textit{blow up} at $p$ replaces the chart $U$ by two charts with coordinates $(f_1, g_1)$, $(f_2, g_2)$ defined by
\begin{align}
    \label{eq:ncblup}
    \left\{
    \begin{array}{lcl}
         f_1
         &=& \br{f - f_0} \, \br{g - g_0}^{-1}
         ,
         \\[2mm]
         g_1 
         &=& g - g_0, 
    \end{array}
    \right.
    &&&
    \left\{
    \begin{array}{lcl}
         f_2
         &=& f - f_0
         ,
         \\[2mm]
         g_2 
         &=& \br{f - f_0}^{-1} \, \br{g - g_0}
         .
    \end{array}
    \right.
\end{align}
The point $p$ is then replaced by the element $E = \pbr{g_1 = 0} \cup \pbr{f_2 = 0}$ called an \textit{exceptional set}.
\end{defn}

\begin{rem}
\label{rem:ncblup}
The order of multiplication in the non-commutative setting matters and may be chosen to suit the structure of a given system. In particular, the relations between $f$ and $g$---such as $f g = \lambda \, g f$ in quantum settings---can affect the form of the blow-up.
\end{rem}

\begin{exmp}
Let $\mathcal{R} = \mathbb{C}_{\lambda} \abr{f, g}$ be the coordinate ring of the quantum plane, with relation $f g = \lambda \, g f$ for $\lambda \in \mathbb{C}\setminus\pbr{0, 1}$. Let $p = \br{0,0}$. The first chart in \eqref{eq:ncblup} yields
\begin{align}
    &&
    f_1
    &= f \, g^{-1},
    &
    g_1
    &= g
    .
    &&
\end{align}
This transformation is well-defined in the localization where $g$ is non-zero. The exceptional set is $E = \pbr{f = 0}$.
\end{exmp}

\begin{defn}
\label{def:ncdiv}
We call a \textit{non-commutative divisor} ${D}$ a formal $\mathbb{Z}$-linear combination of formal curves $\mathcal{C}_i$:
\begin{align}
    {D}
    &= m_1 \mathcal{C}_1 + m_2 \mathcal{C}_2 + \dots + m_k \mathcal{C}_k
    .
\end{align}
The divisor ${D}$ is \textit{effective} if all $m_i \geq 0$.
\end{defn}

Similarly to commutative case, we give
\begin{defn}
\label{def:ncdivclass}
Two divisors $D_1$ and $D_2$ are \textit{linearly equivalent}, $D_1 \sim D_2$, if their defining formal curves belong to the same linear space of non-commutative bihomogeneous polynomials. The equivalence class is called the \textit{divisor class} $\mathcal{D}$.
\end{defn}

\subsubsection{Picard groups and Cremona isometries}
\label{sec:picard}
Suppose we have $n$ points on $X_0$ and denote by $\mathcal{X}_{\nc}$ the resulting non-commutative space after performing blow-ups at these points. 

{Let ${H}_1$ and ${H}_2$ stand for the divisors of the formal lines $f = 0$ and $g = 0$ respectively, and $\mathcal{E}_i$ are the classes of the exceptional sets arising from the blow-up at the points $p_i$, $i = 1, \dots, n$.
\begin{defn}
\label{def:ncpic}
A \textit{non-commutative Picard group} of $\mathcal{X}_{\nc}$ denoted $\Pic\br{\mathcal{X}_{\nc}}$ is a free $\mathbb{Z}$-module group generated by the classes $\mathcal{H}_1$, $\mathcal{H}_2$, $\mathcal{E}_i$, $i = 1, \dots, n$:
\begin{align}
    \Pic \br{\mathcal{X}_{\nc}}
    &= \mathbb{Z} \mathcal{H}_1 
    \oplus \mathbb{Z} \mathcal{H}_2 
    \oplus \mathbb{Z} \mathcal{E}_1 
    \oplus \dots 
    \oplus \mathbb{Z} \mathcal{E}_n
    .
\end{align}
\end{defn}}

Let us equip the Picard group $\Pic\br{\mathcal{X}_{\nc}}$ with an intersection form.
\begin{defn}
\label{def:intformnc}
An \textit{intersection form} on $\Pic\br{\mathcal{X}_{\nc}}$ is the symmetric bilinear map
\begin{align}
    \br{- \, \cdot \, -}
    &: \, \Pic\br{\mathcal{X}_{\nc}} \times \Pic\br{\mathcal{X}_{\nc}} \to \mathbb{Z}
\end{align}
defined by the relations
\begin{align}
    &&
    \br{\mathcal{H}_i \cdot \mathcal{H}_j}
    &= 1 - \delta_{ij},
    &
    \br{\mathcal{H}_i \cdot \mathcal{E}_j}
    &= 0,
    &
    \br{\mathcal{E}_i \cdot \mathcal{E}_j}
    &= - \delta_{ij}
    ,
    &&
\end{align}
where $\delta_{ij}$ denotes the standard Kronecker delta.
\end{defn}

Let $V = \Pic\br{\mathcal{X}_{\nc}} \otimes \mathbb{Q}$ and consider the inner product $\abr{-,-} : V \times V \to \mathbb{Q}$ such that $\abr{\lambda, \mu} = - \br{\lambda, \mu}$ for any $\lambda$, $\mu \in V$. Consider the elements which we call \textit{simple roots}
\begin{align}
    &&
    \delta_0
    &= \mathcal{E}_1 - \mathcal{E}_2
    ,
    &&&
    \delta_1
    &= \mathcal{H}_1 - \mathcal{H}_2
    ,
    &&&
    \delta_2 
    &= \mathcal{H}_2 - \mathcal{E}_{1} - \mathcal{E}_{2}
    ,
    &&&
    \delta_i
    &= \mathcal{E}_{i - 1} - \mathcal{E}_i
    ,
    &
    i &= 3, \dots, n
    .
    &&
\end{align}
The matrix $C = \br{\abr{{\delta}_i, {\delta}_j}}_{0 \leq i,j \leq n}$ coincides with the generalized Cartan matrix of type $E_n^{(1)}$. Consider the element $-\mathcal{K}_{\mathcal{X_{\nc}}} $ which we call the \textit{null root}
\begin{align}
    \label{eq:ncantidiv}
    -\mathcal{K}_{\mathcal{X_{\nc}}} 
    &= 2 \mathcal{H}_1 + 2 \mathcal{H}_2 - \mathcal{E}_1 - \dots - \mathcal{E}_n
    .
\end{align}
Thus, similar to the commutative case, we have the root lattice $Q(E_8^{(1)}) \subset \Pic\br{\mathcal{X}_{\nc}} \subset V$. Note that for $n = 8$ the decomposition \eqref{eq:ddec} also holds.

\begin{defn}
The null root $-\mathcal{K}_{\mathcal{X_{\nc}}} $ is called the \textit{non-commutative anti-canonical divisor class}.
\end{defn}

\begin{defn}
A non-commutative divisor class $\mathcal{D}$ is of \textit{canonical type} if for any irreducible component $\mathcal{D}_i$ of $\mathcal{D}$ we have $\br{- \mathcal{K}_{\mathcal{X_{\nc}}} \cdot \mathcal{D}_i} = 0$.
\end{defn}

\begin{defn}
The space $\mathcal{X}_{\nc}$ is called a \textit{non-commutative version of the generalized Halphen surface of index zero} if it has a unique non-commutative anti-canonical divisor class of canonical type.
\end{defn}

\begin{defn}
\label{def:crisnc}
\textit{Cremona isometries} are automorphisms of $\Pic\br{\mathcal{X}_{\nc}}$ such that they preserve the intersection form, the element $-\mathcal{K}_{\mathcal{X}_{\nc}}$, and effectiveness of each effective divisor of $\Pic(\mathcal{X}_{\nc})$.
\end{defn}

\subsubsection{From surfaces to discrete dynamics: first steps}
\label{sec:ncsurftosys}

Let us fix $n = 8$ and consider the space $\mathcal{X}_{\nc}$ obtained by blowing up of $X_0 = \mathbb{P}^1_{\nc}\times\mathbb{P}^1_{\nc}$ at eight points $p_i$, $i = 1, \dots, 8$. This leads to the Picard group $\Pic(\mathcal{X}_{\nc})$ containing a lattice of type~$E_8^{(1)}$. Similarly to the commutative case, we consider two sublattices $Q(R) = \spn_\mathbb{Z}\pbr{\delta_i}$ and $Q(R^\bot) = \spn_\mathbb{Z}\pbr{\alpha_j}$ in $Q(E_8^{(1)})$, where $\abr{\delta_i \cdot \alpha_j} = 0$ for any $i$, $j$. We call $R$ and $R^\bot$ the \textit{surface type} and the \textit{symmetry type} of $\mathcal{X}_{\nc}$, respectively. Affine Weyl groups associated with them acts on the Picard lattice $\Pic\br{\mathcal{X_{\nc}}}$ by reflections $s_{\alpha_i}$ with respect to the simple roots $\alpha_i$:
\begin{align}
    \label{eq:ncrefl}
    &&
    s_{\alpha_i}(\mathcal{L}) 
    &= \mathcal{L} 
    + (\alpha_i \cdot \mathcal{L}) \, \alpha_i,
    &
    \mathcal{L} 
    &\in \Pic(\mathcal{X}_{\nc})
    ,
    &&
\end{align}
where $\br{- \, \cdot \, -}$ stands for the intersection form given in Definition \ref{def:intformnc}.

Discrete dynamical systems of Painlevé type in the non-commutative setting arise from a \textit{translation operator} $T$ in the extended affine Weyl group $\widetilde W(R^{\bot})$ associated with the symmetry root system $R^{\perp}$ if the group acts on $\Pic(\mathcal{X}_{\nc})$ as Cremona isometries. The action \eqref{eq:ncrefl} can be lifted to the non-commutative functions $(f,g)$ and results to an \textit{extended non-commutative birational representation} of the affine Weyl group, while non-commutative discrete dynamics reads as follows
\begin{align}
    T \br{f, g; \alpha_j}
    &= \br{\bar f, \, \bar g; \, \bar \alpha_j}
    .
\end{align}

\begin{defn}
A \textit{discrete non-commutative Painlevé equation} is a discrete dynamical system on the family $\mathcal{X}_{\nc}$ induced by a translation in $Q(R^\bot)$ if $\widetilde W(R^{\bot})$ acts on $\Pic(\mathcal{X}_{\nc})$ as Cremona isometries.
\end{defn}

\medskip
Although the entire framework of Sakai’s theory can, in principle, be formulated in the non-commutative setting, we do not aim to use it directly for the classification of non-commutative analogues of discrete Painlevé equations. Instead, the goal of our generalization is to describe the formal geometric structures that emerge from the non-commutative analogue of the sixth $q$-Painlevé equation, denoted by \ref{eq:qPA3}, and obtained via the method developed in \cite{bobrova2024affine}. Furthermore, the definitions presented above do not provide a clear path toward treating elliptic discrete dynamical systems in the non-commutative framework. In spite of the fact that a non-commutative analogue of the elliptic discrete Painlevé equation has been discussed in \cite{okounkov2015noncommutative}, this work does not contain an explicit description of the dynamics in concrete coordinates.
We also note that a recent attempt to formalize a non-commutative geometry for discrete equations appears in \cite{rains2025noncommutative}, where a module-theoretic approach is proposed. However, the connection between that framework and concrete birational dynamics remains to be clarified.

\subsection{Discrete systems}
\label{sec:ncdsys}

In this section, we develop a non-commutative algebraic framework for discrete integrable systems, with a focus on Painlevé-type equations. We begin by formulating non-commutative ordinary discrete equations (O$\Delta$Es) over the division ring $\mathcal{R}$, using a shift operator to define discrete dynamics. These systems naturally split into autonomous, additive ($d$-type), and multiplicative ($q$-type) classes, depending on how the parameters evolve under the shift. Although a natural extension would include elliptic-type systems, we do not treat this case here due to the lack of known non-commutative examples.

We then discover the existence of first integrals, which are preserved under certain maps. This leads to explicit constructions of first integrals for specific families of systems, including the non-commutative analog of the sixth $q$-Painlevé equation. 

Finally, we briefly discuss the affine Weyl group approach, which we use in order to derive a non-commutative analog for the sixth $q$-Painlevé equation.

\subsubsection{Non-commutative ordinary discrete equations}
\label{sec:ncodde}

In this subsection we introduce the basic setup for non-commutative ordinary discrete equations necessary for this work. 

Let $\mathcal{R}$ be a division ring with non-commutative elements ${\rm f} = \pbr{f_1, f_2, \dots, f_k}$ and ${\rm b} = \pbr{b_1, b_2, \dots, b_l}$ be the set of elements from the center $\mathcal{Z}(\mathcal{R})$ of $\mathcal{R}$. We will call $f_i$ \textit{functions} and $b_j$ \textit{parameters}. There is a natural involution on $\mathcal{R}$ called the transposition.  

\begin{defn}
\label{def:tau}
The \textit{transposition} $\tau : \, \mathcal{R} \to \mathcal{R}$ is an involutive $\mathbb{C}$-linear map such that
\begin{align}
    &&
    \tau\br{f_i}
    &= f_i,
    &
    \tau\br{b_j}
    &= b_j,
    &
    \tau\br{F_1 \, F_2}
    &= \tau\br{F_2} \, \tau\br{F_1}, 
    \quad
    F_1, \, F_2 \in \mathcal{R}
    .
    &&
\end{align}
\end{defn}
\begin{rem}
\label{rem:taumat}
In case $\Mat_n\br{\mathcal{R}}$, the $\tau$-action on the matrices $M = \br{m_{ij}}$, $i, j = 1, \dots, n$ extends as
\begin{align}
    \tau\br{m_{ij}}
    &= \br{\tau\br{m_{ji}}}
    .
\end{align}
\end{rem}

\begin{exmp}
$\tau\br{b \, f_1 \, f_2 \, \sigma_2} = - b \, f_2 \, f_1 \,  \sigma_2$, where $\sigma_2$ is a standard Pauli matrix. 
\end{exmp}

Consider $f_i$ and $b_j$ depending on the index $n \in \mathbb{Z}$ and write $f_{i, n}$, $b_{j, n}$ and ${\rm f}_n$, ${\rm b}_n$, respectively. We define an evolution on $\mathcal{R}$ by using a shift operator.
\begin{defn}
Let $T: \mathcal{R} \to \mathcal{R}$ be a $\mathbb{C}$-linear operator called a \textit{shift operator} which acts on the set of functions and parameters by a shift with respect to $n$:
\begin{align}
    &&
    T(b_{j, n}) 
    &= b_{j, n + 1}
    ,
    &
    T(f_{i, n}) 
    &= f_{i, n + 1}
    .
    &&
\end{align}
\end{defn}
An iteration of $T$ we write as $T^m(f_{i, n}) = f_{i, n + m}$ and $T^m(b_{j, n}) = b_{j, n + m}$. In case when $m = \pm 1$, we will often omit the index $n$ and write $T(f_i) = \bar f_i$, $T^{-1} (f_i) = \underline{f} {}_i$ for a function $f_i \in \mathcal{R}$ and $T(b_j) = \bar b_j$, $T^{-1} (b_j) = \underline{b}{}_j$ for a parameter $b_j \in \mathcal{Z(}\mathcal{R})$.

\medskip 
Consider the elements $F_i({\rm f}_n, {\rm b}_n) \in \mathcal{R}$ and $G_j ({\rm b}_n) \in \mathcal{Z}\br{\mathcal{R}}$ which depend on the sets ${\rm f}_n$ and ${\rm b}_n$. 
\begin{defn}
\label{def:ncdiscrsys}
A set of relations of the form
\begin{align}
    \label{eq:ncdiscrsys}
    &&
    T\br{f_{i, n}}
    &= F_i \br{{\rm f}_n, {\rm b}_n},
    &
    T\br{b_{j, n}}
    &= G_j \br{{\rm b}_n},
    &
    i
    &= 1, \dots, k,
    &
    j
    &= 1, \dots, l
    &&
\end{align}
is called a \textit{system of non-commutative ordinary discrete equations} (O$\Delta$Es).
\end{defn}
\begin{rem}
Note that sometimes the system \eqref{eq:ncdiscrsys} can be reduced to a single relation called a \textit{non-commutative analogue for a $k$-th order {\rm O$\Delta$E}}. 
\end{rem}
\begin{exmp}
Let $k = 4$ and $l = 1$. The system
\begin{gather}
    \label{eq:ncS4_sys}
    \begin{aligned}
    &\begin{aligned}
    T\br{f_{1,n}}
    &= f_{2,n},
    &&&
    T\br{f_{2,n}}
    &= f_{3,n}
    ,
    &&&
    T\br{f_{3,n}}
    &= f_{4,n},
    \end{aligned}
    \\[1mm]
    &\begin{aligned}
    T\br{f_{4,n}}
    &= b_{1,n}^2 \, f_{2,n} + b_{1,n}^2 \, f_{3,n} \, \br{
    f_{1,n}^{-1} - f_{4,n}^{-1}
    } \, f_{3,n}
    ,
    &&&
    T(b_{1,n})
    &= b_{1,n}
    \end{aligned}
    \end{aligned}
\end{gather}
can be rewritten as the following fourth order non-commutative O$\Delta$E for $f_n := f_{4,n}$
\begin{align}
    \label{eq:ncS4}
    f_{n + 4}
    &= b_{1,n}^2 \, f_{n + 1} + b_{1,n}^2 \, f_{n + 2} \, \br{
    f_{n}^{-1} - f_{n + 3}^{-1}
    } \, f_{n + 2}
    ,
\end{align}
which is called a \textit{non-commutative Somos-4 equation} \cite{bobrova2023non}.
\end{exmp}

Below we define first integrals of the system \eqref{eq:ncdiscrsys}. 

\begin{defn}
An element $I = I \br{{\rm f}_n, {\rm b}_n} \in \mathcal{R}$ is called a \textit{first integral} for system \eqref{eq:ncdiscrsys} if $T\br{I} = I$.
\end{defn}
\begin{exmp}
$I = b_{1,n}^{-1} \, \br{b_{1,n}^2 \, f_{2,n} \, f_{3,n}^{-1} + f_{3,n} \, f_{1,n}^{-1} + f_{4,n} \, f_{2,n}^{-1}}$ is a first integral of \eqref{eq:ncS4_sys}.
\end{exmp}

\medskip
We are interested in three specific types of $G_j({\rm b}_n)$, which determine the type of the discrete dynamics. Let $q = q\br{{\rm b}_n}$ be a central element of $\mathcal{R}$ such that $T(q) = q$.
\begin{defn}
\label{def:ncdiscrsys_types}
Let $G_j({\rm b}_n)$ be one of the following forms
\begin{itemize}
\item $G_j\br{{\rm b}_n} = b_{j, n}$ for all $j = 1, \dots, l$,
\vspace{1mm}

\item $G_j\br{{\rm b}_n} = b_{j, n} \pm 1$ for at least one $j$ and $G_i\br{{\rm b}_n} = b_{i, n}$ for the remaining $i \neq j$,
\vspace{1mm}

\item $G_j\br{{\rm b}_n} = q^{\varepsilon_j} \, b_{j, n}$, $\varepsilon_j = \pm 1$ for at least one $j$ and $G_i\br{{\rm b}_n} = b_{i, n}$ for the remaining $i \neq j$.
\vspace{1mm}
\end{itemize}
Then, the system \eqref{eq:ncdiscrsys} is called \textit{autonomous}, \textit{non-autonomous of additive type} ($d$-type), \textit{non-autonomous of multiplicative type} ($q$-type), respectively.
\end{defn}

\begin{rem}
Here we do not consider a non-commutative version of discrete elliptic systems.
\end{rem}

\begin{defn}
If $G_j({\rm b}_n) = b_{j, n}$, $b_{j, n}$ is called a \textit{constant parameter}.
\end{defn}

\begin{exmp}
The system \eqref{eq:ncS4_sys} is an autonomous system. The following system (eq. d-P$(E_7)$ in \cite{bobrova2024affine})
\begin{align}  
    \label{eq:dPE7sys}
    &&
    \bar f_1
    &= - f_{1} - b_{1} \, f_{2}^{-1}
    ,
    &
    \bar f_2
    &= - f_{2} + b_{2}
    + 2 \br{- f_{1} - b_{1} \, f_{2}^{-1}}^2,
    &
    \bar b_1
    &= b_{1} - 1
    &&
\end{align}
is of non-autonomous additive type. It has the first integral $I = f_{1} \, f_{2} - f_{2} \, f_{1}$. The~\ref{eq:qPA3} system constructed in Subsection \ref{sec:qPA3_affW} is an example of non-autonomous multiplicative type. 
\end{exmp}

\medskip
Though not the focus of this paper, discrete non-commutative systems can be represented via Lax pairs, which are considered as a criterion of integrability. For the completeness, we present their definitions here. 

We introduce a parameter $\lambda$ such that $T\br{\lambda} = \lambda$ and equip $\mathcal{R}$ with a derivation $d_{\lambda}$ which is a $\mathbb{C}$-linear map satisfying the Leibnitz rule and such that $d_{\lambda} \br{f_{i, n}} = 0$ and $d_{\lambda} \br{b_{j, n}} = 0$ for any $i = 1, \dots, k$ and $j = 1, \dots, l$. 
Let $\mathbf{\Psi}_n (\lambda)$, $\mathbf{L}_n (\lambda) = \mathbf{L}_n \br{\lambda, {\rm f}_n, {\rm b}_n}$, $\mathbf{M}_n \br{\lambda} = \mathbf{M}_n \br{\lambda, {\rm f}_n, {\rm b}_n}$ be $r \times r$ matrices  over $\mathcal{R}$. 
\begin{defn}
Let the autonomous system \eqref{eq:ncdiscrsys} be equivalent to the compatibility condition
\begin{align}
    \label{eq:disos}
    \mathbf{L}_{n + 1}(\lambda) \, \mathbf{M}_n(\lambda)
    = \mathbf{M}_n(\lambda) \, \mathbf{L}_n(\lambda)
\end{align}
of the following linear system
\begin{align}
    &&&&
    \mathbf{L}_n\br{\lambda} \, \mathbf{\Psi}_n(\lambda)
    &= \lambda \, \mathbf{\Psi}_n(\lambda),
    &
    \mathbf{\Psi}_{n + 1} (\lambda)
    &= \mathbf{M}_n(\lambda) \, \mathbf{\Psi}_n(\lambda)
    .
    &&&&
\end{align}
Then matrices $\mathbf{L}_n = \mathbf{L}_n (\lambda)$, $\mathbf{M}_n = \mathbf{M}_n (\lambda)$ are called a {\textit{discrete isospectral Lax pair}} for the system \eqref{eq:ncdiscrsys}.
\end{defn}

\begin{exmp}
Set $\varphi_n := b^{-1} \,\, f_{n + 2} \, f_{n}^{-1}$ and $\psi_n := b \,\, f_n \, f_{n + 1}^{-1}$. The equation \eqref{eq:ncS4} has the following Lax pair
\begin{align}
    &&
    \mathbf{L}_n (\lambda)
    &= 
    \begin{pmatrix}
    \lambda \, \br{
    \lambda^2 + \psi_{n + 1} + \varphi_{n + 1}
    }
    & 
    \br{
    \lambda^2 + \psi_{n + 1}
    } \, \varphi_n 
    \\[0.9mm]
    \lambda^2 + \psi_n
    & 
    \lambda \, \phi_n 
    \end{pmatrix}
    ,
    &
    \mathbf{M}_n (\lambda)
    &= 
    \begin{pmatrix}
        \lambda & \varphi_n
        \\[0.9mm]
        1 & 0
    \end{pmatrix}
    .
    &&
\end{align}
\end{exmp}

\begin{defn}
If the non-autonomous system \eqref{eq:ncdiscrsys} of $d$-type is equivalent to the compatibility condition
\begin{align}
    \label{eq:disom_d}
    d_\lambda \mathbf{M}_n (\lambda)
    + \mathbf{M}_n(\lambda) \, \mathbf{L}_n(\lambda)
    = \mathbf{L}_{n + 1}(\lambda) \, \mathbf{M}_n(\lambda)
\end{align}
of the following linear system
\begin{align}
    &&&&
    d_{\lambda} \, \mathbf{\Psi}_n(\lambda)
    &= \mathbf{L}_n (\lambda) \, \mathbf{\Psi}_n(\lambda),
    &
    \mathbf{\Psi}_{n + 1} (\lambda)
    &= \mathbf{M}_n(\lambda) \, \mathbf{\Psi}_n(\lambda)
    ,
    &&&&
\end{align}
then matrices $\mathbf{L}_n = \mathbf{L}_n (\lambda)$, $\mathbf{M}_n = \mathbf{M}_n (\lambda)$ are called an {\textit{isomonodromic Lax pair of $d$-type}} for the system \eqref{eq:ncdiscrsys}.
\end{defn}
\begin{exmp}
The system \eqref{eq:dPE7sys} has the following Lax pair
\begin{align}
    \mathbf{L}_n (\lambda)
    &= 
    \begin{pmatrix}
        1 & 0 \\[0.9mm]
        0 & -1
    \end{pmatrix}
    \lambda^2
    + 
    \begin{pmatrix}
        0 & 1 \\[0.9mm]
        2 f_{2} & 0
    \end{pmatrix}
    \lambda
    + 
    \begin{pmatrix}
        - f_{2} + \tfrac12 b_2 & - f_{1} \\[0.9mm]
        2 f_{2} f_{1} + 2 b_{1} & f_{2} - \tfrac12 b_2
    \end{pmatrix}
    ,
    \\[2mm]
    \mathbf{M}_n (\lambda)
    &= 
    \begin{pmatrix}
        - 2 & 0 \\[0.9mm] 0 & 0
    \end{pmatrix}
    \lambda
    + 
    \begin{pmatrix}
        - 2 f_{1} & - 1 \\[0.9mm] - 2 \bar f_{2} & 0
    \end{pmatrix}
    .
\end{align}
\end{exmp}

\begin{defn}
If the non-autonomous system \eqref{eq:ncdiscrsys} of $q$-type is equivalent to the compatibility condition
\begin{align}
    \label{eq:disom_q}
    \mathbf{M}_n (q \, \lambda) \, \mathbf{L}_n (\lambda)
    = \mathbf{L}_{n + 1} (\lambda) \, \mathbf{M}_n(\lambda)
\end{align}
of the following linear system
\begin{align}
    &&&&
    \mathbf{\Psi}_n(q \, \lambda)
    &= \mathbf{L}_n (\lambda) \, \mathbf{\Psi}_n(\lambda),
    &
    \mathbf{\Psi}_{n + 1} (\lambda)
    &= \mathbf{M}_n(\lambda) \, \mathbf{\Psi}_n(\lambda)
    ,
    &&&&
\end{align}
then matrices $\mathbf{L}_n = \mathbf{L}_n (\lambda)$, $\mathbf{M}_n = \mathbf{M}_n (\lambda)$ are called an {\textit{isomonodromic Lax pair of $q$-type}} for the system \eqref{eq:ncdiscrsys}.
\end{defn}

\begin{rem}
Similar to the commutative case, non-commutative discrete systems can be connected with their ``continuous'' analogs via a limiting procedure. One can take the change $z = \varepsilon \, n$ with the commutative parameter $\varepsilon$ supplemented by the maps
\begin{align}
    &&
    f_n 
    &= F,
    &
    f_{n + k}
    &= F 
    + k \varepsilon \, d_z \br{F}
    + \tfrac12 k^2 \varepsilon^2 \, d_z^2 \br{F}
    + O\br{\varepsilon^3}.
    &&
\end{align}
Here $d_z$ is a derivation\footnote{A $\mathbb{C}$-linear map from $\mathcal{R}$ to $\mathcal{R}$ satisfying the Leibnitz rule.} of $\mathcal{R}$ such that $d_z(z) = 1$. The latter must be chosen in such a way that the~limit~exists.
\end{rem}
\begin{exmp}
The change \cite{bobrova2024affine}
\begin{align}
    f_{1,n}
    &= 1 + \varepsilon^2 \, F_1 - \tfrac16 \, \varepsilon^3 \, F_2,
    &
    f_{2,n}
    &= - 2 + 2 \varepsilon^2 \, F_1 + \tfrac23 \, \varepsilon^3 \, F_2,
    &
    b_1
    &= 4 + \tfrac23 \, \varepsilon^4 \, z,
    &
    b_2
    &= - 6 + \tfrac13 \, \varepsilon^4 \, z
\end{align}
transforms the system \eqref{eq:dPE7sys} to the non-commutative first Painlevé equation written as the system
\begin{align}
    &&
    d_z \, F_1
     &= F_2,  
    &
    d_z F_2
    &= 6 F_1^2 + z
    .
    &&
\end{align}
\end{exmp}

\subsubsection{Remarks on first integrals of certain systems}
\label{sec:firstint}

In this section, we study the structure of first integrals of non-commutative discrete systems of the form \eqref{eq:ncdiscrsys}, particularly those connected to Painlevé-type dynamics.

\medskip
As a motivation example, consider the \ref{eq:qPA3} system constructed in Subsection \ref{sec:qPA3_affW}:
\begin{align}
    \tag*{\ref{eq:qPA3}}
    \begin{multlined}
    \underline{f} \, f
    = b_7 b_8 \, \br{
    g + b_6
    } \, \br{
    g + b_8
    }^{-1} \, \br{
    g + b_5
    } \, \br{
    g + b_7
    }^{-1}
    ,
    \hspace{4cm}
    \\[1mm]
    \bar{g} \, g
    = b_3 b_4 \, \br{
    f + b_2
    } \, \br{
    f + b_4
    }^{-1} \, \br{
    f + b_1
    } \, \br{
    f + b_3
    }^{-1}
    .
    \end{multlined}
\end{align} 
Its right-hand sides are non-commutative rational functions in $f$ and $g$ with parameters $b_j$, $j = 1, \dots, 8$. Observe the identity
\begin{align}
    \br{f + b_1} \br{f + b_2}
    = f^2 + b_1 f + f b_2 + b_1 b_2
    = f^2 + f b_1 + b_2 f + b_2 b_1
    = \br{f + b_2} \br{f + b_1},
\end{align}
i.e. $\lbr{f + b_1, f + b_2} = 0$. This yields
\begin{align}
    \br{f + b_1} \br{f + b_2}^{-1}
    = \br{f + b_2}^{-1} \br{f + b_2} \br{f + b_1} \br{f + b_2}^{-1}
    = \br{f + b_2}^{-1} \br{f + b_1}. 
\end{align}
Thus, we generalize these observations in the results below. Let $\mathcal{A} = \mathbb{C} \abr{{\rm f}} = \mathbb{C} \abr{f_1, \dots, f_k}$.

\begin{defn}
\label{def:ncratcan}
An element of the form 
\begin{align}
    \label{eq:ncratcan}
    P\br{F} 
    &= \prod_{1 \leq i \leq n}\br{F + b_i}^{\varepsilon_i}   
    ,
\end{align}
where $F \in \mathcal{A}$, $b_i \in \mathcal{Z} \br{\mathcal{R}}$, $i = 1, \dots, n$, and $\varepsilon_i \in \mathbb{Z}$ is called a \textit{non-commutative rational function in canonical form}. When all $\varepsilon_i = 1$, it is called a \textit{non-commutative polynomial in canonical form}. 
\end{defn}
\begin{lem}
\label{thm:ncpoly}
Let $P(F) = \prod_{1 \leq i \leq n} \br{F + b_i}$ be polynomial in $F \in \mathcal{A}$, where all $b_i \in \mathcal{Z} \br{\mathcal{R}}$. Then $P(F)$ is invariant under the permutations of $b_i$ for any $i = 1, \dots, n$.
\end{lem}\begin{proof}
Follows by induction, using the associative and distributive laws.
\end{proof}
\begin{corl}
\label{thm:ncrat}
Let $P\br{F} = \prod_{1 \leq i \leq n}\br{F + b_i}^{\varepsilon_i}$, where $F \in \mathcal{A}$, $b_i \in \mathcal{Z} \br{\mathcal{R}}$, $i = 1, \dots, n$, and $\varepsilon_i \in \mathbb{Z}$. Then $P(F)$ is invariant under the permutations of $b_i$ for any $i = 1, \dots, n$.
\end{corl}

Thanks to the lemma and its corollary, one can investigate first integrals of certain discrete systems. We~set $k = 2$ and denote $f := f_1$ and $g := f_2$. 
Due to known examples of discrete systems of the Painlevé-type, we are interested in the systems of one of the following forms
\begin{align}
\label{eq:dissys_pt}
    &&
    &
    \left\{
    \begin{array}{lcl}
     f \, {\bar f}
     &=& P_1 \br{\bar g},  
     \\[2mm]
     \bar g \, g
     &=& P_2 \br{f},  
    \end{array}
    \right.
    &&&
    &
    \left\{
    \begin{array}{rcl}
     f \, \bar f
     &=& P_1 \br{g},  
     \\[2mm]
     \bar g + g
     &=& P_2 \br{\bar f},  
    \end{array}
    \right.
    &&&
    &
    \left\{
    \begin{array}{lcl}
    \bar f + f
     &=& P_1 \br{g},  
     \\[2mm]
     \bar g + g
     &=& P_2 \br{\bar f}
     ,
    \end{array}
    \right.
    &&
\end{align}
where $P_1$ and $P_2$ satisfy Definition \ref{def:ncratcan}. 
\medskip

We now apply the statements above to this discrete systems. 
\begin{prop}
\label{thm:firstint_qcase}
Consider the first system in \eqref{eq:dissys_pt}, i.e. 
\begin{align}
    \label{eq:ncdsys_qtype}
    &&
    f \, \bar{f} 
    &= P_1 (\bar g),
    &
    \bar{g} \, g
    &= P_2 (f), 
    &&
\end{align}
and two maps
\begin{align}   
    &&
    T&: (f, g) \mapsto (\bar{f}, \bar{g}),
    &
    i&: (f, g) \mapsto (f^{-1}, g^{-1})
    .
    &&
\end{align}
Then the element $I (f, g) = f \, g^{-1} f^{-1} g$ is invariant under the map $\widetilde{T} = i \circ T$. 
\end{prop}
\begin{proof}
Indeed, since $\br{f \, \bar f} \, \br{f \, \bar f}^{-1} = \br{f \, \bar f}^{-1} \, \br{f \, \bar f} = 1$ and $P_1 \br{\bar g}$ satisfies Corollary \ref{thm:ncrat}, we have four identities
\begin{align}
    \br{f \, \bar f} \, \bar g \, \br{f \, \bar f}^{-1}
    &= \bar g,
    &
    \br{f \, \bar f}^{-1} \, \bar g \, \br{f \, \bar f}
    &= \bar g,
    &
    \br{f \, \bar f} \, \bar g^{-1} \, \br{f \, \bar f}^{-1}
    &= \bar g^{-1},
    &
    \br{f \, \bar f}^{-1} \, \bar g^{-1} \, \br{f \, \bar f}
    &= \bar g^{-1}.
\end{align}
In fact, they are equivalent to each other, so that it is enough to consider only the first one. Similarly, one can use the second equation of the system. As a result, we get two relations
\begin{align}
    \label{eq:fg_rels}
    \bar f \, \bar g \, \bar f^{-1}
    &= f^{-1} \bar g \, f,
    &
    g \, f \, g^{-1}
    &= \bar g^{-1} \, f \, \bar g
\end{align}
and their inverses
\begin{align}
    \bar f \, \bar g^{-1} \, \bar f^{-1}
    &= f^{-1} \bar g^{-1} \, f,
    &
    g \, f^{-1} \, g^{-1}
    &= \bar g^{-1} \, f^{-1} \, \bar g
    .
\end{align}
Taking the first identity in \eqref{eq:fg_rels} and simplifying it by using the second relation, one can get
\begin{align}
    \bar f \, \bar g \, \bar f^{-1}
    &= f^{-1} \bar g \, f
    = \br{\bar g^{-1} f}^{-1} f
    = \br{g \, f \, g^{-1} \bar g^{-1}}^{-1} f
    = \bar g \, g \, f^{-1} g^{-1} f
    &
    \Leftrightarrow&&
    \bar g^{-1} \bar f \, \bar g \, \bar f^{-1}
    &= g \, f^{-1} g^{-1} \, f.
\end{align}
Let $I (f, g) = g^{-1} f \, g \, f^{-1}$, then the latter can be rewritten in the form $T \br{I(f, g)} = I (f^{-1}, g^{-1})$ and, therefore, $\widetilde T \br{I(f, g)} = I (f, g)$.
Similarly, one can start with the second relation in \eqref{eq:fg_rels} and simplify it by using the first one:
\begin{align}
    g \, f \, g^{-1}
    &= \br{\bar g^{-1} f} \, \bar g
    = \br{f \, \bar f \, \bar g^{-1} \bar f^{-1}} \, \bar g
    &\Leftrightarrow&&
    f^{-1} g \, f \, g^{-1}
    &= \bar f \, \bar g^{-1} \bar f^{-1} \bar g
    .
\end{align}
Set $J (f, g) = f \, g^{-1} f^{-1} g$, then the latter reads as $T \br{J (f, g)} = J \br{f^{-1}, g^{-1}}$
and, hence, $\widetilde T \br{J (f, g)} = J (f, g)$. Finally, note that $I (f, g) = \br{J(f, g)}^{-1}$. 
\end{proof}

\begin{rem}
\label{rem:firstint_qcase}
The element $I = I(f, g)$ is preserved under each second iteration of the $T$-map, i.e. 
$$T^2 \br{I (f, g)} = I (f, g).$$
\end{rem}

\begin{rem}
\label{rem:kaw}
Our \ref{eq:qPA3} is a generalization of the matrix analog \eqref{eq:qPA3_mat} of the sixth $q$-Painlevé equation obtained in \cite{kawakami2020q} (see Theorem 5.2 therein). Note that, in the matrix case, the matrix variables $F$ and $G$ of the Kawakami system should satisfy the additional relation given by \eqref{eq:qPA3_matK} \cite[(4.34)]{kawakami2020q}, which is equivalent to the first integral given in Proposition \ref{thm:firstint_qcase} if one works with the matrix algebra. 
\end{rem}

\medskip
Now let us proceed with studying the first integrals related to the second system in~\eqref{eq:dissys_pt}.
\begin{prop}
\label{thm:firstint_qdcase}
Consider the second system in \eqref{eq:dissys_pt}, 
\begin{align}
    \label{eq:ncdsys_qdtype}
    &&
    f \, \bar{f} 
    &= P_1 (g),
    &
    \bar{g} + g
    &= P_2 (\bar f), 
    &&
\end{align}
and two mappings
\begin{align}   
    &&
    T
    &: (f, g) \mapsto (\bar{f}, \bar{g}),
    &
    \sigma_g
    &: (f, g) \mapsto (f, f^{-1} \, g \, f)
    .
    &&
\end{align}
Then the element $I (f, g) = g - f \, g \, f^{-1}$ is invariant under the map $\widetilde{T} = \sigma_g \circ T$. 
\end{prop}
\begin{proof}
The proof follows by substitution and manipulation using properties of $P_1$, $P_2$ and relations similar to those used in Proposition \ref{thm:firstint_qcase}. Indeed, we have an identity
\begin{align}
    \label{eq:fgrel_qd}
    &&
    \br{f \, \bar f}^{-1} g \, \br{f \, \bar f}
    &= g
    &\Leftrightarrow&&
    f^{-1} g \, f
    &= \bar f \, g \, \bar f^{-1}
    .
    &&
\end{align}
The second equation in \eqref{eq:ncdsys_qdtype} gives us
\begin{align}
    &&
    \bar g + g 
    &= P_2 \br{\bar f}
    &\Leftrightarrow&&
    \bar f \, \bar g \, \bar f^{-1}
    + \bar f \, g \, \bar f^{-1}
    &= P_2 \br{\bar f},
    &&
\end{align}
or, after the substitution \eqref{eq:fgrel_qd} into it,
\begin{align}
    &&
    \bar f \, \bar g \, \bar f^{-1}
    + f^{-1} \, g \, f
    &= P_2 \br{\bar f}.
    &&
\end{align}
Taking the difference of the second equation in \eqref{eq:ncdsys_qdtype} and the latter, we obtain
\begin{align}
    \bar g 
    - \bar f \, \bar g \, \bar f^{-1}
    &= f^{-1} \br{ 
    g - f \, g \, f^{-1}
    } \, f
    .
\end{align}
Setting $I \br{f, g} = g - f \, g \, f^{-1}$, it can be rewritten as
\begin{align}
    T \br{I\br{f, g}}
    &= f^{-1} \, I \br{f, g} \, f
    ,
\end{align}
or, after using the map $\sigma_g(f, g) = \br{f, f \, g f^{-1}}$, 
\begin{align}
    \sigma_g \br{T\br{I(f, g)}}
    &= \sigma_g \br{f^{-1} g \, f - g}
    = g - f \, g \, f^{-1}
    = I (f, g)
    ,
\end{align}
i.e. $I = I (f, g)$ is a first integral w.r.t. $\widetilde T$-dynamics.
\end{proof}

\begin{rem}
The element $I = I(f, g)$ is invariant under the $T^2$-action if $\lbr{f, \, \bar f \,} = 0$, since  
\begin{align}
    T^2 \br{I (f, g)} 
    &= T\br{f^{-1} \, I (f, g) \, f} 
    = \br{f \, \bar f}^{-1}\, \br{g - f \, g \, f^{-1}} \, \br{f \bar f} 
    \\
    &= \br{f \, \bar f}^{-1} g \br{f \bar f}
    - f \, \br{\bar f \, f}^{-1} g \, \br{\bar f \, f} \, f^{-1}
    = I\br{f,g}
    .
\end{align}
\end{rem}

\begin{rem}
Consider the system symmetric to \eqref{eq:ncdsys_qdtype}:
\begin{align}
    \label{eq:ncdsys_qdtype2}
    &&
    \bar{f} +  f
    &= P_1 (\bar g),
    &
    \bar g \, g
    &= P_2 (f)
    .
    &&
\end{align}
Then, one can formulate a similar to Proposition \ref{thm:firstint_qdcase} statement. Namely, the element $I (f, g) = f - g^{-1} f \, g$ is a first integral of the map $\widetilde T = \sigma_f \circ T$, where $\sigma_f \br{f, g} = \br{g^{-1} f \, g, \, g}$ and $T$ is defined by \eqref{eq:ncdsys_qdtype2}. Moreover, $I = I (f, g)$ is preserved under the $T^2$-action if $\lbr{\, g, \, \bar g \,} = 0$. 
\end{rem}

\medskip
Finally, let us consider the system of the third type in~\eqref{eq:dissys_pt}.
\begin{prop}
\label{thm:firstint_dcase}
For the third non-commutative discrete system given in \eqref{eq:dissys_pt}, i.e.
\begin{align}
    \label{eq:ncdsys_dtype}
    &&
    \bar f + {f} 
    &= P_1 (g),
    &
    \bar{g} + g
    &= P_2 (\bar f), 
    &&
\end{align}
the element $I (f, g) = f \, g - g \, f$ is a first integral. 
\end{prop}
\begin{proof}
The proof is again given by combining the relations arising from the system. To be precise, the first equation of the system \eqref{eq:ncdsys_dtype} leads to 
\begin{align}
    &&
    \bar f + f 
    &= P_1 (g)
    &\Leftrightarrow&
    &
    g \, \bar f \, g^{-1} + g \, f \, g^{-1}
    &= P_1 (g)
    .
    &&
\end{align}
Their difference reads as
\begin{align}
    \label{eq:grel}
    g \, \bar f \, g^{-1} + g \, f \, g^{-1}
    - \bar f - f
    &= 0
    &\Leftrightarrow&
    &
    g \, \bar f + g \, f
    - \bar f \, g - f \, g
    &= 0
    .
\end{align}
Similar arguments about the second equation in \eqref{eq:ncdsys_dtype} leads to
\begin{align}
    \label{eq:frel}
    \bar f \, \bar g \, \bar f^{-1} + \bar f \, g \, \bar f^{-1}
    - \bar g - g
    &= 0
    &\Leftrightarrow&
    &
    \bar f \, \bar g + \bar f \, g
    - \bar g \, \bar f - g \, \bar f
    &= 0
    .
\end{align}
Taking the sum of \eqref{eq:grel} and \eqref{eq:frel}, one arrives at
\begin{align}
    &&
    \bar f \, \bar g
    - \bar g \, \bar f
    - f \, g
    + g \, f
    &= 0
    &\Leftrightarrow&
    &
    T \br{f \, g - g \, f}
    = f \, g - g \, f
    ,
    &&
\end{align}
where $T (f, g) = \br{\bar f, \bar g}$, i.e. $I (f, g) = f \, g - g \, f$ is a first integral of the system \eqref{eq:ncdsys_dtype}.
\end{proof}

\begin{rem}
The system \eqref{eq:ncdsys_dtype} is invariant under the $\tau$-action.
\end{rem}

\begin{prop}
\label{thm:degfirstint}
There exists a degeneration of the first integrals
\begin{align}
    &&
    I_1(f, g)
    &= g^{-1} f \, g \, f^{-1}
    &
    &\to&
    I_2(f, g)
    &= g - f \, g \, f^{-1}
    &
    &\to&
    I_3 \br{f,g}
    &= f \, g - g \, f
    .
    &&
\end{align}
\end{prop}
\begin{proof}
Indeed, one should consider a formal Taylor series of an element $h \in \mathcal{R}$ with a small parameter $\varepsilon$. Then, the degeneration procedure is given by the following formula
\begin{align}
    h
    &= 1 + \varepsilon \, H,
\end{align}
which yields $h^{-1} = 1 - \varepsilon \, H + O\br{\varepsilon}$. Note that in the case when $I_k$, $k = 1, 2, 3$ are first integrals, one can make a shift and a rescaling by a non-zero constant that does not affect the dynamics. Thus, we have the chain of identities:
\begin{align}
    I_1 (f, G)
    &\sim \br{1 - \varepsilon \, G} \, f \, \br{1 + \varepsilon \, G} \, f^{-1}
    = 1 - \varepsilon \, \br{G - f \, g \, f^{-1}} + O \br{\varepsilon},
    \\[2mm]
    I_2 (F, g)
    &\sim g - \br{1 + \varepsilon \, F} \, g \, \br{1 - \varepsilon \, F}
    = - \varepsilon \, \br{F \, g - g \, F} + O \br{\varepsilon}
    .
    \\[-1.cm]
\end{align}
\end{proof}

\medskip
The discrete $d$-Painlevé systems obtained in the paper \cite{bobrova2024affine} (see Appendix A therein) are one of the forms \eqref{eq:dissys_pt}. Thanks to Proposition \ref{thm:firstint_dcase}, the systems d-P$\br{D_5}$, d-P$\br{D_6'}$, d-P$\br{E_6}$, and d-P$\br{E_7}$ have first integral $I \br{f, g} = f \, g - g \, f$. Regarding the remaining systems, we have the following
\begin{prop}
\label{thm:firstint_qdP}
The systems {\rm{d-P}}$\br{D_4}$, {\rm{d-P}}$\br{D_6}$, {\rm{d-P}}$\br{E_6'}$ have the first integral $I \br{f, g} = g - f^{-1} \, g \, f$, while the systems {\rm{d-P}}$\br{D_5'}$ and {\rm{d-P}}$\br{D_7}$ have the first integral $I \br{f, g} = f - g \, f \, g^{-1}$. 
\end{prop}

\subsubsection{Affine Weyl groups and discrete systems}
\label{sec:afftosys}

In Section \ref{sec:sakai}, we discussed how discrete dynamical systems can be constructed from configurations of eight points on $\mathbb{P}^1 \times \mathbb{P}^1$. This geometric approach provides a powerful and systematic framework for classifying discrete Painlevé equations in the commutative setting. Some first examples of discrete Painlevé equations did not arise from geometry, but, for instance, from studying the symmetry groups of the classical differential Painlevé equations \cite{fokas1993continuous}, due to the remarkable observation made by K. Okamoto. 

Specifically, the symmetries of classical differential Painlevé equations generate a group isomorphic to an extended affine Weyl group $\widetilde W \br{C}$, where $C$ is a generalized Cartan matrix (see \cite{okam1}, \cite{okam2}, \cite{okam3}, \cite{okam4}). Much like the constructions described in Section \ref{sec:sakai}, one can consider translations in this affine Weyl group to generate discrete time evolutions on the associated root lattice. This leads naturally to additive-type discrete Painlevé equations, where the parameters increment additively \cite{fokas1993continuous}.

Thanks to the translation elements in affine Weyl groups, one can define discrete dynamics purely algebraically. A foundational work in this direction is the paper by M. Noumi and Y. Yamada \cite{noumi1998affine} (see~also~\cite{shi2022translations}), which shows how an extended birational representations of affine Weyl groups naturally give rise to discrete systems. In some cases, these birational representations arise from the symmetries of certain differential systems, but more often they are postulated. A systematic way to obtain them is provided by the surface theory.

Since the approach introduced by N. Noumi and Y. Yamada is algebraic rather than geometric, it can be adapted to the non-commutative setting with relatively few modifications. In particular, this adaptation was carried out in \cite{bobrova2024affine}, where non-commutative analogues for several additive-type Painlevé equations were constructed. Here, we briefly recall the key aspects of this method. In Subsection \ref{sec:qPA3_affW}, we apply this technique to derive a non-commutative discrete system of multiplicative type. Under commutative reduction, this system recovers the well-known sixth $q$-Painlevé equation \cite{jimbo1996q}. We refer to this non-commutative system as \ref{eq:qPA3}, and in Subsection \ref{sec:qPA3surf_nc} we explain this notation by studying the associated non-commutative geometric structure.

\medskip
Let us begin with a generalized Cartan matrix $C = (c_{ij})$ indexed by $i, j \in I := \pbr{0, 1, \dots, n}$, corresponding to an affine root system. We denote the sets of simple roots and simple co-roots by $\Delta = \{\alpha_0, \dots, \alpha_n \}$, $\Delta^\vee = \{\alpha_0^\vee, \dots, \alpha_n^\vee \}$, where $\alpha_i$ and $\alpha_j^\vee$ belong to the center $\mathcal{Z}(\mathcal{R})$ of $\mathcal{R}$.
These sets form bases of dual vector spaces $V$ and $V^*$, and span the root and co-root lattices
\begin{align}
    &&
    Q
    &:= \spn_\mathbb{Z} \, \Delta,
    &
    Q^\vee
    &:= \spn_\mathbb{Z} \, \Delta^\vee.
    &&
\end{align}
We define the natural pairing $\abr{\, \cdot \, , \cdot \,}: Q \times Q^\vee \to \mathbb{Z}$ by $\abr{\alpha_i, \alpha_j^\vee} = c_{ij}$ and $\alpha_i^\vee = 2 \alpha_i / (\alpha_i, \alpha_i)$. 

The associated Weyl group $W = W(C)$ is generated by simple reflections $s_i$, $i \in I$:
\begin{align}
    W(C)
    &= \abr{
    s_0, s_1, \dots, s_n \, 
    \big| \, 
    s_i^2
    = 1, \,
    (s_i \, s_j)^{m_{ij}}
    = 1
    },
\end{align}
{where the exponents $m_{ij}$ are determined from the product $c_{ij} c_{ji}$ according to the standard table}
\vspace{-2mm}
\begin{table}[H]
    \centering
    \begin{tabular}{c||ccccc}
         $c_{ij} c_{ji}$
         & 0
         & 1
         & 2
         & 3
         & $\geq 4$
         \\
         \hline\hline
         $m_{ij}$
         & 2 
         & 3
         & 4
         & 6
         & $\infty$
    \end{tabular}
\end{table}
\vspace{-2mm}
These relations are the \emph{generalized braid {\rm(}Coxeter{\rm)} relations}, i.e.
\begin{align}
    \underbrace{s_i s_j s_i \cdots}_{m_{ij}\ \text{factors}}
    =
    \underbrace{s_j s_i s_j \cdots}_{m_{ij}\ \text{factors}}.
\end{align}

Each reflection $s_i$ acts on $Q$ via
\begin{align}
    s_i (\alpha_j)
    &= \alpha_j - \abr{\alpha_i, \alpha_j^\vee} \, \alpha_i
    = \alpha_j - c_{ij} \, \alpha_i.
\end{align}

These group actions extend naturally to automorphisms of the field $\mathbb{C} (\alpha) = \mathbb{C} (\alpha_i, \, i \in I)$ of rational functions in $\alpha_i$. 
In this way, $\mathbb{C}(\alpha)$ becomes a left $W$-module. 
\medskip

We now extend this representation to a larger field
\begin{align}
    \mathcal{K}\br{\alpha, f}
    := \mathbb{C} \br{\alpha} \abr{f_i \in \mathcal{R} \, | \, i \in I}
    ,
\end{align}
consisting of rational functions in the $\alpha_i$ and new elements $f_i \in \mathcal{R}$, which are \textit{non-commutative rational functions}. To ensure compatibility with the Weyl group structure, we must define the action of each $s_i \in W$ on $f_j$ so that the full action on $\mathcal{K}\br{\alpha, f}$ preserves the group structure of $W$.

\medskip
A key feature of affine Weyl groups is the existence of translation elements, often referred to as \textit{Kac translations} $t_{\mu} \in W$, where $\mu$ belongs to the lattice part $M$. 
For an untwisted affine root system, this lattice $M$ can be identified with the root lattice $Q_0$ of the underlying finite root system. Recall that the affine Weyl group decomposes as a semi-direct product $W = W_0 \ltimes M$,
where $W_0$ is the finite Weyl subgroup. The lattice part $M \simeq Q_0 $ acts as a group of translations (shift operators) on the root lattice. Since the null root $ \delta $ is $ W $-invariant, it is often fixed to a constant to serve as a scaling parameter in the action of $M$.

Suppose we have now extended the affine Weyl group action from $\mathbb{C}\br{\alpha}$ to the larger field $\mathcal{K}\br{\alpha, f}$, and treat it as a $W$-module. Each $t_{\mu} \in M$ defines a discrete evolution by acting on the functions $f_i$
\begin{align}
    &&
    \label{eq:dys}
    t_{\mu} (f_i)
    &= F_{\mu, i} (\alpha, f),
    &
    F_{\mu, i}(\alpha, f) 
    &\in \mathcal{K}\br{\alpha, f}
    .
    &&
\end{align}
This set together with the action of $t_{\mu}$ on $\alpha_j$ can be considered as a \textit{discrete dynamical system} (cf. with Definition~\ref{def:ncdiscrsys}). The elements $\alpha_j$ and $f_i$ play the role of parameters (possibly evolving under $t_{\mu}$) and the dynamical functions respectively. Similarly to Definition \ref{def:ncdiscrsys_types}, the action of $t_\mu$ on the $\alpha_j$ determines the type of the resulting system, and, thus, the latter can be classified into autonomous, additive ($d$-equations), and multiplicative ($q$-equations) types. As we have already mentioned, we do not consider elliptic-type systems here, as concrete non-commutative examples in this class are not yet available in the literature. 

\begin{rem}
Note that this construction can be also carried out within the extended affine Weyl group $\widetilde W$. 
\end{rem}

\section{A non-commutative analog of the sixth \texorpdfstring{$q$-Painlevé}{q-Painlevé} system}
\label{sec:qPA3}

In this section, we focus on a non-commutative analogue of the sixth $q$-Painlevé equation, denoted \ref{eq:qPA3}. Our goal is to construct and analyze this system from the view point given in Section \ref{sec:ncsurfandsys}. Thanks to the approach discussed in Subsection \ref{sec:afftosys}, we first derive this system using an extended birational representation of an affine Weyl group of type $D_5^{(1)}$. Next, we interpret the resulting system within the framework of non-commutative surface theory, showing how it fits naturally into a geometric context analogous to Sakai's surface theory. Moreover, we use it to obtain the extended birational representation given in Theorem~\ref{thm:WD5_nc}. Finally, in Subsection \ref{sec:coal}, we examine a coalescence cascade starting from the \ref{eq:qPA3} and descending to lower $q$-Painlevé systems as well as to $d$-Painlevé cases.
Notably, since the base points in our construction are central elements, they still can be permuted by automorphisms of the associated Dynkin diagrams. This makes it possible to perform the coalescence procedure using point configurations analogous to those in the commutative case. 

\subsection{From affine Weyl group to discrete system}
\label{sec:qPA3_affW}

Here we apply the method developed in \cite{bobrova2024affine} and briefly discussed in Subsection \ref{sec:afftosys} in order to derive a non-commutative analog of the sixth $q$-Painlevé equation, which corresponds to the $A_3^{(1)}$/$D_5^{(1)}$ surface/symmetry type in the Sakai classification. While the paper \cite{bobrova2024affine} gives examples of additive-type systems, below we use this approach to the multiplicative setting by considering multiplicative roots instead of additive ones. 

\medskip
We begin with the Cartan matrix $C = (c_{ij})$, where $i, j \in I := \{0, 1, \dots, 5\}$, of the affine type $D_5^{(1)}$. The associated Dynkin diagram $\Gamma(C)$ is shown below (with diagram automorphisms $\pi_1$, $\pi_2$, and $\pi_3 := \pi_1 \pi_2 \pi_1$):
\begin{table}[H]
    \vspace{1mm}
    \begin{tabular}{ccc}
    $C
    = 
    \begin{pmatrix}
    2 & 0 & -1 & 0 & 0 & 0
    \\[0.9mm]
    0 & 2 & -1 & 0 & 0 & 0
    \\[0.9mm]
    - 1 & - 1 & 2 & -1 & 0 & 0
    \\[0.9mm]
    0 & 0 & -1 & 2 & -1 & -1
    \\[0.9mm]
    0 & 0 & 0 & -1 & 2 & 0
    \\[0.9mm]
    0 & 0 & 0 & -1 & 0 & 2
    \end{pmatrix}
    $
    & \hspace{2cm} &
    \begin{tabular}{c}
    \\[-9mm]
    \scalebox{0.85}{\tikzset{every picture/.style={line width=0.75pt}} 

\begin{tikzpicture}[x=0.75pt,y=0.75pt,yscale=-1,xscale=1]

\draw    (72.36,155.62) -- (72.6,155.91) ;
\draw [shift={(73.84,157.48)}, rotate = 231.7] [color={rgb, 255:red, 0; green, 0; blue, 0 }  ][line width=0.75]    (6.56,-2.94) .. controls (4.17,-1.38) and (1.99,-0.4) .. (0,0) .. controls (1.99,0.4) and (4.17,1.38) .. (6.56,2.94)   ;
\draw    (73.25,63.89) ;
\draw [shift={(74.19,62.91)}, rotate = 134.08] [color={rgb, 255:red, 0; green, 0; blue, 0 }  ][line width=0.75]    (6.56,-2.94) .. controls (4.17,-1.38) and (1.99,-0.4) .. (0,0) .. controls (1.99,0.4) and (4.17,1.38) .. (6.56,2.94)   ;

\draw  [fill={rgb, 255:red, 0; green, 0; blue, 0 }  ,fill opacity=1 ] (74.15,160.6) .. controls (74.15,158.17) and (76.13,156.19) .. (78.56,156.19) .. controls (81,156.19) and (82.98,158.17) .. (82.98,160.6) .. controls (82.98,163.04) and (81,165.02) .. (78.56,165.02) .. controls (76.13,165.02) and (74.15,163.04) .. (74.15,160.6) -- cycle ;
\draw  [fill={rgb, 255:red, 0; green, 0; blue, 0 }  ,fill opacity=1 ] (74.15,59.73) .. controls (74.15,57.29) and (76.13,55.32) .. (78.56,55.32) .. controls (81,55.32) and (82.98,57.29) .. (82.98,59.73) .. controls (82.98,62.17) and (81,64.14) .. (78.56,64.14) .. controls (76.13,64.14) and (74.15,62.17) .. (74.15,59.73) -- cycle ;
\draw    (129.44,110.35) -- (82.33,157.46) ;
\draw    (81.67,63.24) -- (128.78,110.35) ;
\draw  [fill={rgb, 255:red, 0; green, 0; blue, 0 }  ,fill opacity=1 ] (124.59,110.35) .. controls (124.59,107.91) and (126.56,105.94) .. (129,105.94) .. controls (131.44,105.94) and (133.41,107.91) .. (133.41,110.35) .. controls (133.41,112.79) and (131.44,114.76) .. (129,114.76) .. controls (126.56,114.76) and (124.59,112.79) .. (124.59,110.35) -- cycle ;
\draw    (244.79,54.05) -- (245.41,54.52) ;
\draw [shift={(247,55.74)}, rotate = 217.51] [color={rgb, 255:red, 0; green, 0; blue, 0 }  ][line width=0.75]    (6.56,-2.94) .. controls (4.17,-1.38) and (1.99,-0.4) .. (0,0) .. controls (1.99,0.4) and (4.17,1.38) .. (6.56,2.94)   ;
\draw    (83.55,54.97) ;
\draw [shift={(82.43,56.06)}, rotate = 315.69] [color={rgb, 255:red, 0; green, 0; blue, 0 }  ][line width=0.75]    (6.56,-2.94) .. controls (4.17,-1.38) and (1.99,-0.4) .. (0,0) .. controls (1.99,0.4) and (4.17,1.38) .. (6.56,2.94)   ;
\draw  [fill={rgb, 255:red, 0; green, 0; blue, 0 }  ,fill opacity=1 ] (246.22,59.35) .. controls (246.22,56.91) and (248.2,54.94) .. (250.64,54.94) .. controls (253.07,54.94) and (255.05,56.91) .. (255.05,59.35) .. controls (255.05,61.79) and (253.07,63.76) .. (250.64,63.76) .. controls (248.2,63.76) and (246.22,61.79) .. (246.22,59.35) -- cycle ;
\draw  [fill={rgb, 255:red, 0; green, 0; blue, 0 }  ,fill opacity=1 ] (246.22,160.22) .. controls (246.22,157.79) and (248.2,155.81) .. (250.64,155.81) .. controls (253.07,155.81) and (255.05,157.79) .. (255.05,160.22) .. controls (255.05,162.66) and (253.07,164.64) .. (250.64,164.64) .. controls (248.2,164.64) and (246.22,162.66) .. (246.22,160.22) -- cycle ;
\draw    (246.87,62.86) -- (200.2,109.53) ;
\draw    (200.2,109.97) -- (246.87,156.64) ;
\draw  [fill={rgb, 255:red, 0; green, 0; blue, 0 }  ,fill opacity=1 ] (195.79,110.31) .. controls (195.79,107.88) and (197.76,105.9) .. (200.2,105.9) .. controls (202.64,105.9) and (204.61,107.88) .. (204.61,110.31) .. controls (204.61,112.75) and (202.64,114.73) .. (200.2,114.73) .. controls (197.76,114.73) and (195.79,112.75) .. (195.79,110.31) -- cycle ;
\draw    (195.79,110.31) -- (133.41,110.35) ;
\draw  [draw opacity=0][dash pattern={on 0.84pt off 2.51pt}] (77.62,160.86) .. controls (62.91,146.93) and (54.14,129.39) .. (54.14,110.33) .. controls (54.14,64.96) and (103.84,28.17) .. (165.14,28.17) .. controls (226.45,28.17) and (276.14,64.96) .. (276.14,110.33) .. controls (276.14,129.39) and (267.38,146.93) .. (252.67,160.86) -- (165.14,110.33) -- cycle ; \draw  [dash pattern={on 0.84pt off 2.51pt}] (77.62,160.86) .. controls (62.91,146.93) and (54.14,129.39) .. (54.14,110.33) .. controls (54.14,64.96) and (103.84,28.17) .. (165.14,28.17) .. controls (226.45,28.17) and (276.14,64.96) .. (276.14,110.33) .. controls (276.14,129.39) and (267.38,146.93) .. (252.67,160.86) ;  
\draw    (257.44,64.07) -- (257.2,63.77) ;
\draw [shift={(255.96,62.2)}, rotate = 51.7] [color={rgb, 255:red, 0; green, 0; blue, 0 }  ][line width=0.75]    (6.56,-2.94) .. controls (4.17,-1.38) and (1.99,-0.4) .. (0,0) .. controls (1.99,0.4) and (4.17,1.38) .. (6.56,2.94)   ;
\draw    (256.55,155.79) ;
\draw [shift={(255.61,156.77)}, rotate = 314.08] [color={rgb, 255:red, 0; green, 0; blue, 0 }  ][line width=0.75]    (6.56,-2.94) .. controls (4.17,-1.38) and (1.99,-0.4) .. (0,0) .. controls (1.99,0.4) and (4.17,1.38) .. (6.56,2.94)   ;

\draw (46.26,156) node [anchor=north west][inner sep=0.75pt]  [font=\large]  {$\alpha _{1}$};
\draw (46.27,55.58) node [anchor=north west][inner sep=0.75pt]  [font=\large,rotate=-0.45]  {$\alpha _{0}$};
\draw (120.6,85.08) node [anchor=north west][inner sep=0.75pt]  [font=\large]  {$\alpha _{2}$};
\draw (158.61,40.38) node [anchor=north west][inner sep=0.75pt]  [font=\large,rotate=-0.07]  {$\pi _{1}$};
\draw (264.86,155.59) node [anchor=north west][inner sep=0.75pt]  [font=\large]  {$\alpha _{4}$};
\draw (264.87,55.2) node [anchor=north west][inner sep=0.75pt]  [font=\large,rotate=-0.45]  {$\alpha _{5}$};
\draw (191.8,85.1) node [anchor=north west][inner sep=0.75pt]  [font=\large]  {$\alpha _{3}$};
\draw (65.61,116.01) node [anchor=north west][inner sep=0.75pt]  [font=\large,rotate=-270.13]  {$\pi _{2}$};
\draw (264.19,103.68) node [anchor=north west][inner sep=0.75pt]  [font=\large,rotate=-90.13]  {$\pi _{3}$};

\end{tikzpicture}}
    \end{tabular}
    \end{tabular}
    \vspace{0mm}
\end{table}
Let $\Delta := \{ \alpha_0, \alpha_1, \dots, \alpha_5 \}$ be the set of simple roots. The extended affine Weyl group $\widetilde W (D_5^{(1)})$ is defined by the generators $s_i$ and the diagram automorphisms as follows
\begin{gather}
    \label{eq:WD5}
    \begin{gathered}
    \widetilde{W}(D_5^{(1)})
    = \abr{s_0, s_1, s_2, s_3, s_4, s_5; \pi_1, \pi_2},
    \\[2mm]
    \begin{aligned}
    s_i^2
    &= 1,
    &&&&&
    (s_i \, s_{j})^2
    &= 1 \,\,\, (c_{ij} = 0),
    &&&&&
    (s_i \, s_{j})^3
    &= 1\,\,\, (c_{ij} = -1),
    &&&&&
    i, j
    &= 0, 1, 2, 3, 4, 5,
    \end{aligned}
    \\[1mm]
    \begin{aligned}
    \pi_k^2
    &= 1,
    &&&&&
    \pi_1 s_{\{0, 1, 2, 3, 4, 5\}}
    &= s_{\{5, 4, 3, 2, 1, 0\}} \pi_1,
    &&&&&
    \pi_2 s_{\{0, 1\}}
    &= s_{\{1, 0\}} \pi_2,
    &&&&&
    k
    &= 1, 2.
    \end{aligned}
    \end{gathered}
\end{gather}
Note that the $\widetilde W(D_5^{(1)})$ acts on the root lattice $Q := \spn_\mathbb{Z} \, \Delta$ by reflections:
\begin{align}
    \label{eq:refll}
    s_i (\alpha_j)
    &= \alpha_j - c_{ij} \, \alpha_i.
\end{align}
Now we define multiplicative roots $a = \br{a_0, a_1, \dots, a_5}$ and the parameter $q := a_0 a_1 a_2^2 a_3^2 a_4 a_5$. We also introduce eight auxiliary parameters $b_1$, $\dots$, $b_8$:
\begin{gather}
    \begin{aligned}
    b_1
    &:= a_3^2 a_4^{-1} a_5,
    &&&
    b_2
    &:= a_3^2 a_4^3 a_5,
    &&&
    b_3
    &:= a_3^{-2} a_4^{-1} a_5,
    &&&
    b_4
    &:= a_3^{-2} a_4^{-1} a_5^{-3},
    &&&
    b_5
    &:= a_0 a_1^{-1} a_2^{-2},
    \end{aligned}
    \\[2mm]
    \begin{aligned}
    b_6
    &:= a_0^{-3} a_1^{-1} a_2^{-2},
    &&&
    b_7
    &:= a_0 a_1^{-1} a_2^2,
    &&&
    b_8
    &:= a_0 a_1^3 a_2^2.
    \end{aligned}
\end{gather}
We are going to define an extension of a non-commutative birational representation of the extended affine Weyl group on $\mathcal{K} \br{a_i; f, g}$, where $f$, $g$ are elements of $\mathcal{R}$.

\begin{thm}
\label{thm:WD5_nc}
The given formulas define a non-commutative extended birational representation of $\widetilde W(D_5^{(1)})${\rm:}
\begin{gather}
    \begin{aligned}
    s_i(a_j)
    &= a_j \, a_i^{- c_{ij}},
    &&&
    \pi_1 \br{a_{\{0, 1, 2, 3, 4, 5\}}}
    &= a_{\{5, 4, 3, 2, 1, 0\}}^{-1},
    &&&
    \pi_2 \br{a_{\{0, 1, 2, 3, 4, 5\}}}
    &= a_{\{1, 0, 2, 3, 4, 5\}}^{-1},
    \end{aligned}
    \\[2mm]
    \begin{aligned}
    s_2(f)
    &= f \, \br{b_7^{- \frac12} g + b_7^{\frac12}} 
    \br{b_5^{-\frac12} g + b_5^{\frac12}}^{-1},
    &&&
    s_3(g)
    &= \br{
    b_3^{-\frac12} f + b_3^{\frac12}
    } \, \br{
    b_1^{- \frac12} f + b_1^{\frac12}
    }^{-1} \, g,
    \end{aligned}
    \\[2mm]
    \begin{aligned}
    \pi_1 (f)
    &= g^{-1},
    &&&
    \pi_1 (g)
    &= f^{-1},
    &&&
    \pi_2 (f)
    &= f^{-1}
    .
    \end{aligned}
\end{gather}
\end{thm}
\begin{proof}
By direct computation, one can verify that the affine Weyl group structure of type $D_5^{(1)}$ is preserved. 
\end{proof}

\begin{rem}
When $\mathcal{R}$ is a commutative ring of rational functions, these theorem gives an extended birational representation equivalent to those presented in the paper \cite{sakai2001rational}.
\end{rem}

Proceeding to the discrete dynamics, we define the translation operator
\begin{align}
    \label{eq:trop}
    T
    &:= \pi_3 s_3 s_5 s_4 s_3 \,\, \pi_2 s_2 s_0 s_1 s_2
    ,
\end{align}
which acts on $\mathcal{K} (a_i; f, g)$ and preserves the parameter $q$, i.e. $\bar q = q$. Then, the discrete system evolves as
\begin{align}
    T \br{
    f, \, g, \, q; \, a_0, \, a_1, \, a_2, \, a_3, \, a_4, \, a_5
    }
    &= \br{
    \bar f, \, \bar g, \, q; \, a_0, \, a_1, \, q^{-1} a_2, \, q \, a_3, \, a_4, \, a_5
    },
\end{align}
where 
\begin{align}
    \tag*{\ref{eq:qPA3}}
    \begin{multlined}
    \underline{f} \, f
    = b_7 b_8 \, \br{
    g + b_6
    } \, \br{
    g + b_8
    }^{-1} \, \br{
    g + b_5
    } \, \br{
    g + b_7
    }^{-1}
    ,
    \hspace{4cm}
    \\[1mm]
    \bar{g} \, g
    = b_3 b_4 \, \br{
    f + b_2
    } \, \br{
    f + b_4
    }^{-1} \, \br{
    f + b_1
    } \, \br{
    f + b_3
    }^{-1}
    .
    \end{multlined}
\end{align}
This defines a non-commutative version of the sixth $q$-Painlevé equation \cite{jimbo1996q}.
\medskip 

In the commutative limit $f \, g = g \, f$, the system \ref{eq:qPA3} recovers the standard sixth $q$-Painlevé equation associated with the $A_3^{(1)}$-surface \cite{sakai2001rational}. Note also that our system is a generalization of the Kawakami matrix sixth $q$-Painlevé system \cite{kawakami2020q} (more detailed explanation of this connection is given in Remark~\ref{rem:kaw}). Due to the formal non-commutative surface theory beyond the \ref{eq:qPA3} discovered in Subsection \ref{sec:qPA3surf_nc}, we use the same surface type as in the commutative case to label the resulting system. 

As we have already mentioned, $q$ is a conserved quantity, i.e. $\bar q = q$, since the parameters $b_i$ evolve as
\begin{gather}
    \begin{aligned}
    s_0(b_5)
    &= b_6,
    &&&
    s_1(b_7)
    &= b_8,
    \end{aligned}
    \\[1mm]
    \begin{aligned}
    s_2(b_1)
    &= a_2^2 \, b_1,
    &&&
    s_2(b_2)
    &= a_2^2 \, b_2,
    &&&
    s_2(b_3)
    &= a_2^{-2} \, b_3,
    &&&
    s_2(b_4)
    &= a_2^{-2} \, b_4,
    &&&
    s_2(b_5)
    &= b_7,
    \end{aligned}
    \\[1mm]
    \begin{aligned}
    s_3(b_1)
    &= b_3,
    &&&
    s_3(b_5)
    &= a_3^{-2} \, b_5,
    &&&
    s_3(b_6)
    &= a_3^{-2} \, b_6,
    &&&
    s_3(b_7)
    &= a_3^{2} \, b_7,
    &&&
    s_3(b_8)
    &= a_3^{2} \, b_8,
    \end{aligned}
    \\[1mm]
    \begin{aligned}
    s_4(b_1)
    &= b_2,
    &&&
    s_5(b_3)
    &= b_4,
    \end{aligned}
    \\[1mm]
    \begin{aligned}
    \pi_1(b_1)
    &= b_7^{-1},
    &&&
    \pi_1(b_2)
    &= b_8^{-1},
    &&&
    \pi_1(b_3)
    &= b_5^{-1},
    &&&
    \pi_1(b_4)
    &= b_6^{-1},
    \end{aligned}
    \\[1mm]
    \begin{aligned}
    \pi_2(b_1)
    &= b_1^{-1},
    &&&
    \pi_2(b_2)
    &= b_2^{-1},
    &&&
    \pi_2(b_3)
    &= b_3^{-1},
    &&&
    \pi_2(b_4)
    &= b_4^{-1},
    &&&
    \pi_2(b_5)
    &= b_7,
    &&&
    \pi_2(b_6)
    &= b_8,
    \end{aligned}
\end{gather}
and, therefore,
\begin{align}
    T(b_1)
    &= q^2 \, b_1,
    &
    T(b_2)
    &= q^2 \, b_2,
    &
    T(b_3)
    &= q^{-2} \, b_3,
    &
    T(b_4)
    &= q^{-2} \, b_4,
    \\[1mm]
    T(b_5)
    &= q^2 \, b_5,
    &
    T(b_6)
    &= q^2 \, b_6,
    &
    T(b_7)
    &= q^{-2} \, b_7,
    &
    T(b_8)
    &= q^{-2} \, b_8.
\end{align}

\medskip
Recall that for the system of the form \eqref{eq:ncdsys_qtype}, the element $I (f, g) = f \, g^{-1} f^{-1} g$ is preserved either under the $T^2$-action or the $i \circ T$-action, where $i \br{f, g} = \br{f^{-1}, g^{-1}}$ (see Proposition \ref{thm:firstint_qcase} and Remark \ref{rem:firstint_qcase}). 

\subsection{From discrete system to surface theory}
\label{sec:qPA3surf_nc}

In this subsection, we illustrate how the discrete non-commutative system \ref{eq:qPA3}, derived via an extended birational representation of the affine Weyl group (see Subsection~\ref{sec:qPA3_affW}), gives rise to a non-commutative space $\mathcal{X}_{\nc}$ within the framework of the non-commutative Sakai theory developed in Subsection~\ref{sec:ncgeom}. Specifically, we describe how the system determines a point configuration, a sequence of blow-ups, and a corresponding Picard lattice structure, yielding an extended birational representation of the extended affine Weyl group $\widetilde W (D_5^{(1)})$.

\subsubsection{Point configuration}
The dynamical system \ref{eq:qPA3} gives us eight points $p_i = \br{f_i, g_i}$, $i = 1, \dots, 8$, on the non-commutative space $X_0 = \mathbb{P}_{\nc}^1 \times \mathbb{P}_{\nc}^1$. Let us choose the point configuration as it is given on Figure \ref{fig:pointconf_A3}. 

\begin{minipage}[l]{0.39\linewidth}
\begin{align}
    \\[-2.5mm]
    p_1
    &= (- b_1, 0),
    &&&
    p_2 
    &= (- b_2, 0),
    \\[3mm]
    p_3
    &= (- b_3, \infty),
    &&&
    p_4
    &= (- b_4, \infty),
    \\[3mm]
    p_5
    &= (0, - b_5),
    &&&
    p_6
    &= (0, - b_6),
    \\[3mm]
    p_7
    &= (\infty, - b_7),
    &&&
    p_8
    &= (\infty, - b_8).
    \\[3mm]
    \phantom{}
\end{align}
\end{minipage}
\begin{minipage}[l]{0.59\linewidth}
\begin{figure}[H]
    \centering
    \scalebox{0.8}{\tikzset{every picture/.style={line width=0.75pt}} 

\begin{tikzpicture}[x=0.75pt,y=0.75pt,yscale=-1,xscale=1]

\draw    (90.8,20.92) -- (90.8,210.7) ;
\draw  [fill={rgb, 255:red, 0; green, 0; blue, 0 }  ,fill opacity=1 ] (87.46,145.54) .. controls (87.46,143.75) and (88.92,142.29) .. (90.71,142.29) .. controls (92.5,142.29) and (93.96,143.75) .. (93.96,145.54) .. controls (93.96,147.33) and (92.5,148.79) .. (90.71,148.79) .. controls (88.92,148.79) and (87.46,147.33) .. (87.46,145.54) -- cycle ;
\draw  [fill={rgb, 255:red, 0; green, 0; blue, 0 }  ,fill opacity=1 ] (87.26,85.32) .. controls (87.26,83.53) and (88.71,82.08) .. (90.5,82.08) .. controls (92.3,82.08) and (93.75,83.53) .. (93.75,85.32) .. controls (93.75,87.12) and (92.3,88.57) .. (90.5,88.57) .. controls (88.71,88.57) and (87.26,87.12) .. (87.26,85.32) -- cycle ;

\draw    (220.38,21.42) -- (220.38,211.2) ;
\draw  [fill={rgb, 255:red, 0; green, 0; blue, 0 }  ,fill opacity=1 ] (217.05,146.04) .. controls (217.05,144.25) and (218.5,142.79) .. (220.29,142.79) .. controls (222.09,142.79) and (223.54,144.25) .. (223.54,146.04) .. controls (223.54,147.83) and (222.09,149.29) .. (220.29,149.29) .. controls (218.5,149.29) and (217.05,147.83) .. (217.05,146.04) -- cycle ;
\draw  [fill={rgb, 255:red, 0; green, 0; blue, 0 }  ,fill opacity=1 ] (216.84,85.82) .. controls (216.84,84.03) and (218.3,82.58) .. (220.09,82.58) .. controls (221.88,82.58) and (223.33,84.03) .. (223.33,85.82) .. controls (223.33,87.62) and (221.88,89.07) .. (220.09,89.07) .. controls (218.3,89.07) and (216.84,87.62) .. (216.84,85.82) -- cycle ;

\draw    (250.33,50.5) -- (60.55,50.5) ;
\draw  [fill={rgb, 255:red, 0; green, 0; blue, 0 }  ,fill opacity=1 ] (125.71,47.16) .. controls (127.5,47.16) and (128.95,48.62) .. (128.95,50.41) .. controls (128.95,52.2) and (127.5,53.66) .. (125.71,53.66) .. controls (123.92,53.66) and (122.46,52.2) .. (122.46,50.41) .. controls (122.46,48.62) and (123.92,47.16) .. (125.71,47.16) -- cycle ;
\draw  [fill={rgb, 255:red, 0; green, 0; blue, 0 }  ,fill opacity=1 ] (185.93,46.96) .. controls (187.72,46.96) and (189.17,48.41) .. (189.17,50.21) .. controls (189.17,52) and (187.72,53.45) .. (185.93,53.45) .. controls (184.13,53.45) and (182.68,52) .. (182.68,50.21) .. controls (182.68,48.41) and (184.13,46.96) .. (185.93,46.96) -- cycle ;

\draw    (250.33,180.75) -- (60.55,180.75) ;
\draw  [fill={rgb, 255:red, 0; green, 0; blue, 0 }  ,fill opacity=1 ] (125.71,177.41) .. controls (127.5,177.41) and (128.95,178.87) .. (128.95,180.66) .. controls (128.95,182.45) and (127.5,183.91) .. (125.71,183.91) .. controls (123.92,183.91) and (122.46,182.45) .. (122.46,180.66) .. controls (122.46,178.87) and (123.92,177.41) .. (125.71,177.41) -- cycle ;
\draw  [fill={rgb, 255:red, 0; green, 0; blue, 0 }  ,fill opacity=1 ] (185.93,177.21) .. controls (187.72,177.21) and (189.17,178.66) .. (189.17,180.46) .. controls (189.17,182.25) and (187.72,183.7) .. (185.93,183.7) .. controls (184.13,183.7) and (182.68,182.25) .. (182.68,180.46) .. controls (182.68,178.66) and (184.13,177.21) .. (185.93,177.21) -- cycle ;

\draw (71.78,213.74) node [anchor=north west][inner sep=0.75pt]    {\Large$g=0$};
\draw (200.67,216.7) node [anchor=north west][inner sep=0.75pt]    {\Large$g=\infty $};
\draw (15.03,171.77) node [anchor=north west][inner sep=0.75pt]    {\Large$f=0$};
\draw (7,42.02) node [anchor=north west][inner sep=0.75pt]    {\Large$f=\infty $};
\draw (62.71,141.89) node [anchor=north west][inner sep=0.75pt]    {\Large$p_{1}$};
\draw (62.7,81.56) node [anchor=north west][inner sep=0.75pt]    {\Large$p_{2}$};
\draw (230.71,141.9) node [anchor=north west][inner sep=0.75pt]    {\Large$p_{3}$};
\draw (230.7,81.6) node [anchor=north west][inner sep=0.75pt]    {\Large$p_{4}$};
\draw (118.26,156.56) node [anchor=north west][inner sep=0.75pt]    {\Large$p_{5}$};
\draw (178.71,156.34) node [anchor=north west][inner sep=0.75pt]    {\Large$p_{6}$};
\draw (118.26,62.48) node [anchor=north west][inner sep=0.75pt]    {\Large$p_{7}$};
\draw (178.71,62.25) node [anchor=north west][inner sep=0.75pt]    {\Large$p_{8}$};

\end{tikzpicture}}
    \caption{The \ref{eq:qPA3} point configuration}
    \label{fig:pointconf_A3}
\end{figure}
\end{minipage}
\\[2mm]

Note that the coordinate components of the points belong to the center of $\mathcal{R}$. Thus, the order of the points on the non-commutative space $X_0$ is not essential, since they can be permuted by the automorphisms of a Dynkin diagram. As a result, the point configuration does make sense in this non-commutative framework. 

\subsubsection{Surface type}
In order to construct the space $\mathcal{X}_{\nc}$, we proceed with the resolution procedure at $p_i$, $i = 1, \dots, 8$ shown on Figure \ref{fig:pointconf_A3}. Note that these points are given in different affine charts. We are going to demonstrate how Definition \ref{def:blups} is applied in order to resolve the point $p_1 = \br{f_1, g_1} = \br{- b_1, 0}$. Consider the following transformation
\begin{align}
    &&
    f_1
    &= F_1 - b_1,
    &
    g_1
    &= G_1 \, F_1,
    &&
\end{align}
where $(F_1, G_1)$ corresponds to a new affine chart. It gives the system
\begin{align}
    \underline{f}
    &= b_7 b_8 \, \br{F_1 G_1 + b_6} \,
    \br{F_1 G_1 + b_8}^{-1} 
    \br{F_1 G_1 + b_5} \,
    \br{F_1 G_1 + b_7}^{-1}
    \br{F_1 - b_1}^{-1}
    ,
    &
    \\[2mm]
    \bar{g}
    &= b_3 b_4 \, \br{F_1 + b_2 - b_1} \, 
    \br{F_1 + b_4 - b_1}^{-1} F_1 \,
    \br{F_1 + b_3 - b_1}^{-1} 
    \br{G_1 \, F_1}^{-1}
    \\
    &= b_3 b_4 \, \br{F_1 + b_2 - b_1} \, 
    \br{F_1 + b_4 - b_1}^{-1} 
    \br{F_1 + b_3 - b_1}^{-1} 
    G_1^{-1}
    ,
\end{align}
which does not have an inaccessible point $p_1$. Note that here we have used Corollary \ref{thm:ncrat}. 

Similarly, one can proceed with the remaining points and, as a result, obtains the space $\mathcal{X}_{\nc}$ schematically shown on Figure \ref{fig:ratsurf_A3}. Each point $p_i$, $i = 1, \dots, 8$ is replaced with the exceptional set $\mathcal{E}_i$, while ${H}_1$, ${H}_2$ stand for the lines $f = 0$, $g = 0$, respectively. A decomposition of the anti-canonical class $\mathcal{K}_{\mathcal{X}_{\nc}}$ reads as
\begin{align}
    - \mathcal{K}_{\mathcal{X}_{\nc}} 
    &= \delta_0 + \delta_1 + \delta_2 + \delta_3
    ,
\end{align}
where its irreducible components are given below.
\newline
\begin{minipage}[l]{0.39\linewidth}
\begin{align}
    \\[-2.5mm]
    \delta_0
    &= \mathcal{H}_2 - \mathcal{E}_{12},
    &&&
    \delta_1
    &= \mathcal{H}_1 - \mathcal{E}_{56},
    \\[3mm]
    \delta_2
    &= \mathcal{H}_2 - \mathcal{E}_{34},
    &&&
    \delta_3
    &= \mathcal{H}_1 - \mathcal{E}_{78}.
\end{align}
\begin{figure}[H]
    \centering
    \scalebox{0.7}{\tikzset{every picture/.style={line width=0.75pt}} 

\begin{tikzpicture}[x=0.75pt,y=0.75pt,yscale=-1,xscale=1]

\draw  [fill={rgb, 255:red, 0; green, 0; blue, 0 }  ,fill opacity=1 ] (201.72,106.66) .. controls (203.42,104.92) and (206.21,104.9) .. (207.96,106.6) .. controls (209.7,108.31) and (209.72,111.1) .. (208.02,112.84) .. controls (206.31,114.58) and (203.52,114.61) .. (201.78,112.9) .. controls (200.04,111.2) and (200.01,108.4) .. (201.72,106.66) -- cycle ;
\draw  [fill={rgb, 255:red, 0; green, 0; blue, 0 }  ,fill opacity=1 ] (59.06,108.11) .. controls (60.77,106.37) and (63.56,106.34) .. (65.3,108.04) .. controls (67.04,109.75) and (67.07,112.54) .. (65.37,114.28) .. controls (63.66,116.03) and (60.87,116.05) .. (59.13,114.35) .. controls (57.39,112.64) and (57.36,109.85) .. (59.06,108.11) -- cycle ;
\draw  [fill={rgb, 255:red, 0; green, 0; blue, 0 }  ,fill opacity=1 ] (129.67,36.06) .. controls (131.37,34.32) and (134.17,34.29) .. (135.91,36) .. controls (137.65,37.7) and (137.68,40.5) .. (135.97,42.24) .. controls (134.26,43.98) and (131.47,44.01) .. (129.73,42.3) .. controls (127.99,40.59) and (127.96,37.8) .. (129.67,36.06) -- cycle ;
\draw    (199.72,109.62) -- (67.11,111.59) ;
\draw    (132.82,39.15) -- (62.22,111.2) ;
\draw  [fill={rgb, 255:red, 0; green, 0; blue, 0 }  ,fill opacity=1 ] (130.26,107.52) .. controls (131.97,105.78) and (134.76,105.75) .. (136.5,107.45) .. controls (138.24,109.16) and (138.27,111.95) .. (136.56,113.69) .. controls (134.86,115.43) and (132.06,115.46) .. (130.32,113.76) .. controls (128.58,112.05) and (128.55,109.26) .. (130.26,107.52) -- cycle ;
\draw    (132.82,39.15) -- (204.87,109.75) ;

\draw (125.33,12.4) node [anchor=north west][inner sep=0.75pt]  {\LARGE$\delta_{0}$};
\draw (43.16,121.4) node [anchor=north west][inner sep=0.75pt]  {\LARGE$\delta_{1}$};
\draw (127.33,121.4) node [anchor=north west][inner sep=0.75pt]  {\LARGE$\delta_{2}$};
\draw (205.61,121.4) node [anchor=north west][inner sep=0.75pt]  {\LARGE$\delta_{3}$};

\end{tikzpicture}}
\end{figure}
\end{minipage}
\begin{minipage}[l]{0.59\linewidth}
\begin{figure}[H]
    \centering
    \scalebox{0.8}{\tikzset{every picture/.style={line width=0.75pt}} 

\begin{tikzpicture}[x=0.75pt,y=0.75pt,yscale=-1,xscale=1]

\draw    (100.8,20.92) -- (100.8,210.7) ;
\draw    (230.38,21.42) -- (230.38,211.2) ;
\draw    (260.33,50.5) -- (70.55,50.5) ;
\draw    (260.33,180.75) -- (70.55,180.75) ;
\draw [color={rgb, 255:red, 255; green, 99; blue, 71 }  ,draw opacity=1 ][line width=1.5]    (135.71,165.38) -- (135.71,195.58) ;
\draw [color={rgb, 255:red, 255; green, 99; blue, 71 }  ,draw opacity=1 ][line width=1.5]    (195.93,164.78) -- (195.93,195.38) ;
\draw [color={rgb, 255:red, 255; green, 99; blue, 71 } ,draw opacity=1 ][line width=1.5]    (135.76,35) -- (135.76,65.4) ;
\draw [color={rgb, 255:red, 255; green, 99; blue, 71 }  ,draw opacity=1 ][line width=1.5]    (195.93,35) -- (195.93,65.2) ;
\draw [color={rgb, 255:red, 255; green, 99; blue, 71 } ,draw opacity=1 ][line width=1.5]    (115.45,85.32) -- (84.25,85.32) ;
\draw [color={rgb, 255:red, 255; green, 99; blue, 71 }  ,draw opacity=1 ][line width=1.5]    (115.45,145.54) -- (84.65,145.54) ;
\draw [color={rgb, 255:red, 255; green, 99; blue, 71 } ,draw opacity=1 ][line width=1.5]    (246.85,85.32) -- (215.65,85.32) ;
\draw [color={rgb, 255:red, 255; green, 99; blue, 71 } ,draw opacity=1 ][line width=1.5]    (246.85,146) -- (215.65,146) ;

\draw (69.78,213.74) node [anchor=north west][inner sep=0.75pt]    {\Large$H_{2} -E_{12}$};
\draw (200.67,216.7) node [anchor=north west][inner sep=0.75pt]    {\Large$H_{2} -E_{34}$};
\draw (4,173.77) node [anchor=north west][inner sep=0.75pt]    {\Large$H_{1} -E_{56}$};
\draw (4,44.02) node [anchor=north west][inner sep=0.75pt]    {\Large$H_{1} -E_{78}$};
\draw (59.71,137.89) node [anchor=north west][inner sep=0.75pt]    {\Large$E_{1}$};
\draw (59.7,77.56) node [anchor=north west][inner sep=0.75pt]    {\Large$E_{2}$};
\draw (252.71,137.9) node [anchor=north west][inner sep=0.75pt]    {\Large$E_{3}$};
\draw (252.7,77.6) node [anchor=north west][inner sep=0.75pt]    {\Large$E_{4}$};
\draw (127.26,147.4) node [anchor=north west][inner sep=0.75pt]    {\Large$E_{5}$};
\draw (187.71,147.4) node [anchor=north west][inner sep=0.75pt]    {\Large$E_{6}$};
\draw (127.26,70.4) node [anchor=north west][inner sep=0.75pt]    {\Large$E_{7}$};
\draw (187.71,70.4) node [anchor=north west][inner sep=0.75pt]    {\Large$E_{8}$};

\end{tikzpicture}}
    \caption{The space $\mathcal{X}_{\nc}$ of the \ref{eq:qPA3}}
    \label{fig:ratsurf_A3}
\end{figure}
\end{minipage}
\vspace{2mm}
\newline
As we mentioned above, the intersection matrix of the irreducible components $\delta_i$ gives us the type of the Dynkin diagram, which is of $A_3^{(1)}$ type in this case. 

\begin{thm}
\label{thm:qPA3_blups}
One can construct a sequence of non-commutative blow-ups that resolves all the base points of~the {\rm\ref{eq:qPA3}} system.
\end{thm}
\begin{proof}
The statement is proved by a simple case-by-case analysis of substitutions of the form \eqref{eq:ncblup} or its transpose in the sense of $\tau$-action (see Remark~\ref{rem:ncblup}). 
\end{proof}

\subsubsection{Symmetry type}
Next, we identify the orthogonal complement of the surface type root lattice using the orthogonality condition (see Subsection~\ref{sec:ncsurftosys}):
\begin{align}
    \label{eq:orthcond}
    &&
    \abr{\alpha_i \, \cdot \, \delta_j}
    &= 0
    &
    \forall \,\,
    i, j
    .
    &&
\end{align}
Taking the root $\alpha$ in the following form
\begin{align}
    \alpha
    = h_1 \, \mathcal{H}_1 + h_2 \, \mathcal{H}_2
    + e_1 \, \mathcal{E}_1 
    + \dots + e_8 \, \mathcal{E}_8,
\end{align}
the condition \eqref{eq:orthcond} leads to the constraints
\begin{align}
    &&
    h_1 + e_1 + e_2 
    &= 0,
    &
    h_2 + e_5 + e_6
    &= 0,
    &
    h_1 + e_3 + e_4
    &= 0,
    &
    h_2 + e_7 + e_8
    &= 0
    .
    &&
\end{align}
Since we want to obtain an extended birational representation as it is given in Theorem \ref{thm:WD5_nc}, we choose the coefficients $h_1$, $h_2$, and $e_k$ with $k = 2, 4, 6, 8$ as basis and pick the following labeling of the roots:
\begin{align}
    \alpha_0
    &= \mathcal{E}_8 - \mathcal{E}_7,
    &
    \alpha_1
    &= \mathcal{E}_6 - \mathcal{E}_5,
    &
    \alpha_2
    &= \mathcal{H}_2 - \mathcal{E}_{57},
    &
    \alpha_3
    &= \mathcal{H}_1 - \mathcal{E}_{13},
    &
    \alpha_4 
    &= \mathcal{E}_4 - \mathcal{E}_3,
    &
    \alpha_5
    &= \mathcal{E}_2 - \mathcal{E}_1.
\end{align}
This root system is of the $D_5^{(1)}$ type. The anti-canonical class $- \mathcal{K}_{\mathcal{X}_{\nc}}$ decomposes as
\begin{align}
    - \mathcal{K}_{\mathcal{X}_{\nc}}
    &= \alpha_0 + \alpha_1 
    + 2 \alpha_2 + 2 \alpha_3 
    + \alpha_4 + \alpha_5
    .
\end{align}
Note that our choice is inessential since the roots are commutative ones and can be mixed up by using the automorphisms of the Dynkin diagram. 

\subsection{From surface theory to birational representation}
\label{sec:qPA3surftoW_nc}

Now we proceed with the birational representation of the corresponding Weyl group $\widetilde W ({D_5^{(1)}})$. We will adopt the following guiding principle formulated in commutative setting as \cite{kajiwara2017geometric}: for~each element $s$ of the affine Weyl group $\widetilde W$, $s(f)$ and $s(g)$ should be rational functions in the class $s\br{\mathcal{H}_1}$ and $s\br{\mathcal{H}_2}$, respectively.

Generally speaking, the action of the affine Weyl group on the functions $f$, $g$ can be obtained as follows. The reflections on the basis of the Picard lattice $\Pic\br{\mathcal{X}_{\nc}}$ correspond to pencils of the biquadratic curves $\mathcal{C}$ passing through certain points $p_i$, $i = 1, \dots, k$. This action lifts to the $(f, g)$-coordinated by considering the projective coordinates on the pencil. Technically, one needs to choose a basis for the pencil
\begin{align}
    &&
    \lambda_1 \, A\br{f, g}
    + \lambda_2 \, B\br{f, g}
    &= 0
    &\Leftrightarrow&&
    A\br{f, g}
    + \lambda \, B\br{f, g}
    &= 0,
    &&
\end{align}
where $A\br{f, g}$, $B\br{f, g}$ are two fixed non-commutative biquadratic polynomials, while $\lambda = \lambda_1^{-1} \lambda_2$ parametrizes the pencil. Then, the projective coordinate on the pencil reads accordingly to Definition \ref{def:mobius}~as
\begin{align}
    \tilde f
    &= \br{
    a \, A\br{f, g} + b \, B\br{f, g}
    } \, \br{
    c \, A\br{f, g} + d \, B\br{f, g}
    }^{-1}
    .
\end{align}
The coefficients $a$, $b$, $c$, $d \in \mathcal{R}$ can be determined by investigating the image of appropriate points or divisors.

Note that together with the formal $(2,2)$-curve 
\begin{gather}
    m_{00} \, f^2 \, g^2
    + m_{01} \, f^2 \, g
    + m_{02} \, f^2
    + m_{10} \, f \, g^2
    + m_{11} \, f \, g
    + m_{12} \, f
    + m_{20} \, g^2
    + m_{21} \, g
    + m_{22}
    = 0
\end{gather}
one can consider its $\tau$-version
\begin{gather}
    \label{eq:taucurve}
    m_{00} \, g^2 \, f^2
    + m_{10} \, g \, f^2
    + m_{20} \, f^2
    + m_{01} \, g^2 \, f
    + m_{11} \, g \, f
    + m_{21} \, f
    + m_{02} \, g^2
    + m_{12} \, g
    + m_{22}
    = 0
    ,
\end{gather}
Let us also stress that $\tau$ is an involution, i.e. $\tau^2 = \text{id}$. 

Here we will not present the whole computations for the extended birational representation related to the \ref{eq:qPA3} system and will focus only on two the most non-trivial cases, that are $s_2(f)$ and $s_3(g)$, in order to demonstrate the procedure. The remaining cases can be computed in a similar manner. Miraculously, the computations for the $s_3(g)$ case involves the $\tau$-action, so that we need to first consider the formal $\tau$-curve and then apply $\tau$ to the projective coordinate on a corresponding formal pencil. At this stage, we are not able to explain why exactly this procedure gives us the correct formulae, however, formally, $\tau^2$ gives a trivial action. 

\medskip 
Recall that the reflection action on the Picard group $\Pic\br{\mathcal{X}_{\nc}}$, defined by equation~\eqref{eq:refl}, gives the symmetry action of $\widetilde W ({D_5^{(1)}})$ on the root lattice.
Thus, the non-trivial reflections read as
\begin{gather}
    \begin{aligned}
    s_0: \,\, 
    \mathcal{E}_8
    &\leftrightarrow \mathcal{E}_7,
    &&&&&\,\,
    s_1: \,\, 
    \mathcal{E}_6 
    &\leftrightarrow \mathcal{E}_5,
    &&&&&\,\,
    s_4: \,\, 
    \mathcal{E}_4 
    &\leftrightarrow \mathcal{E}_3,
    &&&&&\,
    s_5: \,\, 
    \mathcal{E}_2 
    &\leftrightarrow \mathcal{E}_1,
    \end{aligned}
    \\[2mm]
    \begin{aligned}
    s_2: \,\, 
    \mathcal{H}_1
    &\leftrightarrow \mathcal{H}_1 + \mathcal{H}_2 - \mathcal{E}_{57},
    &&&&&&&\,\,\,
    \mathcal{E}_5 
    &\leftrightarrow 
    \mathcal{H}_2 - \mathcal{E}_{7},
    &&&&&&&\,\,\,
    \mathcal{E}_7
    &\leftrightarrow 
    \mathcal{H}_2 - \mathcal{E}_{5},
    \end{aligned}
    \\[2mm]
    \begin{aligned}
    s_3: \,\, 
    \mathcal{H}_2
    &\leftrightarrow \mathcal{H}_1 + \mathcal{H}_2 - \mathcal{E}_{13},
    &&&&&&&\,\,\,
    \mathcal{E}_1 
    &\leftrightarrow 
    \mathcal{H}_1 - \mathcal{E}_{3},
    &&&&&&&\,\,\,
    \mathcal{E}_3
    &\leftrightarrow 
    \mathcal{H}_1 - \mathcal{E}_{1}.
    \end{aligned}
\end{gather}
One can verify that this action are Cremona isometries in the sense of Definition \ref{def:crisnc}. To lift them to the functions $f$ and $g$, we use formal biquadratic curves and Möbius transformations (see Definitions~\ref{def:nccurve} and~\ref{def:mobius}). Let us consider the cases $s_2(f)$ and $s_3(g)$ separately. The remaining actions can be found similarly.

\medskip
\textbf{\textbullet \,\, $s_2\br{f}$ case.}
Recall that $s_2\br{\mathcal{H}_1} = \mathcal{H}_1 + \mathcal{H}_2 - \mathcal{E}_{57}$. Hence, we consider a formal $(1,1)$-curve containing the points $p_5 = \br{f_5, g_5} = \br{0, - b_5}$ and $p_7 = \br{f_7, g_7} = \br{\infty, - b_7}$. Substituting these conditions into the curve
\begin{align}
    m_{11} \, f \, g
    + m_{12}  \, f
    + m_{21} \, g
    + m_{22}
    &= 0
    ,
\end{align}
we find that $m_{22} = b_5 \, m_{21}$ and $m_{12} = b_7 \, m_{11}$. Therefore, the curve $\mathcal{C}$ passing via $p_5$ and $p_7$ reads as
\begin{align}
    m_{11} \, f \,  
    \br{g + b_7}
    + m_{21} \, \br{g + b_5}
    &= 0
    .
\end{align}
Let us take two representative curves $A\br{f, g} = m_{11} \, f \, \br{g + b_7}$ and $B\br{f, g} = \br{g + b_5}$ which span the pencil. The projective coordinate on it is
\begin{align}
    s_2 \br{f}
    &= \br{
    a \, 
    m_{11} \, f \, 
    \br{g + b_7}
    + b \, \br{g + b_5}
    } \, \br{
    c \, m_{11} \, f \,
    \br{g + b_7}
    + d \, \br{
    g + b_5
    }
    }^{-1}
    .
\end{align}
As we mentioned above, the unknown coefficients can be determined by investigating the images of divisors. For instance, requiring $\pbr{f = 0} \leftrightarrow \pbr{f = 0}$ and $\pbr{F = 0} \leftrightarrow \pbr{F = 0}$, one can find that $b = 0$ and $c = 0$, respectively. Indeed,
\begin{align}
    &&
    s_2 \br{f = 0}
    &= 0 
    &
    \Leftrightarrow&
    &
    b \, \br{
    g + b_5
    } \, \br{
    g + b_5
    }^{-1} d^{-1}
    &= 0
    &
    \Leftrightarrow&
    &
    b 
    &= 0
    ,
    &&
\end{align}
and since
\begin{align}
    s_2 (f)
    &= a \, m_{11} \, f \, \br{g + b_7} \, \br{
    c \, m_{11} \, f \, 
    \br{g + b_7}
    + d \, \br{g + b_5}
    }^{-1}
    \\[2mm]
    &= a \, \br{
    c
    + d \, \br{g + b_5} \, \br{g + b_7}^{-1} \, f^{-1} \, m_{11}^{-1} 
    }^{-1}
    ,
\end{align}
we have
\begin{align}
    &&
    s_2 \br{F}
    &= \br{s_2(f)}^{-1}
    = \br{
    c
    + d \, \br{g + b_5} \, \br{g + b_7}^{-1} \, F \, m_{11}^{-1}
    } \, a^{-1}
    .
\end{align}
Hence,
\begin{align}
    &&
    s_2 \br{F = 0}
    &= 0,
    &\Leftrightarrow&&
    c \, a^{-1}
    &= 0
    &\Leftrightarrow&&
    c
    &= 0
    .
    &&
\end{align}
If we set $a = 1$, $m_{11} = b_7^{- \frac12}$, and $d = b_5^{-\frac12}$, the resulting expression,
\begin{align}
    s_2(f)
    &= a \, m_{11} \,  \, 
    (g + b_7) \, (g + b_5)^{-1} d^{-1}
    ,
\end{align}
turns into
\begin{align}
    s_2(f)
    &= f \,
    \br{
    b_7^{-\frac12} g + b_7^{\frac12}
    } \, \br{
    b_5^{- \frac12} g + b_5^{\frac12}
    }^{-1}
    .
\end{align} 

\medskip
\textbf{\textbullet \,\, $s_3\br{g}$ case.} Computations are similar to the previous ones except of involving the $\tau$-action. Let us take the formal curve of the form \eqref{eq:taucurve}:
\begin{gather}
    m_{00} \, g^2 \, f^2
    + m_{10} \, g \, f^2
    + m_{20} \, f^2
    + m_{01} \, g^2 \, f
    + m_{11} \, g \, f
    + m_{21} \, f
    + m_{02} \, g^2
    + m_{12} \, g
    + m_{22}
    = 0.
\end{gather}
Since $s_3 \br{\mathcal{H}_2} = \mathcal{H}_1 + \mathcal{H}_2 - \mathcal{E}_{13}$, a~formal $(1,1)$-curve $\mathcal{C}$ passing through the points $p_1 = \br{f_1, g_1} = \br{- b_1, 0}$ and $p_3 = \br{f_3, g_3} = \br{- b_3, \infty}$ is given by
\begin{align}
    m_{11} \, g \, \br{f + b_3}
    +  m_{21} \, \br{f + b_1}
    &= 0
    .
\end{align}
Let us choose the basis of this pencil to be $A \br{f, g} = m_{11} \, g \, (f + b_3)$ and $B\br{f, g} = (f + b_1)$. Then, the~corresponding projective coordinate reads
\begin{align}
    s_3(g)
    &= \br{
    a \, m_{11} \, g \, (f + b_3)
    + b \, (f + b_1)
    } \, \br{
    c \, m_{11} \, g \, (f + b_3)
    + d \, (f + b_1)
    }^{-1}
    .
\end{align}
The parameters $b$ and $c$ can be found by using the conditions $\pbr{g = 0} \leftrightarrow \pbr{g = 0}$ and $\pbr{G = 0} \leftrightarrow \pbr{G = 0}$, respectively. Namely, 
\begin{align}
    &&
    s_3 \br{g = 0}
    &= 0 
    &\Leftrightarrow&&
    b \, \br{f + b_1} \, \br{f + b_1}^{-1} d^{-1}
    &= 0
    &\Leftrightarrow&&
    b
    &= 0,
    &&
\end{align}
and, since
\begin{align}
    \br{s_3 (G)}^{-1}
    = 
    s_3 (g)
    &= a \, m_{11} \, g \, (f + b_3) \, \br{
    c \, m_{11} \, g \, (f + b_3)
    + d \, (f + b_1)
    }^{-1}
    \\[2mm]
    &= a \, \br{
    c 
    + d \, (f + b_1) \, (f + b_3)^{-1} \, G \, m_{11}^{-1}
    }^{-1}
    ,
\end{align}
we have
\begin{align}
    &&
    s_3 \br{G = 0}
    &= 0 
    &\Leftrightarrow&&
    c \, a^{-1}
    &= 0
    &\Leftrightarrow&&
    c
    &= 0
    .
    &&
\end{align}
Thus, we arrive at the expression
\begin{align}
    s_3 (g)
    &= a \, m_{11} \, g (f + b_3) \, 
    \br{f + b_1}^{-1} d^{-1}
    = g \, \br{
    b_3^{-\frac12} f + b_3^{\frac12}
    } \, 
    \br{b_1^{-\frac12} f + b_1^{\frac12}}^{-1}
    ,
\end{align}
where $m_{11} = b_{3}^{-\frac12}$, $d = b_1^{-\frac12}$, $a = 1$ were chosen.  

In order to obtain formulas from Theorem \ref{thm:WD5_nc}, we first apply $\tau$ and then permute the factors containing only $f$, thanks to Corollary \ref{thm:ncrat}:
\begin{align}
    \tau\br{s_3 \br{g}}
    &= s_3 \br{\tau\br{g}}
    = s_3 \br{g}
    \\[2mm]
    &= 
    \tau\br{g \, \br{
    b_3^{-\frac12} f + b_3^{\frac12}
    } \, 
    \br{b_1^{-\frac12} f + b_1^{\frac12}}^{-1}}
    = \br{\tau{\br{b_1^{-\frac12} f + b_1^{\frac12}}}}^{-1}
    \tau\br{b_3^{-\frac12} f + b_3^{\frac12}} \, 
    \tau\br{g}
    \\[2mm]
    &= \br{b_1^{-\frac12} f + b_1^{\frac12}}^{-1}
    \br{
    b_3^{-\frac12} f + b_3^{\frac12}
    } \, g
    = \br{
    b_3^{-\frac12} f + b_3^{\frac12}
    } \, \br{b_1^{-\frac12} f + b_1^{\frac12}}^{-1} g
    .
\end{align}
Note that here we used that $\tau$ is a linear map that commutes with the reflections $s_i$ and with taking an inverse. As a result, we arrive at the desired formula.

\begin{thm}
\label{thm:WD5_ratsurf}
The birational representation of $\widetilde W\br{D_5^{(1)}}$ described in Theorem~{\rm\ref{thm:WD5_nc}} arises from automorphisms of the Picard lattice $\Pic\br{\mathcal{X_{\nc}}}$ determined by the point configuration shown in Figure~{\rm\ref{fig:pointconf_A3}}.
\end{thm}

\subsection{A coalescence}
\label{sec:coal}

This section is devoted to a degeneration of the \ref{eq:qPA3} system to lower $q$-Painlevé equations and to the \ref{eq:dPD4} system obtained in the paper \cite{bobrova2024affine}. 

The first part is given in Subsection \ref{sec:qdeg} and contains new examples of non-commutative versions of the $q$-Painlevé equations which we also listed in Appendix \ref{app:qP}. Due to the fact that the \ref{eq:qPA3} can be associated with a point configuration (see Subsection \ref{sec:qPA3surf_nc}), we consider the degeneration process as a coalescence cascade of the point configurations similar to the commutative ones (see Figure \ref{pic:qPA3deg}). Recall that in the commutative case, each degeneration step corresponds to merging or sending base points on the $\mathbb{P}^1 \times \mathbb{P}^1$ surface to special positions as $\br{0, \, 0}$, $\br{0, \, \infty}$, $\br{\infty, \, 0}$, or $\br{\infty, \, \infty}$, using a small parameter $\varepsilon$. We repeat the same procedure in the non-commutative case, thanks to the fact that all base points belong to the center $\mathcal{Z}(\mathcal{R})$ of $\mathcal{R}$. For~instance, in order to get the \ref{eq:qPA4} system from the \ref{eq:qPA3} system, one needs to send $p_4 = \br{- b_4, \infty}$ and $p_8 = \br{\infty, \, - b_8}$ to $\br{\infty, \, \infty}$. It can be achieved by making the change $b_4 = \varepsilon^{-1} \, B_4$, $b_8 = \varepsilon^{-1} \, B_8$ and taking the limit $\varepsilon \to 0$, where $B_4$ and $B_8$ are new parameters. The latter will be written as $b_4 \mapsto \varepsilon \, b_4$, $b_8 \mapsto \varepsilon \, b_8$, where we assume that $B_4 := \varepsilon \, b_4$, $B_8 := \varepsilon \, b_8$, but not mention it explicitly in the limiting system, hopping that it will not lead to misunderstanding. 

\medskip
Sometimes, in order to take a limit, we need to make a rescaling of the functions and parameters. In particular, one can implement an inessential parameter $a \in \mathcal{Z}\br{\mathcal{R}}$ into the \ref{eq:qPA3} system by using the maps
\begin{align}
    &&
    f_n
    &\mapsto a^{-n} \, f_n,
    &&&
    g_n
    &\mapsto a^{-n} \, g_n,
    &&&
    b_k
    &\mapsto a^{-n} \, b_k,
    &
    k = 1, 2, \dots, 8,
    &&
\end{align}
and obtains the system
\begin{align}
    \begin{aligned}
    \label{eq:qPA3_a}
    \underline{f} \, f
    &= a \, b_7 b_8 \,
    \br{g + b_6} \, 
    \br{g + b_8}^{-1}
    \br{g + b_5} \,
    \br{g + b_7}^{-1},
    \\[2mm]
    \bar g \, g
    &= a^{-1} \, b_3 b_4 \,
    \br{f + b_2} \,
    \br{f + b_4}^{-1}
    \br{f + b_1} \,
    \br{f + b_3}^{-1}
    .
    \end{aligned}
\end{align}
We will keep this inessential parameter $a$ in the initial \ref{eq:qPA3} system in order to make possible to take the limit in certain cases. Once the limit is taken, we set $a = 1$ in the resulting system. 

Moreover, since the parameter $q$ should be a conserved quantity of the dynamics, i.e. $\bar q = q$, we will authonomize some parameters $b_j$ in certain cases. 

\medskip
Subsection \ref{sec:ddeg} presents the degeneration \ref{eq:qPA3} $\to$ \ref{eq:dPD4} which connects the \ref{eq:qPA3} system with the non-commutative $d$-Painlevé systems obtained in the paper \cite{bobrova2024affine}. Note that the limiting system~\eqref{eq:dPD4fromqPA3} is equivalent to those from \cite{bobrova2024affine} due to the existence of the first integrals (see Proposition \ref{thm:firstint_qdP}).

\subsubsection{Multiplicative cases}
\label{sec:qdeg}

We begin with the degeneration scheme of the non-commutative sixth $q$-Painlevé system to lower $q$-discrete equations. The starting system is \ref{eq:qPA3}, associated with the $A_3^{(1)}$-surface type (see Subsection~\ref{sec:qPA3surf_nc}), where the inessential parameter $a$ is implemented. 
\begin{figure}[H]
    \centering
    \begin{tabular}{ccccccc}
        \small\ref{eq:qPA3}
        \\
        \scalebox{0.55}{\tikzset{every picture/.style={line width=0.75pt}} 

\begin{tikzpicture}[x=0.75pt,y=0.75pt,yscale=-1,xscale=1]

\draw    (90.8,20.92) -- (90.8,210.7) ;
\draw  [fill={rgb, 255:red, 0; green, 0; blue, 0 }  ,fill opacity=1 ] (87.46,145.54) .. controls (87.46,143.75) and (88.92,142.29) .. (90.71,142.29) .. controls (92.5,142.29) and (93.96,143.75) .. (93.96,145.54) .. controls (93.96,147.33) and (92.5,148.79) .. (90.71,148.79) .. controls (88.92,148.79) and (87.46,147.33) .. (87.46,145.54) -- cycle ;
\draw  [fill={rgb, 255:red, 0; green, 0; blue, 0 }  ,fill opacity=1 ] (87.26,85.32) .. controls (87.26,83.53) and (88.71,82.08) .. (90.5,82.08) .. controls (92.3,82.08) and (93.75,83.53) .. (93.75,85.32) .. controls (93.75,87.12) and (92.3,88.57) .. (90.5,88.57) .. controls (88.71,88.57) and (87.26,87.12) .. (87.26,85.32) -- cycle ;

\draw    (220.38,21.42) -- (220.38,211.2) ;
\draw  [fill={rgb, 255:red, 0; green, 0; blue, 0 }  ,fill opacity=1 ] (217.05,146.04) .. controls (217.05,144.25) and (218.5,142.79) .. (220.29,142.79) .. controls (222.09,142.79) and (223.54,144.25) .. (223.54,146.04) .. controls (223.54,147.83) and (222.09,149.29) .. (220.29,149.29) .. controls (218.5,149.29) and (217.05,147.83) .. (217.05,146.04) -- cycle ;
\draw  [fill={rgb, 255:red, 0; green, 0; blue, 0 }  ,fill opacity=1 ] (216.84,85.82) .. controls (216.84,84.03) and (218.3,82.58) .. (220.09,82.58) .. controls (221.88,82.58) and (223.33,84.03) .. (223.33,85.82) .. controls (223.33,87.62) and (221.88,89.07) .. (220.09,89.07) .. controls (218.3,89.07) and (216.84,87.62) .. (216.84,85.82) -- cycle ;

\draw    (250.33,50.5) -- (60.55,50.5) ;
\draw  [fill={rgb, 255:red, 0; green, 0; blue, 0 }  ,fill opacity=1 ] (125.71,47.16) .. controls (127.5,47.16) and (128.95,48.62) .. (128.95,50.41) .. controls (128.95,52.2) and (127.5,53.66) .. (125.71,53.66) .. controls (123.92,53.66) and (122.46,52.2) .. (122.46,50.41) .. controls (122.46,48.62) and (123.92,47.16) .. (125.71,47.16) -- cycle ;
\draw  [fill={rgb, 255:red, 0; green, 0; blue, 0 }  ,fill opacity=1 ] (185.93,46.96) .. controls (187.72,46.96) and (189.17,48.41) .. (189.17,50.21) .. controls (189.17,52) and (187.72,53.45) .. (185.93,53.45) .. controls (184.13,53.45) and (182.68,52) .. (182.68,50.21) .. controls (182.68,48.41) and (184.13,46.96) .. (185.93,46.96) -- cycle ;

\draw    (250.33,180.75) -- (60.55,180.75) ;
\draw  [fill={rgb, 255:red, 0; green, 0; blue, 0 }  ,fill opacity=1 ] (125.71,177.41) .. controls (127.5,177.41) and (128.95,178.87) .. (128.95,180.66) .. controls (128.95,182.45) and (127.5,183.91) .. (125.71,183.91) .. controls (123.92,183.91) and (122.46,182.45) .. (122.46,180.66) .. controls (122.46,178.87) and (123.92,177.41) .. (125.71,177.41) -- cycle ;
\draw  [fill={rgb, 255:red, 0; green, 0; blue, 0 }  ,fill opacity=1 ] (185.93,177.21) .. controls (187.72,177.21) and (189.17,178.66) .. (189.17,180.46) .. controls (189.17,182.25) and (187.72,183.7) .. (185.93,183.7) .. controls (184.13,183.7) and (182.68,182.25) .. (182.68,180.46) .. controls (182.68,178.66) and (184.13,177.21) .. (185.93,177.21) -- cycle ;

\draw (64.71,141.89) node [anchor=north west][inner sep=0.75pt]    {\LARGE$p_{1}$};
\draw (64.7,81.56) node [anchor=north west][inner sep=0.75pt]    {\LARGE$p_{2}$};
\draw (230.71,141.9) node [anchor=north west][inner sep=0.75pt]    {\LARGE$p_{3}$};
\draw (230.7,81.6) node [anchor=north west][inner sep=0.75pt]    {\LARGE$p_{4}$};
\draw (118.26,156.56) node [anchor=north west][inner sep=0.75pt]    {\LARGE$p_{5}$};
\draw (178.71,156.34) node [anchor=north west][inner sep=0.75pt]    {\LARGE$p_{6}$};
\draw (118.26,62.48) node [anchor=north west][inner sep=0.75pt]    {\LARGE$p_{7}$};
\draw (178.71,62.25) node [anchor=north west][inner sep=0.75pt]    {\LARGE$p_{8}$};

\end{tikzpicture}} \qquad &
        \\
        \scalebox{0.55}{\tikzset{every picture/.style={line width=0.75pt}} 

\begin{tikzpicture}[x=0.75pt,y=0.75pt,yscale=-1,xscale=1]

\draw    (205.8,120.2) -- (205.8,148.85) ;
\draw [shift={(205.8,150.85)}, rotate = 270] [color={rgb, 255:red, 0; green, 0; blue, 0 }  ][line width=0.75]    (10.93,-4.9) .. controls (6.95,-2.3) and (3.31,-0.67) .. (0,0) .. controls (3.31,0.67) and (6.95,2.3) .. (10.93,4.9)   ;

\end{tikzpicture}} &&
        \small\ref{eq:qPA5'} &&
        \small\ref{eq:qPA6'} &&
        \small\ref{eq:qPA7}
        \\
        \scalebox{0.55}{\tikzset{every picture/.style={line width=0.75pt}} 

\begin{tikzpicture}[x=0.75pt,y=0.75pt,yscale=-1,xscale=1]

\draw    (90.8,20.92) -- (90.8,210.7) ;
\draw  [fill={rgb, 255:red, 0; green, 0; blue, 0 }  ,fill opacity=1 ] (87.46,145.54) .. controls (87.46,143.75) and (88.92,142.29) .. (90.71,142.29) .. controls (92.5,142.29) and (93.96,143.75) .. (93.96,145.54) .. controls (93.96,147.33) and (92.5,148.79) .. (90.71,148.79) .. controls (88.92,148.79) and (87.46,147.33) .. (87.46,145.54) -- cycle ;
\draw  [fill={rgb, 255:red, 0; green, 0; blue, 0 }  ,fill opacity=1 ] (87.26,85.32) .. controls (87.26,83.53) and (88.71,82.08) .. (90.5,82.08) .. controls (92.3,82.08) and (93.75,83.53) .. (93.75,85.32) .. controls (93.75,87.12) and (92.3,88.57) .. (90.5,88.57) .. controls (88.71,88.57) and (87.26,87.12) .. (87.26,85.32) -- cycle ;

\draw    (220.38,21.42) -- (220.38,211.2) ;
\draw  [fill={rgb, 255:red, 0; green, 0; blue, 0 }  ,fill opacity=1 ] (217.05,146.04) .. controls (217.05,144.25) and (218.5,142.79) .. (220.29,142.79) .. controls (222.09,142.79) and (223.54,144.25) .. (223.54,146.04) .. controls (223.54,147.83) and (222.09,149.29) .. (220.29,149.29) .. controls (218.5,149.29) and (217.05,147.83) .. (217.05,146.04) -- cycle ;
\draw    (250.33,50.5) -- (60.55,50.5) ;
\draw  [fill={rgb, 255:red, 0; green, 0; blue, 0 }  ,fill opacity=1 ] (125.71,47.16) .. controls (127.5,47.16) and (128.95,48.62) .. (128.95,50.41) .. controls (128.95,52.2) and (127.5,53.66) .. (125.71,53.66) .. controls (123.92,53.66) and (122.46,52.2) .. (122.46,50.41) .. controls (122.46,48.62) and (123.92,47.16) .. (125.71,47.16) -- cycle ;
\draw  [fill={rgb, 255:red, 0; green, 0; blue, 0 }  ,fill opacity=1 ] (220.18,46.96) .. controls (221.97,46.96) and (223.42,48.41) .. (223.42,50.21) .. controls (223.42,52) and (221.97,53.45) .. (220.18,53.45) .. controls (218.38,53.45) and (216.93,52) .. (216.93,50.21) .. controls (216.93,48.41) and (218.38,46.96) .. (220.18,46.96) -- cycle ;
\draw    (250.33,180.75) -- (60.55,180.75) ;
\draw  [fill={rgb, 255:red, 0; green, 0; blue, 0 }  ,fill opacity=1 ] (125.71,177.41) .. controls (127.5,177.41) and (128.95,178.87) .. (128.95,180.66) .. controls (128.95,182.45) and (127.5,183.91) .. (125.71,183.91) .. controls (123.92,183.91) and (122.46,182.45) .. (122.46,180.66) .. controls (122.46,178.87) and (123.92,177.41) .. (125.71,177.41) -- cycle ;
\draw  [fill={rgb, 255:red, 0; green, 0; blue, 0 }  ,fill opacity=1 ] (185.93,177.21) .. controls (187.72,177.21) and (189.17,178.66) .. (189.17,180.46) .. controls (189.17,182.25) and (187.72,183.7) .. (185.93,183.7) .. controls (184.13,183.7) and (182.68,182.25) .. (182.68,180.46) .. controls (182.68,178.66) and (184.13,177.21) .. (185.93,177.21) -- cycle ;

\draw (64.71,141.89) node [anchor=north west][inner sep=0.75pt]    {\LARGE$p_{1}$};
\draw (64.7,81.56) node [anchor=north west][inner sep=0.75pt]    {\LARGE$p_{2}$};
\draw (230.71,141.9) node [anchor=north west][inner sep=0.75pt]    {\LARGE$p_{3}$};
\draw (118.26,156.56) node [anchor=north west][inner sep=0.75pt]    {\LARGE$p_{5}$};
\draw (178.71,156.34) node [anchor=north west][inner sep=0.75pt]    {\LARGE$p_{6}$};
\draw (118.26,62.48) node [anchor=north west][inner sep=0.75pt]    {\LARGE$p_{7}$};
\draw (230.7,62.25) node [anchor=north west][inner sep=0.75pt]    {\LARGE$p_{4,8}$};

\end{tikzpicture}} &
        \scalebox{0.55}{\tikzset{every picture/.style={line width=0.75pt}} 

\begin{tikzpicture}[x=0.75pt,y=0.75pt,yscale=-1,xscale=1]

\draw [color={rgb, 255:red, 0; green, 0; blue, 0 }  ,draw opacity=0 ]   (220.38,21.42) -- (220.38,211.2) ;
\draw    (206.13,120.2) -- (235,120.2) ;
\draw [shift={(237,120.2)}, rotate = 180] [color={rgb, 255:red, 0; green, 0; blue, 0 }  ][line width=0.75]    (10.93,-4.9) .. controls (6.95,-2.3) and (3.31,-0.67) .. (0,0) .. controls (3.31,0.67) and (6.95,2.3) .. (10.93,4.9)   ;

\end{tikzpicture}}
        &
        \scalebox{0.55}{\tikzset{every picture/.style={line width=0.75pt}} 

\begin{tikzpicture}[x=0.75pt,y=0.75pt,yscale=-1,xscale=1]

\draw    (90.8,20.92) -- (90.8,210.7) ;
\draw  [fill={rgb, 255:red, 0; green, 0; blue, 0 }  ,fill opacity=1 ] (87.46,180.54) .. controls (87.46,178.75) and (88.92,177.29) .. (90.71,177.29) .. controls (92.5,177.29) and (93.96,178.75) .. (93.96,180.54) .. controls (93.96,182.33) and (92.5,183.79) .. (90.71,183.79) .. controls (88.92,183.79) and (87.46,182.33) .. (87.46,180.54) -- cycle ;
\draw  [fill={rgb, 255:red, 0; green, 0; blue, 0 }  ,fill opacity=1 ] (87.26,85.32) .. controls (87.26,83.53) and (88.71,82.08) .. (90.5,82.08) .. controls (92.3,82.08) and (93.75,83.53) .. (93.75,85.32) .. controls (93.75,87.12) and (92.3,88.57) .. (90.5,88.57) .. controls (88.71,88.57) and (87.26,87.12) .. (87.26,85.32) -- cycle ;
\draw    (220.38,21.42) -- (220.38,211.2) ;
\draw  [fill={rgb, 255:red, 0; green, 0; blue, 0 }  ,fill opacity=1 ] (217.05,146.04) .. controls (217.05,144.25) and (218.5,142.79) .. (220.29,142.79) .. controls (222.09,142.79) and (223.54,144.25) .. (223.54,146.04) .. controls (223.54,147.83) and (222.09,149.29) .. (220.29,149.29) .. controls (218.5,149.29) and (217.05,147.83) .. (217.05,146.04) -- cycle ;
\draw    (250.33,50.5) -- (60.55,50.5) ;
\draw  [fill={rgb, 255:red, 0; green, 0; blue, 0 }  ,fill opacity=1 ] (125.71,47.16) .. controls (127.5,47.16) and (128.95,48.62) .. (128.95,50.41) .. controls (128.95,52.2) and (127.5,53.66) .. (125.71,53.66) .. controls (123.92,53.66) and (122.46,52.2) .. (122.46,50.41) .. controls (122.46,48.62) and (123.92,47.16) .. (125.71,47.16) -- cycle ;
\draw  [fill={rgb, 255:red, 0; green, 0; blue, 0 }  ,fill opacity=1 ] (220.18,46.96) .. controls (221.97,46.96) and (223.42,48.41) .. (223.42,50.21) .. controls (223.42,52) and (221.97,53.45) .. (220.18,53.45) .. controls (218.38,53.45) and (216.93,52) .. (216.93,50.21) .. controls (216.93,48.41) and (218.38,46.96) .. (220.18,46.96) -- cycle ;
\draw    (250.33,180.75) -- (60.55,180.75) ;
\draw  [fill={rgb, 255:red, 0; green, 0; blue, 0 }  ,fill opacity=1 ] (185.93,177.21) .. controls (187.72,177.21) and (189.17,178.66) .. (189.17,180.46) .. controls (189.17,182.25) and (187.72,183.7) .. (185.93,183.7) .. controls (184.13,183.7) and (182.68,182.25) .. (182.68,180.46) .. controls (182.68,178.66) and (184.13,177.21) .. (185.93,177.21) -- cycle ;

\draw (64.7,81.56) node [anchor=north west][inner sep=0.75pt]    {\LARGE$p_{2}$};
\draw (230.71,141.9) node [anchor=north west][inner sep=0.75pt]    {\LARGE$p_{3}$};
\draw (54.7,156.56) node [anchor=north west][inner sep=0.75pt]    {\LARGE$p_{1,5}$};
\draw (178.71,156.34) node [anchor=north west][inner sep=0.75pt]    {\LARGE$p_{6}$};
\draw (118.26,62.48) node [anchor=north west][inner sep=0.75pt]    {\LARGE$p_{7}$};
\draw (230.7,62.25) node [anchor=north west][inner sep=0.75pt]    {\LARGE$p_{4,8}$};

\end{tikzpicture}} \qquad &
        \scalebox{0.55}{\tikzset{every picture/.style={line width=0.75pt}} 

\begin{tikzpicture}[x=0.75pt,y=0.75pt,yscale=-1,xscale=1]

\draw [color={rgb, 255:red, 0; green, 0; blue, 0 }  ,draw opacity=0 ]   (220.38,21.42) -- (220.38,211.2) ;
\draw    (206.13,120.2) -- (235,120.2) ;
\draw [shift={(237,120.2)}, rotate = 180] [color={rgb, 255:red, 0; green, 0; blue, 0 }  ][line width=0.75]    (10.93,-4.9) .. controls (6.95,-2.3) and (3.31,-0.67) .. (0,0) .. controls (3.31,0.67) and (6.95,2.3) .. (10.93,4.9)   ;

\end{tikzpicture}} &
        \scalebox{0.55}{\tikzset{every picture/.style={line width=0.75pt}} 

\begin{tikzpicture}[x=0.75pt,y=0.75pt,yscale=-1,xscale=1]

\draw    (90.8,20.92) -- (90.8,210.7) ;
\draw  [fill={rgb, 255:red, 0; green, 0; blue, 0 }  ,fill opacity=1 ] (87.46,180.54) .. controls (87.46,178.75) and (88.92,177.29) .. (90.71,177.29) .. controls (92.5,177.29) and (93.96,178.75) .. (93.96,180.54) .. controls (93.96,182.33) and (92.5,183.79) .. (90.71,183.79) .. controls (88.92,183.79) and (87.46,182.33) .. (87.46,180.54) -- cycle ;
\draw    (220.38,21.42) -- (220.38,211.2) ;
\draw  [fill={rgb, 255:red, 0; green, 0; blue, 0 }  ,fill opacity=1 ] (217.05,146.04) .. controls (217.05,144.25) and (218.5,142.79) .. (220.29,142.79) .. controls (222.09,142.79) and (223.54,144.25) .. (223.54,146.04) .. controls (223.54,147.83) and (222.09,149.29) .. (220.29,149.29) .. controls (218.5,149.29) and (217.05,147.83) .. (217.05,146.04) -- cycle ;
\draw    (250.33,50.5) -- (60.55,50.5) ;
\draw  [fill={rgb, 255:red, 0; green, 0; blue, 0 }  ,fill opacity=1 ] (125.71,47.16) .. controls (127.5,47.16) and (128.95,48.62) .. (128.95,50.41) .. controls (128.95,52.2) and (127.5,53.66) .. (125.71,53.66) .. controls (123.92,53.66) and (122.46,52.2) .. (122.46,50.41) .. controls (122.46,48.62) and (123.92,47.16) .. (125.71,47.16) -- cycle ;
\draw  [fill={rgb, 255:red, 0; green, 0; blue, 0 }  ,fill opacity=1 ] (220.18,46.96) .. controls (221.97,46.96) and (223.42,48.41) .. (223.42,50.21) .. controls (223.42,52) and (221.97,53.45) .. (220.18,53.45) .. controls (218.38,53.45) and (216.93,52) .. (216.93,50.21) .. controls (216.93,48.41) and (218.38,46.96) .. (220.18,46.96) -- cycle ;
\draw    (250.33,180.75) -- (60.55,180.75) ;
\draw  [fill={rgb, 255:red, 0; green, 0; blue, 0 }  ,fill opacity=1 ] (185.93,177.21) .. controls (187.72,177.21) and (189.17,178.66) .. (189.17,180.46) .. controls (189.17,182.25) and (187.72,183.7) .. (185.93,183.7) .. controls (184.13,183.7) and (182.68,182.25) .. (182.68,180.46) .. controls (182.68,178.66) and (184.13,177.21) .. (185.93,177.21) -- cycle ;

\draw (230.71,141.9) node [anchor=north west][inner sep=0.75pt]    {\LARGE$p_{3}$};
\draw (44.7,156.56) node [anchor=north west][inner sep=0.75pt]    {\LARGE$p_{1,2,5}$};
\draw (178.71,156.34) node [anchor=north west][inner sep=0.75pt]    {\LARGE$p_{6}$};
\draw (118.26,62.48) node [anchor=north west][inner sep=0.75pt]    {\LARGE$p_{7}$};
\draw (230.7,62.25) node [anchor=north west][inner sep=0.75pt]    {\LARGE$p_{4,8}$};

\end{tikzpicture}} \qquad &
        \scalebox{0.55}{\tikzset{every picture/.style={line width=0.75pt}} 

\begin{tikzpicture}[x=0.75pt,y=0.75pt,yscale=-1,xscale=1]

\draw [color={rgb, 255:red, 0; green, 0; blue, 0 }  ,draw opacity=0 ]   (220.38,21.42) -- (220.38,211.2) ;
\draw    (206.13,120.2) -- (235,120.2) ;
\draw [shift={(237,120.2)}, rotate = 180] [color={rgb, 255:red, 0; green, 0; blue, 0 }  ][line width=0.75]    (10.93,-4.9) .. controls (6.95,-2.3) and (3.31,-0.67) .. (0,0) .. controls (3.31,0.67) and (6.95,2.3) .. (10.93,4.9)   ;

\end{tikzpicture}} &
        \scalebox{0.55}{\tikzset{every picture/.style={line width=0.75pt}} 

\begin{tikzpicture}[x=0.75pt,y=0.75pt,yscale=-1,xscale=1]

\draw    (90.8,20.92) -- (90.8,210.7) ;
\draw  [fill={rgb, 255:red, 0; green, 0; blue, 0 }  ,fill opacity=1 ] (87.46,180.54) .. controls (87.46,178.75) and (88.92,177.29) .. (90.71,177.29) .. controls (92.5,177.29) and (93.96,178.75) .. (93.96,180.54) .. controls (93.96,182.33) and (92.5,183.79) .. (90.71,183.79) .. controls (88.92,183.79) and (87.46,182.33) .. (87.46,180.54) -- cycle ;
\draw    (220.38,21.42) -- (220.38,211.2) ;
\draw    (250.33,50.5) -- (60.55,50.5) ;
\draw  [fill={rgb, 255:red, 0; green, 0; blue, 0 }  ,fill opacity=1 ] (125.71,47.16) .. controls (127.5,47.16) and (128.95,48.62) .. (128.95,50.41) .. controls (128.95,52.2) and (127.5,53.66) .. (125.71,53.66) .. controls (123.92,53.66) and (122.46,52.2) .. (122.46,50.41) .. controls (122.46,48.62) and (123.92,47.16) .. (125.71,47.16) -- cycle ;
\draw  [fill={rgb, 255:red, 0; green, 0; blue, 0 }  ,fill opacity=1 ] (220.18,46.96) .. controls (221.97,46.96) and (223.42,48.41) .. (223.42,50.21) .. controls (223.42,52) and (221.97,53.45) .. (220.18,53.45) .. controls (218.38,53.45) and (216.93,52) .. (216.93,50.21) .. controls (216.93,48.41) and (218.38,46.96) .. (220.18,46.96) -- cycle ;
\draw    (250.33,180.75) -- (60.55,180.75) ;
\draw  [fill={rgb, 255:red, 0; green, 0; blue, 0 }  ,fill opacity=1 ] (185.93,177.21) .. controls (187.72,177.21) and (189.17,178.66) .. (189.17,180.46) .. controls (189.17,182.25) and (187.72,183.7) .. (185.93,183.7) .. controls (184.13,183.7) and (182.68,182.25) .. (182.68,180.46) .. controls (182.68,178.66) and (184.13,177.21) .. (185.93,177.21) -- cycle ;

\draw (44.7,156.56) node [anchor=north west][inner sep=0.75pt]    {\LARGE$p_{1,2,5}$};
\draw (178.71,156.34) node [anchor=north west][inner sep=0.75pt]    {\LARGE$p_{6}$};
\draw (118.26,62.48) node [anchor=north west][inner sep=0.75pt]    {\LARGE$p_{7}$};
\draw (230.7,62.25) node [anchor=north west][inner sep=0.75pt]    {\LARGE$p_{3,4,8}$};

\end{tikzpicture}}
        \\
        \small\ref{eq:qPA4} &
        \scalebox{0.55}{\tikzset{every picture/.style={line width=0.75pt}} 

\begin{tikzpicture}[x=0.75pt,y=0.75pt,yscale=-1,xscale=1]

\draw    (205.8,120.2) -- (232.05,134.84) ;
\draw [shift={(233.8,135.82)}, rotate = 209.15] [color={rgb, 255:red, 0; green, 0; blue, 0 }  ][line width=0.75]    (10.93,-4.9) .. controls (6.95,-2.3) and (3.31,-0.67) .. (0,0) .. controls (3.31,0.67) and (6.95,2.3) .. (10.93,4.9)   ;

\end{tikzpicture}} &
        \small\ref{eq:qPA5} &
        \scalebox{0.55}{\tikzset{every picture/.style={line width=0.75pt}} 

\begin{tikzpicture}[x=0.75pt,y=0.75pt,yscale=-1,xscale=1]

\draw    (205.8,120.2) -- (232.05,134.84) ;
\draw [shift={(233.8,135.82)}, rotate = 209.15] [color={rgb, 255:red, 0; green, 0; blue, 0 }  ][line width=0.75]    (10.93,-4.9) .. controls (6.95,-2.3) and (3.31,-0.67) .. (0,0) .. controls (3.31,0.67) and (6.95,2.3) .. (10.93,4.9)   ;

\end{tikzpicture}} &
        \small\ref{eq:qPA6} &
        \scalebox{0.55}{\tikzset{every picture/.style={line width=0.75pt}} 

\begin{tikzpicture}[x=0.75pt,y=0.75pt,yscale=-1,xscale=1]

\draw    (205.8,120.2) -- (232.05,134.84) ;
\draw [shift={(233.8,135.82)}, rotate = 209.15] [color={rgb, 255:red, 0; green, 0; blue, 0 }  ][line width=0.75]    (10.93,-4.9) .. controls (6.95,-2.3) and (3.31,-0.67) .. (0,0) .. controls (3.31,0.67) and (6.95,2.3) .. (10.93,4.9)   ;

\end{tikzpicture}} &
        \small\ref{eq:qPA7'}
        \\
        &&
        \scalebox{0.55}{\tikzset{every picture/.style={line width=0.75pt}} 

\begin{tikzpicture}[x=0.75pt,y=0.75pt,yscale=-1,xscale=1]

\draw    (90.8,20.92) -- (90.8,210.7) ;
\draw  [fill={rgb, 255:red, 0; green, 0; blue, 0 }  ,fill opacity=1 ] (87.46,145.54) .. controls (87.46,143.75) and (88.92,142.29) .. (90.71,142.29) .. controls (92.5,142.29) and (93.96,143.75) .. (93.96,145.54) .. controls (93.96,147.33) and (92.5,148.79) .. (90.71,148.79) .. controls (88.92,148.79) and (87.46,147.33) .. (87.46,145.54) -- cycle ;
\draw  [fill={rgb, 255:red, 0; green, 0; blue, 0 }  ,fill opacity=1 ] (87.26,85.32) .. controls (87.26,83.53) and (88.71,82.08) .. (90.5,82.08) .. controls (92.3,82.08) and (93.75,83.53) .. (93.75,85.32) .. controls (93.75,87.12) and (92.3,88.57) .. (90.5,88.57) .. controls (88.71,88.57) and (87.26,87.12) .. (87.26,85.32) -- cycle ;

\draw    (220.38,21.42) -- (220.38,211.2) ;
\draw  [fill={rgb, 255:red, 0; green, 0; blue, 0 }  ,fill opacity=1 ] (217.15,180.62) .. controls (217.15,178.83) and (218.61,177.37) .. (220.4,177.37) .. controls (222.19,177.37) and (223.64,178.83) .. (223.64,180.62) .. controls (223.64,182.41) and (222.19,183.87) .. (220.4,183.87) .. controls (218.61,183.87) and (217.15,182.41) .. (217.15,180.62) -- cycle ;
\draw    (250.33,50.5) -- (60.55,50.5) ;
\draw  [fill={rgb, 255:red, 0; green, 0; blue, 0 }  ,fill opacity=1 ] (125.71,47.16) .. controls (127.5,47.16) and (128.95,48.62) .. (128.95,50.41) .. controls (128.95,52.2) and (127.5,53.66) .. (125.71,53.66) .. controls (123.92,53.66) and (122.46,52.2) .. (122.46,50.41) .. controls (122.46,48.62) and (123.92,47.16) .. (125.71,47.16) -- cycle ;
\draw  [fill={rgb, 255:red, 0; green, 0; blue, 0 }  ,fill opacity=1 ] (220.18,46.96) .. controls (221.97,46.96) and (223.42,48.41) .. (223.42,50.21) .. controls (223.42,52) and (221.97,53.45) .. (220.18,53.45) .. controls (218.38,53.45) and (216.93,52) .. (216.93,50.21) .. controls (216.93,48.41) and (218.38,46.96) .. (220.18,46.96) -- cycle ;
\draw    (250.33,180.75) -- (60.55,180.75) ;
\draw  [fill={rgb, 255:red, 0; green, 0; blue, 0 }  ,fill opacity=1 ] (125.71,177.41) .. controls (127.5,177.41) and (128.95,178.87) .. (128.95,180.66) .. controls (128.95,182.45) and (127.5,183.91) .. (125.71,183.91) .. controls (123.92,183.91) and (122.46,182.45) .. (122.46,180.66) .. controls (122.46,178.87) and (123.92,177.41) .. (125.71,177.41) -- cycle ;

\draw (64.71,141.89) node [anchor=north west][inner sep=0.75pt]    {\LARGE$p_{1}$};
\draw (64.7,81.56) node [anchor=north west][inner sep=0.75pt]    {\LARGE$p_{2}$};
\draw (230.71,156.6) node [anchor=north west][inner sep=0.75pt]    {\LARGE$p_{3,6}$};
\draw (118.26,156.56) node [anchor=north west][inner sep=0.75pt]    {\LARGE$p_{5}$};
\draw (118.26,62.48) node [anchor=north west][inner sep=0.75pt]    {\LARGE$p_{7}$};
\draw (230.7,62.25) node [anchor=north west][inner sep=0.75pt]    {\LARGE$p_{4,8}$};

\end{tikzpicture}} &
        \scalebox{0.55}{\tikzset{every picture/.style={line width=0.75pt}} 

\begin{tikzpicture}[x=0.75pt,y=0.75pt,yscale=-1,xscale=1]

\draw [color={rgb, 255:red, 0; green, 0; blue, 0 }  ,draw opacity=0 ]   (220.38,21.42) -- (220.38,211.2) ;
\draw    (206.13,120.2) -- (235,120.2) ;
\draw [shift={(237,120.2)}, rotate = 180] [color={rgb, 255:red, 0; green, 0; blue, 0 }  ][line width=0.75]    (10.93,-4.9) .. controls (6.95,-2.3) and (3.31,-0.67) .. (0,0) .. controls (3.31,0.67) and (6.95,2.3) .. (10.93,4.9)   ;

\end{tikzpicture}} &
        \scalebox{0.55}{\tikzset{every picture/.style={line width=0.75pt}} 

\begin{tikzpicture}[x=0.75pt,y=0.75pt,yscale=-1,xscale=1]

\draw    (90.8,20.92) -- (90.8,210.7) ;
\draw  [fill={rgb, 255:red, 0; green, 0; blue, 0 }  ,fill opacity=1 ] (87.6,180.74) .. controls (87.6,178.95) and (89.05,177.49) .. (90.84,177.49) .. controls (92.64,177.49) and (94.09,178.95) .. (94.09,180.74) .. controls (94.09,182.53) and (92.64,183.99) .. (90.84,183.99) .. controls (89.05,183.99) and (87.6,182.53) .. (87.6,180.74) -- cycle ;
\draw  [fill={rgb, 255:red, 0; green, 0; blue, 0 }  ,fill opacity=1 ] (87.26,85.32) .. controls (87.26,83.53) and (88.71,82.08) .. (90.5,82.08) .. controls (92.3,82.08) and (93.75,83.53) .. (93.75,85.32) .. controls (93.75,87.12) and (92.3,88.57) .. (90.5,88.57) .. controls (88.71,88.57) and (87.26,87.12) .. (87.26,85.32) -- cycle ;
\draw    (220.38,21.42) -- (220.38,211.2) ;
\draw  [fill={rgb, 255:red, 0; green, 0; blue, 0 }  ,fill opacity=1 ] (217.15,180.62) .. controls (217.15,178.83) and (218.61,177.37) .. (220.4,177.37) .. controls (222.19,177.37) and (223.64,178.83) .. (223.64,180.62) .. controls (223.64,182.41) and (222.19,183.87) .. (220.4,183.87) .. controls (218.61,183.87) and (217.15,182.41) .. (217.15,180.62) -- cycle ;
\draw    (250.33,50.5) -- (60.55,50.5) ;
\draw  [fill={rgb, 255:red, 0; green, 0; blue, 0 }  ,fill opacity=1 ] (125.71,47.16) .. controls (127.5,47.16) and (128.95,48.62) .. (128.95,50.41) .. controls (128.95,52.2) and (127.5,53.66) .. (125.71,53.66) .. controls (123.92,53.66) and (122.46,52.2) .. (122.46,50.41) .. controls (122.46,48.62) and (123.92,47.16) .. (125.71,47.16) -- cycle ;
\draw  [fill={rgb, 255:red, 0; green, 0; blue, 0 }  ,fill opacity=1 ] (220.18,46.96) .. controls (221.97,46.96) and (223.42,48.41) .. (223.42,50.21) .. controls (223.42,52) and (221.97,53.45) .. (220.18,53.45) .. controls (218.38,53.45) and (216.93,52) .. (216.93,50.21) .. controls (216.93,48.41) and (218.38,46.96) .. (220.18,46.96) -- cycle ;
\draw    (250.33,180.75) -- (60.55,180.75) ;

\draw (54.71,156.6) node [anchor=north west][inner sep=0.75pt]    {\LARGE$p_{1,5}$};
\draw (64.7,81.56) node [anchor=north west][inner sep=0.75pt]    {\LARGE$p_{2}$};
\draw (230.71,156.6) node [anchor=north west][inner sep=0.75pt]    {\LARGE$p_{3,6}$};
\draw (118.26,62.48) node [anchor=north west][inner sep=0.75pt]    {\LARGE$p_{7}$};
\draw (230.7,62.25) node [anchor=north west][inner sep=0.75pt]    {\LARGE$p_{4,8}$};

\end{tikzpicture}} &
        \scalebox{0.55}{\tikzset{every picture/.style={line width=0.75pt}} 

\begin{tikzpicture}[x=0.75pt,y=0.75pt,yscale=-1,xscale=1]

\draw [color={rgb, 255:red, 0; green, 0; blue, 0 }  ,draw opacity=0 ]   (220.38,21.42) -- (220.38,211.2) ;
\draw    (206.13,120.2) -- (235,120.2) ;
\draw [shift={(237,120.2)}, rotate = 180] [color={rgb, 255:red, 0; green, 0; blue, 0 }  ][line width=0.75]    (10.93,-4.9) .. controls (6.95,-2.3) and (3.31,-0.67) .. (0,0) .. controls (3.31,0.67) and (6.95,2.3) .. (10.93,4.9)   ;

\end{tikzpicture}} &
        \scalebox{0.55}{\tikzset{every picture/.style={line width=0.75pt}} 

\begin{tikzpicture}[x=0.75pt,y=0.75pt,yscale=-1,xscale=1]

\draw    (90.8,20.92) -- (90.8,210.7) ;
\draw  [fill={rgb, 255:red, 0; green, 0; blue, 0 }  ,fill opacity=1 ] (87.6,180.74) .. controls (87.6,178.95) and (89.05,177.49) .. (90.84,177.49) .. controls (92.64,177.49) and (94.09,178.95) .. (94.09,180.74) .. controls (94.09,182.53) and (92.64,183.99) .. (90.84,183.99) .. controls (89.05,183.99) and (87.6,182.53) .. (87.6,180.74) -- cycle ;
\draw    (220.38,21.42) -- (220.38,211.2) ;
\draw  [fill={rgb, 255:red, 0; green, 0; blue, 0 }  ,fill opacity=1 ] (217.15,180.62) .. controls (217.15,178.83) and (218.61,177.37) .. (220.4,177.37) .. controls (222.19,177.37) and (223.64,178.83) .. (223.64,180.62) .. controls (223.64,182.41) and (222.19,183.87) .. (220.4,183.87) .. controls (218.61,183.87) and (217.15,182.41) .. (217.15,180.62) -- cycle ;
\draw    (250.33,50.5) -- (60.55,50.5) ;
\draw  [fill={rgb, 255:red, 0; green, 0; blue, 0 }  ,fill opacity=1 ] (125.71,47.16) .. controls (127.5,47.16) and (128.95,48.62) .. (128.95,50.41) .. controls (128.95,52.2) and (127.5,53.66) .. (125.71,53.66) .. controls (123.92,53.66) and (122.46,52.2) .. (122.46,50.41) .. controls (122.46,48.62) and (123.92,47.16) .. (125.71,47.16) -- cycle ;
\draw  [fill={rgb, 255:red, 0; green, 0; blue, 0 }  ,fill opacity=1 ] (220.18,46.96) .. controls (221.97,46.96) and (223.42,48.41) .. (223.42,50.21) .. controls (223.42,52) and (221.97,53.45) .. (220.18,53.45) .. controls (218.38,53.45) and (216.93,52) .. (216.93,50.21) .. controls (216.93,48.41) and (218.38,46.96) .. (220.18,46.96) -- cycle ;
\draw    (250.33,180.75) -- (60.55,180.75) ;

\draw (44.71,156.6) node [anchor=north west][inner sep=0.75pt]    {\LARGE$p_{1,2,5}$};
\draw (230.71,156.6) node [anchor=north west][inner sep=0.75pt]    {\LARGE$p_{3,6}$};
\draw (118.26,62.48) node [anchor=north west][inner sep=0.75pt]    {\LARGE$p_{7}$};
\draw (230.7,62.25) node [anchor=north west][inner sep=0.75pt]    {\LARGE$p_{4,8}$};

\end{tikzpicture}} \qquad 
    \end{tabular}
    \caption{Degeneration scheme of the \ref{eq:qPA3} system to lower $q$-cases}
    \label{pic:qPA3deg}
\end{figure}

\medskip
\paragraph{\textbullet \,\, \texorpdfstring{\rm{\ref{eq:qPA3} $\to$ \ref{eq:qPA4}}}{qPA3 -> qPA4}}
We want to have in the limit $p_4 = \br{- b_4, \infty} \to \br{\infty, \infty}$ and $p_8 = \br{\infty, - b_8} \to \br{\infty, \infty}$. In order to obtain that, one can use the following transformation with the small parameter $\varepsilon$
\begin{align}
    &&
    b_4
    &\mapsto \varepsilon \, b_4,
    &
    b_8
    &\mapsto \varepsilon \, b_8
    ,
    &&
\end{align}
which leads to
\begin{align}
    \underline{f} \, f
    &= a \, b_7 \, \br{\varepsilon^{-1} b_8} \, 
    \br{g + b_6} \, 
    \br{g + \br{\varepsilon^{-1} b_8}}^{-1}
    \br{g + b_5} \, \br{g + b_7}^{-1}
    \\[1mm]
    &= a\, b_7 b_8 \, \br{g + b_6} \, 
    \br{\varepsilon g + b_8}^{-1}
    \br{g + b_5} \, \br{g + b_7}^{-1}
    ,
    \\[2mm]
    \bar{g} \, g
    &= a^{-1} \, b_3 \, \br{\varepsilon^{-1} b_4} \, 
    \br{f + b_2} \, 
    \br{f + \br{\varepsilon^{-1} b_4}}^{-1} 
    \br{f + b_1} \, \br{f + b_3}^{-1}
    \\[1mm]
    &= a^{-1} \, b_3 b_4 \, 
    \br{f + b_2} \, \br{\varepsilon f + b_4}^{-1} 
    \br{f + b_1} \, \br{f + b_3}^{-1}
    .
\end{align}
The limit $\varepsilon \to 0$ can be taken without specifying the parameter $a$. Thus, we get
\begin{align}
    \tag*{\ref{eq:qPA4}}
    &&
    \underline{f} \, f
    &= b_7 \, \br{g + b_6} \, 
    \br{g + b_5} \, \br{g + b_7}^{-1}
    ,
    &
    \bar{g} \, g
    &= b_3 \, 
    \br{f + b_2} \, 
    \br{f + b_1} \, \br{f + b_3}^{-1}
    .
    &&
\end{align}
Here $q = \br{b_1 \, b_2 \, b_7}^{\frac14} \, \br{b_3 \, b_5 \, b_6}^{-\frac14}$ and $\bar{b}_j = q^2 b_j$, $\bar{b}_k = q^{-2} b_k$ with $j = 1, 2, 5, 6$, $k = 3, 7$, so that $\bar q = q$.

\medskip
\paragraph{\textbullet \,\, \texorpdfstring{\rm{\ref{eq:qPA4} $\to$ \{\ref{eq:qPA5}, \ref{eq:qPA5'}}\}}{qPA4 -> qPA5, qPA5'}}
The \ref{eq:qPA5} corresponds to the case when $p_3 = p_6 = \br{0, \infty}$, i.e. one needs to make the transformation
\begin{align}
    &&
    b_3 
    &\mapsto \varepsilon^{-1} \, b_3,
    &
    b_6 
    &\mapsto \varepsilon \, b_6
    .
    &&
\end{align}
However, to take the limit, we have to use $a \mapsto \varepsilon^{-1} \, a$. Then,
\begin{align}
    \underline{f} \, f
    &= \br{\varepsilon \, a} \, b_7 \br{g + \br{\varepsilon^{-1} b_6}} \,
    \br{g + b_5} \, \br{g + b_7}^{-1}
    = a \, b_7 \, \br{\varepsilon g + b_6} \, \br{g + b_5} \, \br{g + b_7}^{-1}
    ,
    \\[2mm]
    \bar{g} \, g
    &= \br{\varepsilon \, a}^{-1} \, \br{\varepsilon \, b_3} \, 
    \br{f + b_2} \, \br{f + b_1} \, \br{f + \varepsilon b_3}^{-1}
    .
\end{align}
and, therefore, we get the system
\begin{align}
    \tag*{\ref{eq:qPA5}}
    &&
    \underline{f} \, f
    &= b_6 b_7 \, \br{g + b_5} \, \br{g + b_7}^{-1}
    ,
    &
    \bar{g} \, g
    &= b_3 \, 
    \br{f + b_2} \, \br{f + b_1} \, f^{-1}
    ,
    &&
\end{align}
with $q = \br{b_1 \, b_2 \, b_7}^{\frac14} \, \br{b_3 \, b_5 \, b_6}^{-\frac14}$ and $\bar{b}_j = q^2 b_j$, $\bar{b}_k = q^{-2} b_k$, $j = 1, 2, 5, 6$, $k = 3, 7$. Note that $\bar q = q$. 

\medskip
In order to get \ref{eq:qPA5'}, we make $b_1 \mapsto \varepsilon^{-1} \, b_1$ and $b_5 \mapsto \varepsilon^{-1} \, b_5$. Then, in the limit $\varepsilon \to 0$, $p_1 = \br{- b_1, 0} \to \br{0, 0}$ and $p_5 = \br{0, - b_5} \to \br{0, 0}$, while the equations change as follows
\begin{align}
    &&
    \underline{f} \, f
    &= a \, b_7 \, \br{g + b_6} \, \br{g + \varepsilon b_5} \, \br{g + b_7}^{-1},
    &
    \bar{g} \, g
    &= a^{-1} \, b_3 \, \br{f + b_2} \, \br{f + \varepsilon b_1} \, \br{f + b_3}^{-1},
    &&
\end{align}
or, after taking the limit,
\begin{align}
    \tag*{\ref{eq:qPA5'}}
    &&
    \underline{f} \, f
    &= b_7 \, \br{g + b_6} \, g \, \br{g + b_7}^{-1},
    &
    \bar{g} \, g
    &= b_3 \, \br{f + b_2} \, f \, \br{f + b_3}^{-1}.
    &&
\end{align}
Here $q = \br{b_2 \, b_7}^{\frac14} \, \br{b_3 \, b_6}^{-\frac14}$ and $\bar{b}_j = q^2 b_j$, $\bar{b}_k = q^{-2} b_k$ with $j = 2, 6$, $k = 3, 7$, so that $\bar q = q$. 

\medskip
\paragraph{\textbullet \,\, \texorpdfstring{\rm{\ref{eq:qPA5} $\to$ \ref{eq:qPA6}}}{qPA5 -> qPA6}}

In this case $p_1 = p_5 = \br{0, 0}$, that is, $b_1 \mapsto \varepsilon^{-1} \, b_1$, $b_5 \mapsto \varepsilon^{-1} \, b_5$. The limit $\varepsilon \to 0$ gives
\begin{align}
    \tag*{\ref{eq:qPA6}}
    &&
    \underline{f} \, f
    &= b_6 b_7 \, g \, \br{g + b_7}^{-1}
    ,
    &
    \bar{g} \, g
    &= b_3 \, \br{f + b_2}
    &&
\end{align}
with $q = \br{b_2 \, b_7}^{\frac14} \, \br{b_3 \, b_6}^{-\frac14}$ and $\bar{b}_j = q^2 b_j$, $\bar{b}_k = q^{-2} b_k$, where $j = 2, 6$, $k = 3, 7$. Note that $\bar q = q$. 

\medskip
\paragraph{\textbullet \,\, \texorpdfstring{\rm{\ref{eq:qPA5'} $\to$ \{\ref{eq:qPA6}, \ref{eq:qPA6'}}\}}{qPA5' -> qPA6, qPA6'}}
To obtain the \ref{eq:qPA6} from the \ref{eq:qPA5'}, we have to use the inessential parameter as follows $a \mapsto \varepsilon^{-1} \, a$ and then make the change of parameters
\begin{align}
    &&
    b_3
    &\mapsto \varepsilon^{-1} \, b_3,
    &
    b_6
    &\mapsto \varepsilon \, b_6,
    &&
\end{align}
which leads to $p_3 = p_6 = \br{0, \infty}$ in the limit $\varepsilon \to 0$. The resulting system reads as
\begin{align}
    \tag*{\ref{eq:qPA6}}
    &&
    \underline{f} \, f
    &= b_6 b_7 \, g \, \br{g + b_7}^{-1},
    &
    \bar{g} \, g
    &= b_3 \, \br{f + b_2}
    ,
    &&
\end{align}
where $\bar{b}_j = q^2 b_j$, $\bar{b}_k = q^{-2} b_k$ with $j = 2, 6$, $k = 3, 7$ and $q = \br{b_2 \, b_7}^{\frac14} \, \br{b_3 \, b_6}^{-\frac14}$, so that $\bar q = q$. 

\medskip
According to the point configuration for the \ref{eq:qPA6'}, $p_2 = \br{0, 0}$ which corresponds to the transformation $b_2 \mapsto \varepsilon^{-1} \, b_2$. Thus, the \ref{eq:qPA5'} turns into
\begin{align}
    \tag*{\ref{eq:qPA6'}}
    &&
    \underline{f} \, f
    &= b_7 \, \br{g + b_6} \, g \, \br{g + b_7}^{-1},
    &
    \bar{g} \, g
    &= b_3 \, f^2 \, \br{f + b_3}^{-1}
    .
    &&
\end{align}
Here we set $\bar{b}_3 = b_3$, so that the dynamical parameters are $\bar{b}_6 = q^{2} b_6$ and $\bar{b}_7 = q^{-2} b_7$, where {$q = \br{b_6 \, b_7}^{\frac14}$} and, therefore, $\bar q = q$.

\medskip
\paragraph{\textbullet \,\, \texorpdfstring{\rm{\ref{eq:qPA6} $\to$ \ref{eq:qPA7'}}}{qPA6 -> qPA7'}}
The point configuration for the \ref{eq:qPA7'} system can be obtained from the point configuration for the \ref{eq:qPA6} by making $b_2 \mapsto \varepsilon^{-1} \, b_2$, that corresponds to $p_2 = \br{- b_2, 0} \to \br{0, 0}$. One gets
\begin{align}
    \tag*{\ref{eq:qPA7'}}
    &&
    \underline{f} \, f
    &= b_6 b_7 \, g \, \br{g + b_7}^{-1}
    ,
    &
    \bar{g} \, g
    &= b_3 \, f
    ,
    &&
\end{align}
where $\bar b_3 = b_3$, $\bar{b}_6 = q^{2} b_6$, $\bar{b}_7 = q^{-2} b_7$, and {$q = \br{b_6 \, b_7}^{- \frac14}$}. Note that $\bar q = q$.

\medskip
\paragraph{\textbullet \,\, \texorpdfstring{\rm{\ref{eq:qPA6'} $\to$ \{\ref{eq:qPA7}, \ref{eq:qPA7'}}\}}{qPA6' -> qPA7, qPA7'}}

The degeneration \ref{eq:qPA6'} $\to$ \ref{eq:qPA7} corresponds to the transformation $b_3 \mapsto \varepsilon \, b_3$, which leads to $p_3 = \br{- b_3, \infty} \to \br{\infty, \infty}$. Hence, in the limit $\varepsilon \to 0$, the system reads as
\begin{align}
    \tag*{\ref{eq:qPA7}}
    &&
    \underline{f} \, f
    &= b_7 \, \br{g + b_6^{-1}} \, g \, \br{g + b_7}^{-1}
    ,
    &
    \bar{g} \, g
    &= f^2
    ,
    &&
\end{align}
where $\bar{b}_6 = q^{2} b_6$, $\bar{b}_7 = q^{-2} b_7$, and {$q = \br{b_6 \, {b_7}}^{\frac14}$}, so that $\bar q = q$.

\medskip
To get the point configuration for the \ref{eq:qPA7'}, we need to obtain $p_3 = p_6 = \br{0, \infty}$ in the limit $\varepsilon \to 0$, which corresponds to the transformation
\begin{align}
    &&
    b_3
    &\mapsto \varepsilon^{-1} \, b_3,
    &
    b_6
    &\mapsto \varepsilon \, b_6
    .
    &&
\end{align}
To take the limit in the system, the inessential parameter $a$ have to be rescaled as follows $a \mapsto \varepsilon^{-1} a$. Then, after taking the limit, we obtain the system \ref{eq:qPA7'}, where we set $\bar b_3 = b_3$. 

\subsubsection{\texorpdfstring{\rm{\ref{eq:qPA3} $\to$ \ref{eq:dPD4}}}{qPA3 -> dPD4}}
\label{sec:ddeg}

In order to connect the system \ref{eq:qPA3} with the non-commutative $d$-Painlevé systems obtained in \cite{bobrova2024affine}, we consider the degeneration \ref{eq:qPA3} $\to$ \ref{eq:dPD4}. Below, we use capital letters for the new variables and then restore the lowercase letters in the transformed systems.

\medskip
Consider the degeneration data
\begin{gather}
    \begin{aligned}
    f
    &= t^{-\frac12} \, F,
    &&&&&
    g
    &= 1 + \varepsilon \, \br{
    G + B_6 + \tfrac12 Q
    },
    &&&&&
    q
    &= 1 + \tfrac14 \varepsilon \, Q,
    \end{aligned}
    \\[2mm]
    \begin{aligned}
    b_1
    &= - t^{\frac12} \, \br{1 + \varepsilon \, (B_1 + 2 Q)},
    &&&
    b_2
    &= - t^{-\frac12} \, \br{1 + \varepsilon \, (B_2 + 2 Q)},
    &&&
    b_3
    &= - t^{-\frac12} \, (1 + \varepsilon \, B_3),
    \end{aligned}
    \\[2mm]
    \begin{aligned}
    b_4
    &= - t^{\frac12} \, (1 + \varepsilon \, B_4),
    &&&
    b_5
    &= - 1 - \varepsilon \, B_5,
    &&&
    b_6
    &= - 1 - \varepsilon \, B_6,
    &&&
    b_7
    &= - 1 - \varepsilon \, B_7,
    &&&
    b_8
    &= - 1 - \varepsilon \, B_8.
    \end{aligned}
\end{gather}
Substituting this into the \ref{eq:qPA3} system and taking the limit $\varepsilon \to 0$, one gets
\begin{align}
    \begin{multlined}
    f \, \bar{f}
    = t \, \bar{g} \, \br{
    \bar g + b_6 - b_8 + q
    }^{-1} \, \br{
    \bar g + b_6 - b_5
    } \, \br{
    \bar g + b_6 - b_7 + q
    }^{-1}
    ,
    \hspace{4cm}
    \\[1mm]
    g + \underline{g}
    = b_3 + b_4 - 2 b_6 
    + (b_3 - b_2 - q) \, (\underline{f} - 1)^{-1}
    + (b_4 - b_1 - q) \, t \, (\underline{f} - t)^{-1}
    ,
    \end{multlined}
\end{align}
or, equivalently, 
\begin{align}
    \label{eq:dPD4fromqPA3}
    \begin{multlined}
    {f} \, \bar{f}
    = t \, {g} \, \br{
     g + b_6 - b_8 + q
    }^{-1} \, \br{
     g + b_6 - b_5
    } \, \br{
     g + b_6 - b_7 + q
    }^{-1}
    ,
    \hspace{4cm}
    \\[1mm]
    g + \underline{g}
    = b_3 + b_4 - 2 b_6 
    + (b_3 - b_2 - q) \, ({f} - 1)^{-1}
    + (b_4 - b_1 - q) \, t \, ({f} - t)^{-1}
    ,
    \end{multlined}
\end{align}
where the transformation $\underline{f} \mapsto f$ was made. 

\medskip
Recall the form of the \ref{eq:dPD4} system \cite{bobrova2024affine}:
\begin{gather}
    \tag*{\ref{eq:dPD4}}
    \begin{gathered}
    \begin{aligned}
        \begin{aligned}
        \bar \alpha_0 
        &= \alpha_0 - 1,
        &&&&
        \end{aligned} 
        \bar \alpha_2 
        = \alpha_2 + 1,&
        &&&&&
        &
        \bar \alpha_3 
        = \alpha_3 - 1,
        \\
        \begin{aligned}
        f \, \bar f
        = t \, g \, (g + \alpha_2)^{-1} \,
        (g - \alpha_4) \, (g + \alpha_1 + \alpha_2)^{-1},
        \\[1mm]
        \phantom{f}
        \end{aligned}&
        &&&&&
        &
        \begin{aligned}
        \bar g 
        + g
        + \lbr{f^{-1}, \, g \, f}
        = (\alpha_0 + \alpha_3 + \alpha_4 - 2)&
        \\[1mm]
        + \, \bar \alpha_3 \, (\bar f - 1)^{-1}
        + \bar \alpha_0 \, t \, (\bar f - t)^{-1}&
        ,
        \end{aligned}
    \end{aligned}
    \end{gathered}
\end{gather}
where $\alpha_i$, $t$ are central elements of $\mathcal{R}$ and $f$, $g \in \mathcal{R}$. According to Proposition \ref{thm:firstint_qdP}, the commutator $\lbr{f^{-1}, g \, f}$ is a first integral of the \ref{eq:dPD4} system. Indeed, consider the $g$-dynamics:
\begin{align}
    \label{eq:g1}
    \bar g 
    + f^{-1} \, g \, f
    = (\alpha_0 + \alpha_3 + \alpha_4 - 2)
    + (\alpha_3 - 1) \, (\bar f - 1)^{-1}
    + (\alpha_0 - 1) \, t \, (\bar f - t)^{-1}
    .
\end{align}
Since $\lbr{\bar f, \br{\bar f + \alpha}^{\pm 1}} = 0$ for any $\alpha \in \mathcal{Z}\br{\mathcal{R}}$ (see also Corollary \ref{thm:ncrat}), it can be rewritten as
\begin{align}
    \bar f^{-1} \, \bar g \bar f
    + (f \, \bar f)^{-1} \, g \, (f \, \bar f)
    = (\alpha_0 + \alpha_3 + \alpha_4 - 2)
    + (\alpha_3 - 1) \, (\bar f - 1)^{-1}
    + (\alpha_0 - 1) \, t \, (\bar f - t)^{-1}
    .
\end{align}
Using the $f$-dynamics and the fact $[g, g] = 0$, one obtains $\br{f \, \bar f}^{-1} g \, \br{f \, \bar f} = g$. Thus, the latter takes the form
\begin{align}
    \label{eq:g2}
    \bar f^{-1} \, \bar g \bar f
    + g
    = (\alpha_0 + \alpha_3 + \alpha_4 - 2)
    + (\alpha_3 - 1) \, (\bar f - 1)^{-1}
    + (\alpha_0 - 1) \, t \, (\bar f - t)^{-1}
    .
\end{align}
The difference of \eqref{eq:g1} and \eqref{eq:g2} reads as
\begin{align}
    &&
    \bar f^{-1} \, \bar g \bar f + g
    - \bar g - f^{-1} \, g \, f
    &= 0
    &\Leftrightarrow
    &&
    T' \br{\lbr{
    f^{-1}, \, g f
    }
    }
    &= \lbr{
    f^{-1}, \, g \, f
    }
    ,
    &&
\end{align}
where $T'$ stands for the shift operator of the \ref{eq:dPD4} system. The resulting identity means that the value $I (f, g) = \lbr{f^{-1}, \, g \, f}$ is a first integral of the $T'$-map, i.e. one can set $I (f, g) = \gamma$, where $\gamma$ is an arbitrary (probably non-commutative) parameter. By using this fact, the expression $f^{-1} \, g \, f$ can be replaced with $g + \gamma$. Let us also make the transformation $\bar f \mapsto f$ in the \ref{eq:dPD4} system. Then, as a result, the \ref{eq:dPD4} is equivalent to the system given~below
\begin{align}
    \begin{multlined}
    f \, \bar{f}
    = t \, \bar{g} \, \br{
    \bar g + \bar{\alpha}_2
    }^{-1} \, \br{
    \bar g - \alpha_4
    } \, \br{
    \bar g + \alpha_1 + \bar \alpha_2
    }^{-1}
    ,
    \hspace{4cm}
    \\[2mm]
    g + \underline{g} + \gamma
    = (\alpha_0 + \alpha_3 + \alpha_4)
    + \alpha_3 \, (\underline{f} - 1)^{-1}
    + \alpha_0 \, t \, (\underline{f} - t)^{-1}
    .
    \end{multlined}
\end{align}
The correspondence between the parameters reads as follows
\begin{align}
    \alpha_0
    &= b_4 - b_1 - q,
    &
    \alpha_1
    &= b_8 - b_7,
    &
    \alpha_2
    &= b_6 - b_8 - q - 1,
    &
    \alpha_3
    &= b_3 - b_2 - q,
    &
    \alpha_4
    &= b_5 - b_6,
\end{align}
and  $\gamma = 2 q + b_1 + b_2 - b_3 - b_4 - b_5 + b_6$.

\begin{rem}
\label{rem:firstint_qdP}
Recall (see Proposition \ref{thm:firstint_qdP}) that some of the $d$-Painlevé systems obtained in \cite{bobrova2024affine} have the first integral of the form $I (f, g) = \lbr{f^{-1}, g \, f}$ (or $I (f, g) = - \lbr{g^{-1}, f \, g}$). Due to this fact, the list of the $d$-Painlevé systems can be slightly simplified by fixing the level of the first integrals. Therefore, Appendix \ref{app:dP} contains the modified list. There are systems of the form \eqref{eq:ncdsys_dtype}, which have the first integral $I (f, g) = \lbr{f, g}$. Due to Proposition~\ref{thm:degfirstint}, the~degeneration procedure preserves the degeneration of not only the systems, but the first integrals as well.
\end{rem}

\section{Open questions}
\label{sec:openq}
The current work presents an initial attempt to establish a foundation for studying non-commutative analogs of discrete Painlevé equations through surface theory. We have demonstrated its application to the non-commutative analog \ref{eq:qPA3} of the sixth $q$-Painlevé equation and hope that, in a similar way, the remaining systems can be studied as well. Despite the promising structure and results, several important questions remain open and merit further investigation. 

First and foremost, it is unclear how to apply our theory in order to obtain a non-commutative version of the master discrete Painlevé equation of elliptic type. Although such an analog is described in \cite{okounkov2015noncommutative} using derived categories and Sklyanin-type algebras, its formulation in explicit coordinated remains unknown. Once one presents a non-commutative analog for the elliptic function in explicit coordinates, this problem could be solved. 

A related and natural question concerns the classification of all non-commutative analogs of discrete Painlevé equations. Our theory requires further development to provide a systematic approach to this classification problem. Ideally, the list of non-commutative Painlevé systems derived in \cite{bobrova2023classification} should emerge from such a classification via appropriate continuous limits. 

Furthermore, additional structures---such as Lax pairs, Hamiltonians, and Poisson brackets---for the systems presented in Appendices \ref{app:qP} and \ref{app:dP} remain to be constructed and studied. In addition, extended birational representation of the lower $q$-Painlevé systems discussed in this paper should also be derived by using the geometrical approach developed here. 

We intend to address all this problems in forthcoming papers.

    \appendix
\section{\texorpdfstring{$q$}{q}-\Painleve equations}
\label{app:qP}
Here $f$, $g \in \mathcal{R}$, all constant parameters labeling by $b_i$ and $q$ are central elements. 
In all the systems, parameters $b_j$ change according to the rules below, while $q$ is a conserved quantity:
\begin{align}
    &&&&
        \bar b_i 
        &= q^2 \, b_i,
        &&&&
        i 
        = 1, 2, 5, 6,&
        &&&&&
        &
        \bar b_j 
        &= q^{-2} \, b_j,
        &&&&&
        j 
        &= 3, 4, 7, 8.
    &&&&
\end{align}

\begin{gather}
    \tag*{q-P$(A_3)$}
    \label{eq:qPA3}
    \begin{gathered}
        q
        = (b_1 \, b_2 \, b_7 \, b_8)^{\frac14} \, (b_3 \, b_4 \, b_5 \, b_6)^{- \frac14}
        ,
        \\[1mm]
    \begin{aligned}
        \begin{aligned}
        \underline{f} \, f
        = b_7 b_8 \, (g + b_6) \, &(g + b_8)^{-1} 
        \\[1mm]
        &(g + b_5) \, (g + b_7)^{-1}
        ,
        \end{aligned}&
        &&&&&
        &
        \begin{aligned}
        \bar g \, g
        = b_3 b_4 \, (f + b_2) \, &(f + b_4)^{-1}
        \\[1mm]
        &(f + b_1) \, (f + b_3)^{-1}
        .
        \end{aligned}
    \end{aligned}
    \end{gathered}
\end{gather}

\begin{flalign}
    \tag*{q-P$(A_4)$}
    \label{eq:qPA4}
    &&
    \begin{gathered}
        q 
        = (b_1 \, b_2 \, b_7)^{\frac14} \, (b_3 \, b_5 \, b_6)^{- \frac14}
        ,
        \\[1mm]
    \begin{aligned}
        \underline{f} \, f
        = b_7 \, \br{g + b_6} \, \br{g + b_5} \, \br{g + b_7}^{-1}
        ,&
        &&&&&
        &\bar g \, g
        = b_3 \, \br{f + b_2} \, \br{f + b_1} \, \br{f + b_3}^{-1}
        .
    \end{aligned}
    \end{gathered}
    &&
\end{flalign}

\begin{gather}
    \tag*{q-P$(A_5)$}
    \label{eq:qPA5}
    \begin{gathered}
        q 
        = (b_1 \, b_2 \, b_7)^{\frac14} \, (b_3 \, b_5 \, b_6)^{- \frac14}
        ,
        \\[1mm]
    \begin{aligned}
        \underline{f} \, f
        = b_6 b_7 \, \br{g + b_5} \, \br{g + b_7}^{-1}
        ,&
        &&&&&
        &\bar g \, g
        = b_3 \, \br{f + b_2} \, \br{f + b_1} \, f^{-1}
        .
    \end{aligned}
    \end{gathered}
\end{gather}

\begin{gather}
    \tag*{q-P$(A_5)'$}
    \label{eq:qPA5'}
    \begin{gathered}
        q 
        = (b_2 \, b_7)^{\frac14} \, (b_3 \, b_6)^{- \frac14}
        ,
        \\[1mm]
    \begin{aligned}
        \underline{f} \, f
        = b_7 \, \br{g + b_6} \, g \, \br{g + b_7}^{-1}
        ,&
        &&&&&
        &\bar g \, g
        = b_3 \, \br{f + b_2} \, f \, \br{f + b_3}^{-1}
        .
    \end{aligned}
    \end{gathered}
\end{gather}

\begin{gather}
    \tag*{q-P$(A_6)$}
    \label{eq:qPA6}
    \begin{gathered}
        q 
        = (b_2 \, b_7)^{\frac14} \, (b_3 \, b_6)^{- \frac14}
        ,
        \\
    \begin{aligned}
        \underline{f} \, f
        = b_6 b_7 \, g \, \br{g + b_7}^{-1}
        ,&
        &&&&&
        &\bar g \, g
        = b_3 \, \br{f + b_2}
        .
    \end{aligned}
    \end{gathered}
\end{gather}

\begin{gather}
    \tag*{q-P$(A_6)'$}
    \label{eq:qPA6'}
    \begin{gathered}
        q 
        = \br{b_6 \, b_7}^{\frac14}
        ,
        \\[1mm]
    \begin{aligned}
        \underline{f} \, f
        = b_7 \, \br{g + b_6} \, g \, \br{g + b_7}^{-1}
        ,&
        &&&&&
        &\bar g \, g
        = b_3 \, f^2 \, \br{f + b_3}^{-1}
        .
    \end{aligned}
    \end{gathered}
\end{gather}

\begin{gather}
    \tag*{q-P$(A_7)$}
    \label{eq:qPA7}
    \begin{gathered}
        q 
        = \br{b_6 \, b_7}^{\frac14}
        ,
        \\[1mm]
    \begin{aligned}
        \underline{f} \, f
        = b_7 \, \br{g + b_6} \, g \, 
        \br{g + b_7}^{-1}
        ,&
        &&&&&
        &\bar g \, g
        = f^2.
    \end{aligned}
    \end{gathered}
\end{gather}

\begin{gather}
    \tag*{q-P$(A_7)'$}
    \label{eq:qPA7'}
    \begin{gathered}
        q 
        = \br{b_6 \, b_7}^{\frac14}
        ,
        \\[1mm]
    \begin{aligned}
        \underline{f} \, f
        = b_6 b_7 \, g \, \br{g + b_7}^{-1}
        ,&
        &&&&&
        &\bar g \, g
        = b_3 \, f.
    \end{aligned}
    \end{gathered}
\end{gather}
\section{\texorpdfstring{$d$}{d}-\Painleve equations}
\label{app:dP}
Here $f$, $g \in \mathcal{R}$ and all constant parameters labeling by greek letters and $t$ are from the center $\mathcal{Z}(\mathcal{R})$ of $\mathcal{R}$ (see details in \cite{bobrova2024affine}). This list is simplified by using the first integrals described in Propositions~\ref{thm:firstint_dcase} and \ref{thm:firstint_qdP}. 

\begin{gather}
    \tag*{d-P$(D_4)$}
    \label{eq:dPD4}
    \begin{gathered}
    \begin{aligned}
        \begin{aligned}
        \bar \alpha_0 
        &= \alpha_0 - 1,
        &&&&
        \end{aligned} 
        \bar \alpha_2 
        = \alpha_2 + 1,&
        &&&&&
        &
        \bar \alpha_3 
        = \alpha_3 - 1,
        \\
        \begin{aligned}
        f \, \bar f
        = t \, g \, (g + \alpha_2)^{-1} \,
        (g - \alpha_4) \, (g + \alpha_1 + \alpha_2)^{-1},
        \\[1mm]
        \phantom{f}
        \end{aligned}&
        &&&&&
        &
        \begin{aligned}
        \bar g 
        + g
        = (\bar \alpha_0 &+ \, \bar \alpha_3 + \alpha_4)
        \\[1mm]
        &+ \, \bar \alpha_3 \, (\bar f - 1)^{-1}
        + \bar \alpha_0 \, t \, (\bar f - t)^{-1}
        .
        \end{aligned}
    \end{aligned}
    \end{gathered}
\end{gather}

\begin{flalign}
    \tag*{d-P$(D_5)$}
    \label{eq:dPD5}
    &&
    \begin{gathered}
    \begin{aligned}
        \begin{aligned}
        \bar \alpha_0 
        &= \alpha_0 + 1,
        &&&&
        \end{aligned} 
        \bar \alpha_1 
        = \alpha_1- 1,&
        &&&&&
        &
        \begin{aligned}
        \bar \alpha_2 
        &= \alpha_2 + 1,
        &&&&&
        \bar \alpha_3 
        &= \alpha_3 - 1,
        \end{aligned}
        \\
        \bar f + f 
        = 1 - \alpha_2 g^{-1} - \alpha_0 (g + t)^{-1},&
        &&&&&
        &\bar g + g
        = - t + \bar \alpha_1 \bar f^{-1} + \bar \alpha_3 (\bar f - 1)^{-1}.
    \end{aligned}
    \end{gathered}
    &&
\end{flalign}

\begin{gather}
    \tag*{d-P$(D_5)'$}
    \label{eq:dPD5'}
    \begin{gathered}
    \begin{aligned}
        \bar \alpha_2 
        = \alpha_2 - 1,&
        &&&&&
        &\bar \alpha_3 
        = \alpha_3 + 1,
        \\
        \bar f + f 
        = - (\alpha_0 + \alpha_2) - t g - \alpha_2 (g - 1)^{-1},&
        &&&&&
        &g \, \bar g
        = - t^{-1} \bar f (\bar f + \alpha_0) (\bar f - \alpha_3)^{-1}.
    \end{aligned}
    \end{gathered}
\end{gather}

\begin{gather}
    \tag*{d-P$(D_6)$}
    \label{eq:dPD6}
    \begin{gathered}
    \begin{aligned}
        \bar \alpha_1
        = \alpha_1,&
        &&&&&
        &
        \bar \beta_1 
        = \beta_1 - 1,
        \\
        \bar f \, f
        = t + \bar \beta_1 \, t \, \bar g^{-1},&
        &&&&&
        &\bar g
        + g
        = \alpha_1 - \beta_1 + f + t f^{-1}.
    \end{aligned}
    \end{gathered}
\end{gather}

\begin{gather}
    \tag*{d-P$(D_6)'$}
    \label{eq:dPD6'}
    \begin{gathered}
    \begin{aligned}
        \begin{aligned}
        \bar \alpha_0 
        &= \alpha_0 - 1,
        &&&&
        \end{aligned} 
        \bar \alpha_1 
        = \alpha_1 + 1,&
        &&&&&
        &
        \begin{aligned}
        \bar \beta_0 
        &= \beta_0 - 1,
        &&&&&
        \bar \beta_1 
        &= \beta_1 + 1,
        \end{aligned}
        \\
        \bar f + f
        = - \alpha_1 g^{-1} + \beta_1 (1 - g)^{-1},&
        &&&&&
        &\bar g
        + g
        = 1 - (\bar \alpha_1 + \beta_1) \bar f^{-1} - t \bar f^{-2}.
    \end{aligned}
    \end{gathered}
\end{gather}

\begin{gather}
    \tag*{d-P$(D_7)$}
    \label{eq:dPD7}
    \begin{gathered}
    \begin{aligned}
        \bar \alpha_0 
        = \alpha_0 - 1,&
        &&&&&
        &\bar \alpha_1 
        = \alpha_1 + 1,
        \\
        \bar f + f 
        = - \alpha_1 - t g^{-1},&
        &&&&&
        &\bar g \, g
        = t \bar f.
    \end{aligned}
    \end{gathered} 
\end{gather}

\begin{gather}
    \tag*{d-P$(E_6)$}
    \label{eq:dPE6}
    \begin{gathered}
    \begin{aligned}
        \bar \alpha_1 
        = \alpha_1 + 1,&
        &&&&&
        &\bar \alpha_2 
        = \alpha_2 - 1,
        \\
        \bar f
        + f 
        = - t + \bar g - \bar \alpha_2 {\bar{g}}^{-1},&
        &&&&&
        &\bar g
        + g 
        = t + f + \alpha_1 f^{-1}.
    \end{aligned}
    \end{gathered}
\end{gather}

\begin{gather}
    \tag*{d-P$(E_6)'$}
    \label{eq:dPE6'}
    \begin{gathered}
    \begin{aligned}
        \bar \alpha_0 
        = \alpha_0 - 1,&
        &&&&&
        &\bar \alpha_2 
        = \alpha_2 + 1,
        \\
        \bar f \, f
        = - (\bar g - \alpha_1) (\bar g + \alpha_2)^{-1} \bar g,&
        &&&&&
        &\bar g
        + g
        = \alpha_1
        + f t + f^2.
    \end{aligned}
    \end{gathered}
\end{gather}

\begin{gather}
    \tag*{d-P$(E_7)$}
    \label{eq:dPE7}
    \begin{gathered}
    \begin{aligned}
        \bar \alpha_0 
        = \alpha_0 - 1,&
        &&&&&
        &\bar \alpha_1 
        = \alpha_1 + 1,
        \\
        \bar f
        + f 
        = - \alpha_1 g^{-1},&
        &&&&&
        &\bar g
        + g 
        = t + 2 {\bar{f}}^2.
    \end{aligned}
    \end{gathered}
\end{gather}
    
    \bibliographystyle{alpha}
    \bibliography{bib}
\end{document}